\documentclass{aa}

\usepackage{microtype}
\usepackage{amsmath}
\usepackage{color}
\usepackage{graphicx}
\usepackage{hyperref}
\usepackage[varg]{txfonts}
\usepackage{caption}
\usepackage{subcaption}
\usepackage{pifont}
\usepackage{ulem}
\usepackage{orcidlink}

\usepackage{natbib}
\bibpunct{(}{)}{;}{a}{}{,} 
\usepackage{adjustbox}
\usepackage{xargs}                      
\usepackage{enumitem}

\graphicspath{ {images/} }

\newcommand{\range}{\textrm{--}}

\let\orgautoref\autoref
\renewcommand{\autoref}
        {\def\equationautorefname{Eq.}%
         \def\figureautorefname{Fig.}%
         \def\subfigureautorefname{Fig.}%
         \def\sectionautorefname{Sect.}%
         \def\subsectionautorefname{Sect.}%
         \def\subsubsectionautorefname{Sect.}%
         \def\Itemautorefname{item}%
         \def\tableautorefname{Table}%
         \orgautoref}

\begin{document}

   \title{A new understanding of the Gemini-Monoceros X-ray enhancement from discoveries with eROSITA}

   \subtitle{}

   \author{J. R. Knies\inst{1} \and M. Sasaki\inst{1}\orcidlink{0000-0001-5302-1866} \and W. Becker\inst{2,3}\orcidlink{0000-0003-1173-6964}
   \and T. Liu\inst{2}\orcidlink{0000-0002-2941-6734} \and G. Ponti\inst{2,4}\orcidlink{0000-0003-0293-3608} \and P. P. Plucinsky\inst{5}\orcidlink{0000-0003-1415-5823}}

   \institute{Dr. Karl Remeis Observatory and ECAP, Universit\"at Erlangen-N\"urnberg, Sternwartstra{\ss}e 7, D-96049 Bamberg, Germany\label{inst1}
    \and
    Max-Planck-Institut f\"ur extraterrestrische Physik, Gie{\ss}enbachstraße 1, D-85748 Garching, Germany\label{inst2}
     \and
    Max-Planck-Institut f\"ur Radioastronomie, Auf dem H\"ugel 69, 53121 Bonn, Germany, \label{inst3} 
     \and
   INAF-Osservatorio Astronomico di Brera, Via E. Bianchi, 46, 23807 Merate (LC), Italy\label{inst4}
   	 \and
	Harvard-Smithsonian 
Center for Astrophysics, 60 Garden Street, Cambridge, MA 02138 USA \label{inst5}
}

\date{Received January 2024. Accepted 2nd April 2024.}

 \abstract
   {}
   {The Gemini-Monoceros X-ray enhancement is a rich field for studying diffuse X-ray emission and supernova remnants (SNRs). Most SNRs in this part of the sky are notoriously difficult to observe due to their large extent. With the launch of the extended ROentgen Survey with an Imaging Telescope Array (eROSITA) on board the Spektrum-R\"ontgen-Gamma (\textit{SRG}) platform in 2019, we are now able to fully study those objects for the first time with CCD resolution. Many of the SNRs in the vicinity are suspected to be very old remnants, which are severely understudied in X-rays due to numerous observational challenges. In addition, the identification of new faint large SNRs might help to solve the long-standing discrepancy between the observed and expected number of Galactic SNRs. 
  }
   {
We performed a detailed X-ray spectral analysis of the entire Gemini-Monoceros X-ray enhancement and a detailed background analysis of the vicinity, which allowed us to model the background with a high precision inside the X-ray enhancement. 
   We also made use of multiwavelength data to better understand the morphology and to constrain the distances to the different sources. Based on the spectral analysis, we estimated the properties of the sources and calculated a grid of model SNRs to determine the individual SNR properties.}
   {Most of the diffuse plasma of the Monogem Ring SNR is well described by a single nonequilibrium ionization (NEI) component with an average temperature of $kT = 0.14\pm 0.03$\,keV. We obtain an age of $\approx 1.2\cdot 10^5$\,yr - consistent with PSR B0656+14 - for the Monogem Ring and an explosion energy typical for a core-collapse (CC) supernova (SN).
   In the southeast, we found evidence for a significant temperature enhancement and a second plasma component. Our findings show that a scenario of two SNRs at $\approx 300\,$pc is likely, with the new candidate having an age of $\approx 50,000\,$yr.
	 We were also able to improve on previous results for the Monoceros Loop and PKS 0646+06 SNRs by disentangling the foreground diffuse emission of the Monogem Ring SNR. We obtained significantly higher temperatures than previous studies, and for PKS 0646+06 a much lower estimated age of the SNR. 
	  We also found a new SNR candidate G190.4+12.5 which most likely is located at $D > 1.5\,$kpc, expanding into a low density medium at a high distance from the Galactic plane, with an estimated age of $40,000-60,000$\,yr.}
   {}


	\titlerunning{The Gemini-Monoceros X-ray enhancement with eROSITA}
	\authorrunning{Knies et al.}

  \maketitle

\section{Introduction}
The Gemini-Monoceros X-ray enhancement is a field rich in diffuse X-ray sources \citep{introduction_nousek}. Several supernova remnants (SNRs) have been found and studied in the past in this field. The first detection of X-rays from the region was already made as early as with rocket-flight missions \citep{introduction_brunner}. The most prominent source is perhaps the Monogem Ring, a circular X-ray structure with a very large extent of $\sim 25^{\circ}$ centered around $(l, b) \sim (200^{\circ}, 8^{\circ})$ \citep{introduction_nousek}. The first detailed analysis was performed using the R\"ontgensatellit (ROSAT) all-sky survey, where they found low plasma temperatures of $kT = 0.1-0.3\,$keV, and - assuming an SNR in the Sedov expansion phase - an explosion energy of $\sim 0.2\cdot 10^{50}$\,erg, and a relatively high age of $\sim 86,000$\,yr \citep{mg_rosat}. The hypothesis that the Monogem Ring is indeed an old nearby SNR was further established with precise distance measurements to the compact object PSR B0656+14, located close to the approximate center of the SNR at $D \approx 300\,$pc \citep{distance_psr, pulsar_age}. Additionally, optical filaments were found, which are possibly associated with the remnant \citep{optical_filaments_mg}. Toward the Galactic plane, the SNR's shock appears to interact strongly with colder phases of the interstellar medium (ISM), as shown by far-UV observations by \citet{kim_uv}. They also found a ring-like structure in H$\alpha$ anticorrelated with the X-ray emission, which they dubbed the ``Gemini H$\alpha$ ring.'' In our previous study, we showed that this ring was most likely created by a small cluster of massive stars at $D \approx 300\,$pc \citep{Knies2022}. We also studied three regions of the Monogen Ring in X-rays with Suzaku observations, confirming previous results for most parts \citep{mg_paper_2018}. Outside of the Monogem Ring, the pair of two SNRs IC443 and G189.6+03.3 is found, and was recently studied with extended ROentgen Survey with an Imaging Telescope Array \citep[eROSITA, ][]{ic443_erosita}.

Another large SNR was found inside the Monogem Ring in projection at $(l, b) \sim (205.0^{\circ}, 0.5^{\circ})$ with an extent of $\sim 3.5^{\circ}$, named the Monoceros Loop or G205.5+0.5 \citep{einstein_monoceros_loop}. Using \textit{Einstein} Image Proportional Counter (IPC) X-ray data, they found an SNR that apparently exploded in a low-density environment, leading to the age of $\sim 30,000$\,yr and an explosion energy of $\sim 3\cdot 10^{50}$\,erg, despite the large extent. They found a distance of $1.5\,$kpc to be likely, which also suggests that the remnant is located close to stellar cluster NGC 2244 and the Rosette nebula. A radio study by \citet{MC_radio} did indeed find signatures for interactions with the nebula, which further narrows down the distance to the remnant. Nearby, another more distant SNR was found: G206.9+2.3 or PKS 0646+06. Also observed with the \textit{Einstein}  IPC, \citet{pks_einstein} found a very low plasma temperature of only $\sim 0.14\,$keV, and the source to be a distant ($3-11\,$kpc) old remnant with an age of $\sim 60,000\,$yr. The remnant also appears to be emitting in the GeV energy range \citep{pks_gamma}. Earlier radio studies  constrained the distance of G206.9+2.3 to $3-5\,$kpc \citep{MC_distance}. However, X-ray studies of the Monoceros Loop, PKS 0646+06, and the Gemini-Monoceros X-ray enhancement so far either lacked spectral resolution at full coverage, as for the ROSAT data, or coverage at a good spectral resolution, as for the Suzaku data. 

What the numerous studies of sources inside the Gemini-Monoceros X-ray enhancement showcase is that for studying SNRs, this field is truly a treasure trove. This is especially true for studying the evolution of very old remnants, which is usually very difficult at higher distances due to absorption and low surface brightness. Moreover, these faint large SNRs might contribute to the issue of missing SNRs in our Galaxy, compared with model predictions \citep[e.g.,][]{snr_density_thor, snr_density_leahy}. Previous studies in the radio discovered many additional SNRs, but still left a factor of two unaccounted for \citep{snr_density_thor}. Therefore, we need high-quality all-sky (soft) X-ray data to trace these objects, too faint to be discovered in the radio.
With the launch of eROSITA in 2019 and the consequent four all sky surveys (eRASS), we now have the unique opportunity to again study this rich field with unparalleled spectral resolution and coverage \citep{erosita_instrument}.

Our paper is divided into the following sections: In \autoref{sec:data} we introduce the data we used and in \autoref{sec:xray_reduction} we explain the X-ray data reduction procedure. The X-ray and multiwavelength morphology is discussed in \autoref{sec:morphology}, followed by the background analysis and X-ray spectral model in \autoref{sec:background}. In \autoref{sec:spectral_analysis} we present the spectral analysis of the SNRs; in \autoref{sec:properties} we discuss the implications from our results of the spectral analysis and morphology, and compare our findings to SNR model calculations. Finally, in \autoref{sec:summary} we summarize our work.
%
\section{Data}
\label{sec:data}
\subsection{X-ray}
We used X-ray data obtained with 
eROSITA, launched in July 2019 as part of the Spectrum-Roentgen-Gamma (\textit{SRG}) mission \citep{erosita_instrument}. The data were observed from January 2020 to December 2021 during the first four all-sky surveys \citep[eRASS 1-4, the total also called eRASS:4,][]{merloni_catalog}. The data cover the energy range of $\sim 0.2\range 10$~keV with an
energy resolution of $\sim 80$~eV at $1.49$~keV, and a field of view (FoV) averaged spatial resolution of $\sim 26$\arcsec.
We used the data processed with the \texttt{020} pipeline version, which has improved calibration, especially at lower energies, compared to previous processing versions. For the data reduction and analysis, we used the \texttt{eSASS} software\footnote{\url{https://erosita.mpe.mpg.de/edr/DataAnalysis/}} version eSASSuser2112104 \citep{brunner_esass} with the most recent (2021Q4) calibration database available at the time of the data reduction \citep[\texttt{CalDB},][]{esass}.

\subsection{Optical}
To compare the X-ray emission with other wavelengths, we used the H$\alpha$ maps by \citet{halpha_finkbeiner} that cover most of our field with a resolution of $6$\arcmin.
The map was created by combining the Virginia Tech Spectral line Survey \citep[VTSS,][]{VTSS}, Southern H-Alpha Sky Survey \citep[SHASSA,][]{SHASSA}, and the Wisconsin H-Alpha Mapper \citep[WHAM,][]{WHAM}.

We also used the dust absorption $A_v$ data cubes obtained by \citet{dust_lallement}. The data cubes map the interstellar absorption in 3D in a 6 kpc $\times$ 6 kpc $\times$ 0.8 kpc volume around the Sun, derived from the Gaia EDR3 and 2MASS photometric data.



\section{X-ray data analysis}
\label{sec:xray_reduction}
\label{sec:xray_data_reduction}

\subsection{Images}
We first merged all events of the sky tiles covering the Gemini-Monoceros X-ray enhancement and its vicinity. All available data from eRASS1 to eRASS4 were used (eRASS:4 hereafter) and processed with \texttt{evtool}. In total our data include 160 sky tiles. During merging, we applied the strict flag \texttt{0xc00f7f30} to remove bad events\footnote{\url{https://erosita.mpe.mpg.de/edr/DataAnalysis/prod_descript/EventFiles_edr.html\#First_Sub_Point_312}}. We had to de-selected 
all events that were recorded between the times $6.87312\cdot 10^8$\,s and $6.87357\cdot 10^8$\,s (MJD) due to a solar flare, which enhanced the background in a narrow stripe across the field eROSITA
was scanning. We accepted all event patterns from singles to quads with pattern $= 15$.

There are also bad pixels, which are not in the calibration database (\texttt{CalDB}) that have limited impact or are limited to certain specific times. We manually filtered out the events on these pixels to reduce any possible contamination. This additional non-standard filtering reduced the photon statistics by about $0.3\%$. Since our sources of interest are very faint compared to the background, we further investigated the effects by creating images with and without filtering, and compared those. The filtered pixels appear uniformly distributed across the Monogem Ring SNR, with enhancements following the real X-ray emission. Therefore, real events were also filtered out by this procedure, while suspicious structures like stripes in the scanning direction suggesting a time variable effect of the detectors were not found. Therefore we decided to keep those events. Internal tests performed by the eROSITA team showed that contamination by those pixels were constrained in very small stripe-like regions over the sky, showing no spatial overlap with our data\footnote{Private communication, Michael Yeung}.

Next, we created images for three different energy bands with \texttt{evtool}: 0.2-0.4, 0.4-0.8, and 0.8-1.25\,keV. These soft bands were chosen to highlight differences in the relatively soft X-ray emission in the field. We also created exposure maps for each of the bands by using \texttt{expmap}. The vignetting corrected exposure time of the data across the analyzed field after filtering is on average $337 \pm 77$\,s ($1\sigma$). Finally, we created exposure-corrected images by dividing the images with the vignetting corrected exposure maps and removing artifacts from pixel values below zero.
\subsection{Point-source removal}

In order to reduce the contamination from point sources, we used the internal combined eRASS:4 point source catalog to obtain a list of point source candidates. The detection criteria were similar to the eRASS1 catalog, which was published with the first eROSITA data release \citep[3B in][]{merloni_catalog}.
We filtered the catalog with a minimum detection likelihood of 20 in the full $0.2-5.0$\,keV energy band. For each point source candidate, we used a fairly large fixed radius of $46\arcsec$ to fully remove contaminating emission. 
In a second step, we carefully checked by hand if any additional contamination was still left, after applying a cheese mask based on the point source list. As a result, several extended fore-/background sources, where the fixed radius was inappropriate, as well as missed point-like sources, were excluded to minimize possible contamination of other sources.
\subsection{Spectra}
We extracted spectra as well as response files for each analysis region using \texttt{srctool}. We used the combined, filtered eRASS:4 eventfile, additionally applying the \texttt{FLAREGTI} GTI filtering. For the response files, we assumed the point-spread function (PSF) to be a delta function, and a top-hat profile with an initial extent value equal to the approximate radius of the region for the source emission. The accuracy of the response file sampling was chosen as $2$\arcmin\ for the statistically derived region, and varying appropriate sampling size for manually defined regions. Due to the large region sizes used in our analysis, the coarser sampling did not impact the quality of the spectra, while significantly reducing computation times.
For the extraction of the events in the respective region we created and applied masks. Point-source candidates have been also excluded with these masks, as described above.
We did not subtract background spectra from the source, instead we extracted background spectra in the vicinity of the Monogem Ring, as explained in detail below. 

\section{Morphology}
\label{sec:morphology}
\begin{figure*}
	\centering
		\includegraphics[width=0.99\textwidth]{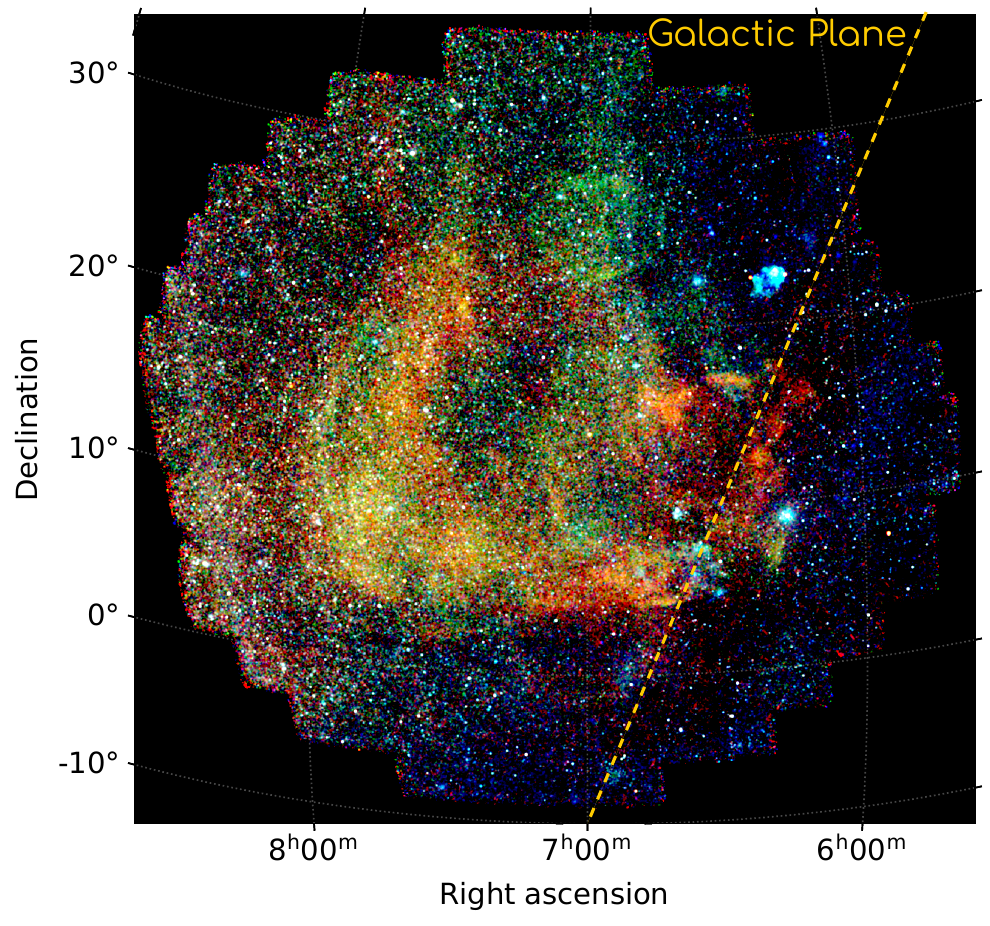}
		\caption{\label{fig:xray_rgb_mosaic}eROSITA X-ray three-color image of the Gemini-Monoceros X-ray enhancement. The energy bands are: 0.2-0.4 keV (red), 0.4-0.8 keV (green), and 0.8-1.25 (blue). The image was created from the combined eRASS:4 data.}
\end{figure*}

\begin{figure*}
\sidecaption
		\includegraphics[width=12cm]{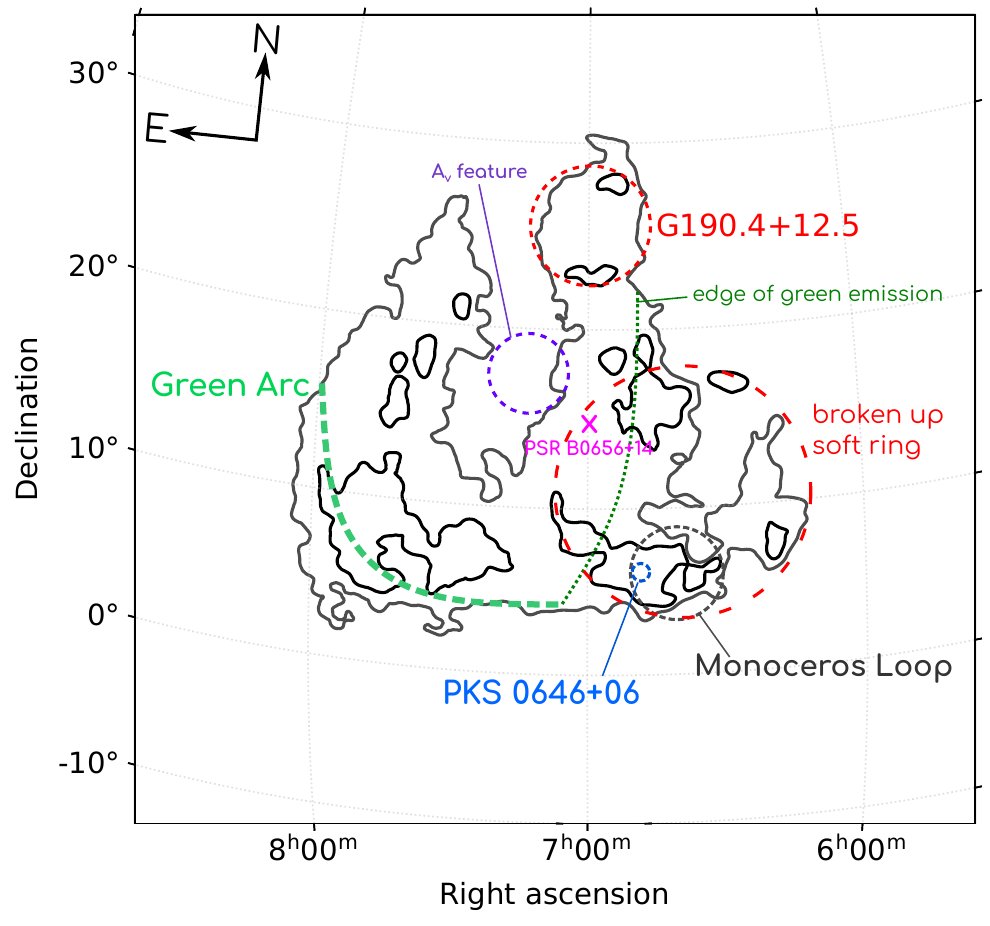}
		\caption{X-ray finding chart based on the eROSITA data. We used the data in the 0.2-1.25\,keV energy range, showing the same frame as \autoref{fig:xray_rgb_mosaic}. Prominent sources and regions mentioned in the text are highlighted for better orientation.}
		\label{fig:finding_chart}
\end{figure*}

\begin{figure}
	\centering
		\includegraphics[width=0.49\textwidth]{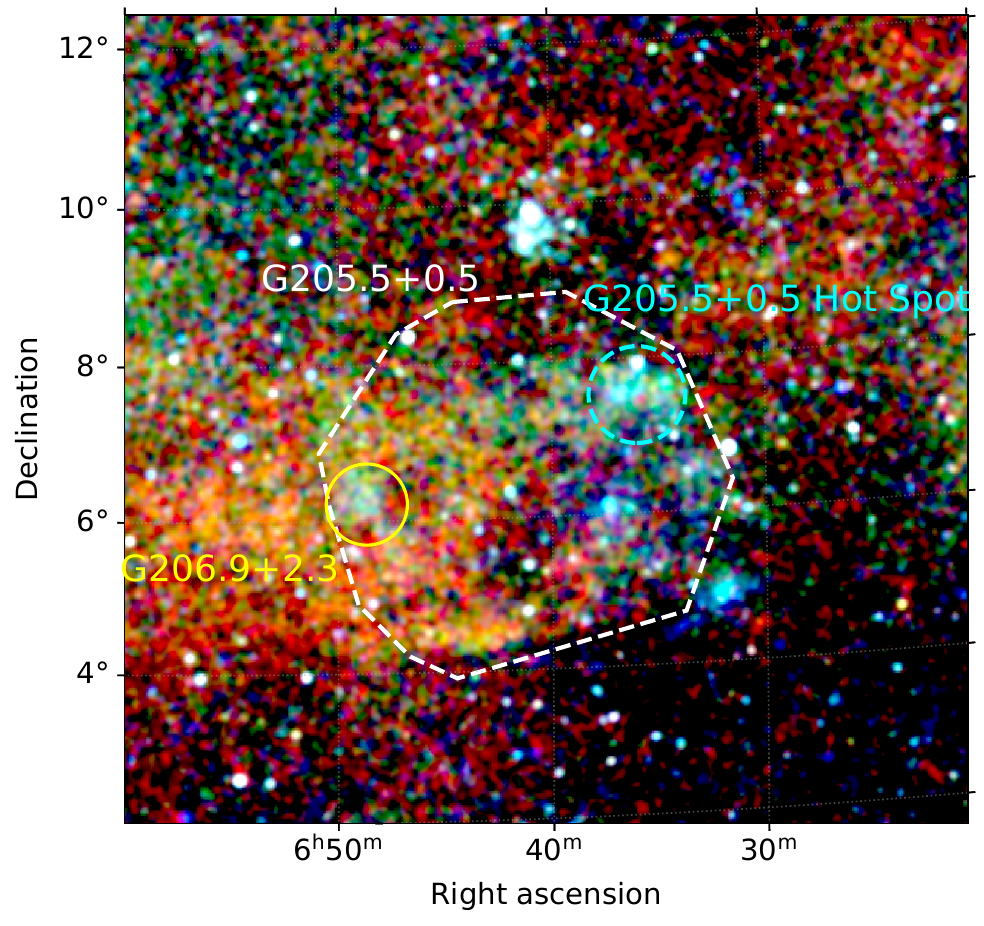}
		\caption{\label{fig:MC_rgb_detail}Detailed view of the vicinity of G205+0.5 (Monoceros Loop SNR). The energy bands and data are similar to \autoref{fig:xray_rgb_mosaic}. The extent of the Monoceros Loop is shown with a dashed white region, while the hot spot is shown in cyan. PKS 0646+06 (G206.9+2.3) is shown in yellow \citep{snr_catalog_green}.}
\end{figure}

\paragraph{X-rays:}
The three-color large-scale image of the Gemini-Monoceros X-ray enhancement shown in \autoref{fig:xray_rgb_mosaic} reveals a highly complex structure of the diffuse emission. To the east (left-handed) side we observe a half ring-like emission with a radius of $\sim 10^{\circ}$ (green arc, \autoref{fig:finding_chart}). The emission appears to be harder toward to southeast (i.e., more green), with the softer (red) emission more pronounced toward the center, west, and north. The green emission appears almost ring-like, while the soft emission seems to have a more complex morphology. To the west (right) we observe a very soft emission with a broken-up structure in most parts (broken up soft ring, \autoref{fig:finding_chart}).
 The center of the diffuse emission is significantly darker, which could indicate limb-brightening of the deformed expanding shell of the SNR. We measure an increase in brightness by 50\% when comparing the darkest central region with the shell.
 In the southwest, we find the more distant Monoceros Loop and PKS 0646+06 SNRs as shown in  \autoref{fig:MC_rgb_detail}, accompanied by harder X-ray emission compared to the surroundings (gray and blue circle, \autoref{fig:finding_chart}). The soft (red) emission of the Monogem Ring SNR and the more distant SNRs appear to be overlapping in projection. Toward the north, we observe an almost perfect circular diffuse emission structure, only visible in the medium band (G190.4+12.5 hereafter, \autoref{fig:finding_chart}).


\begin{figure}
	\centering
		\includegraphics[width=0.49\textwidth]{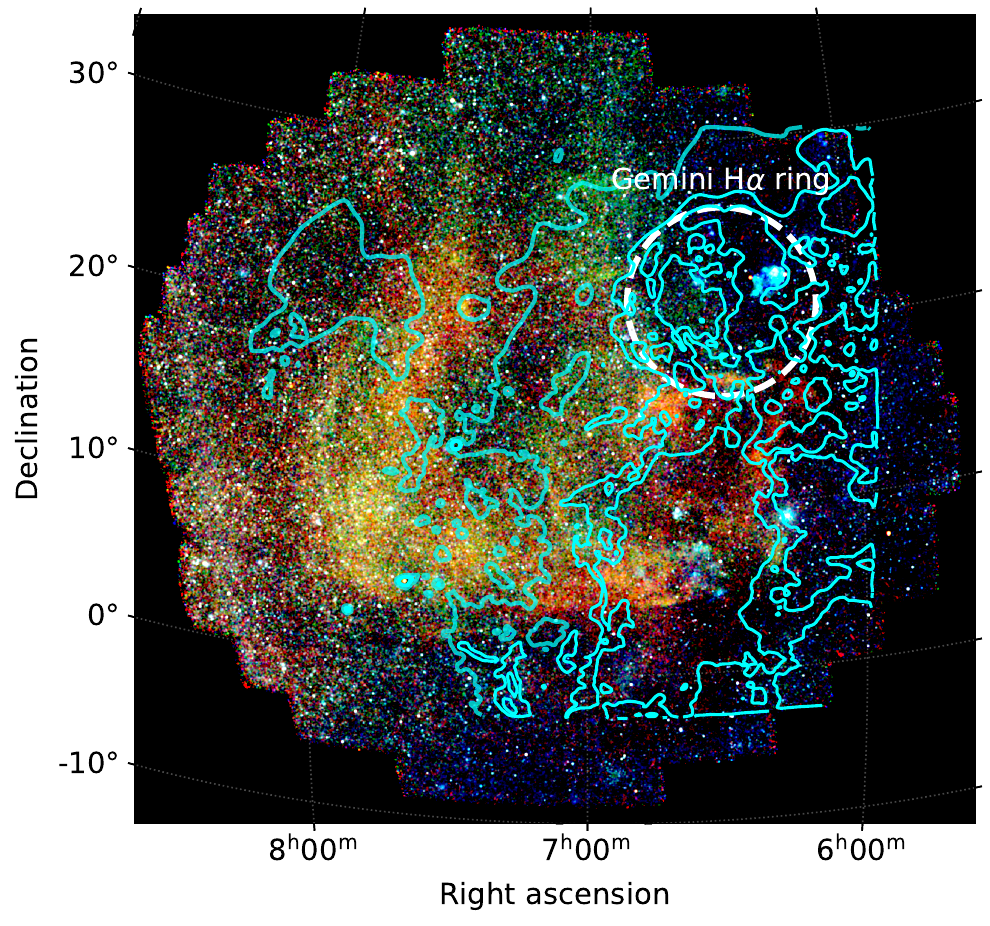}
		\caption{\label{fig:xray_rgb_mosaic_halpha}Three-color X-ray mosaic as in \autoref{fig:xray_rgb_mosaic} with H$\alpha$ contours overlaid in cyan \citep{halpha_finkbeiner}. The lowest contour level has a lower transparency for better visibility. The contour levels are 3.86, 8.11, and 12.35 Rayleighs, corresponding to the 40\%-82.5\% percentiles of the H$\alpha$ image.}
\end{figure}

\begin{figure*}
	\centering
	\begin{subfigure}[t]{0.33\textwidth}
		\includegraphics[width=1.0\textwidth]{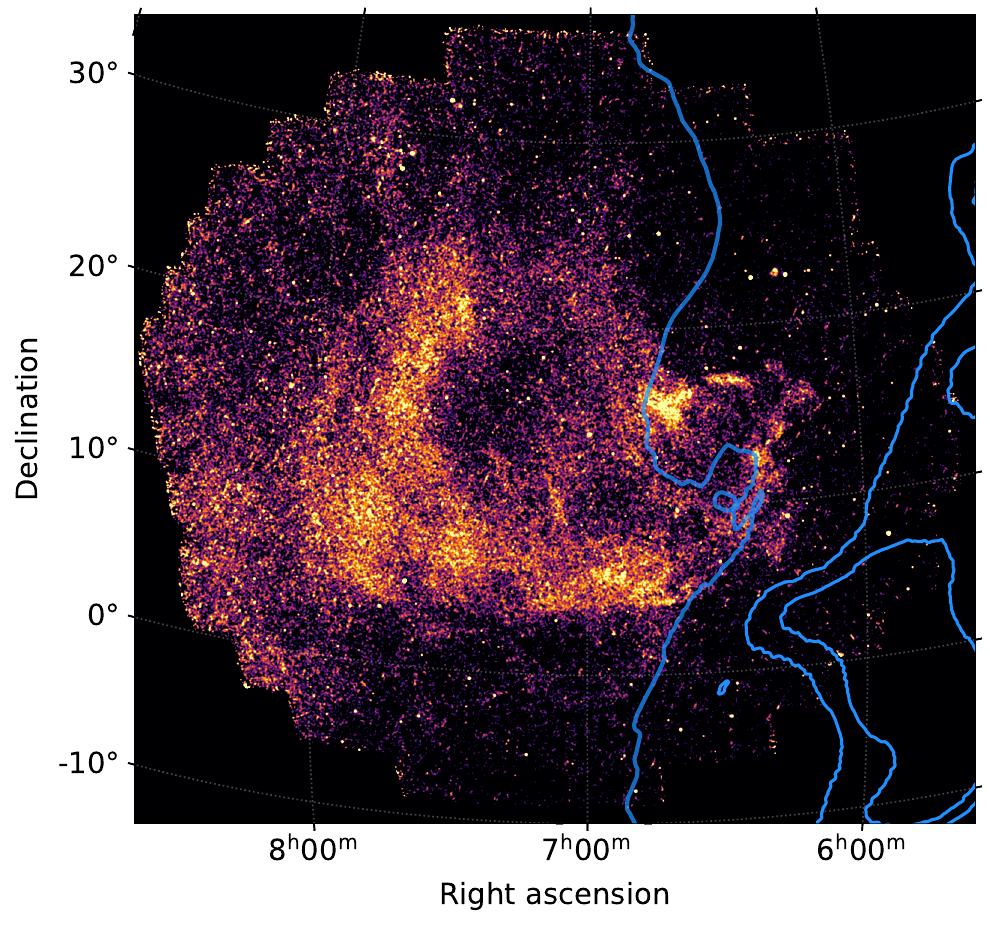}
		\caption{\label{fig:soft_gal_dust_300-600pc}}
\end{subfigure}
\hfill
\begin{subfigure}[t]{0.33\textwidth}
	\centering
		\includegraphics[width=1.0\textwidth]{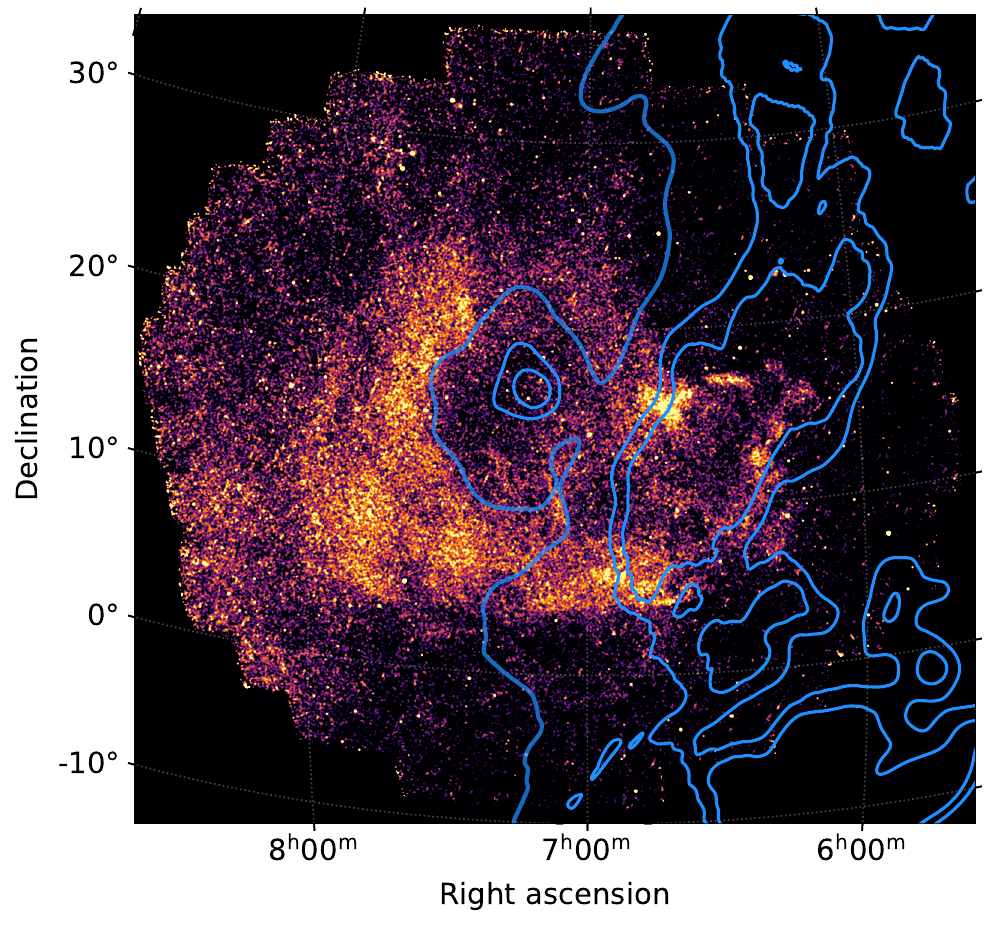}
		\caption{\label{fig:soft_gal_dust_600-900pc}}
		 \end{subfigure}
		 \hfill
		\begin{subfigure}[t]{0.33\textwidth}
		\includegraphics[width=1.0\textwidth]{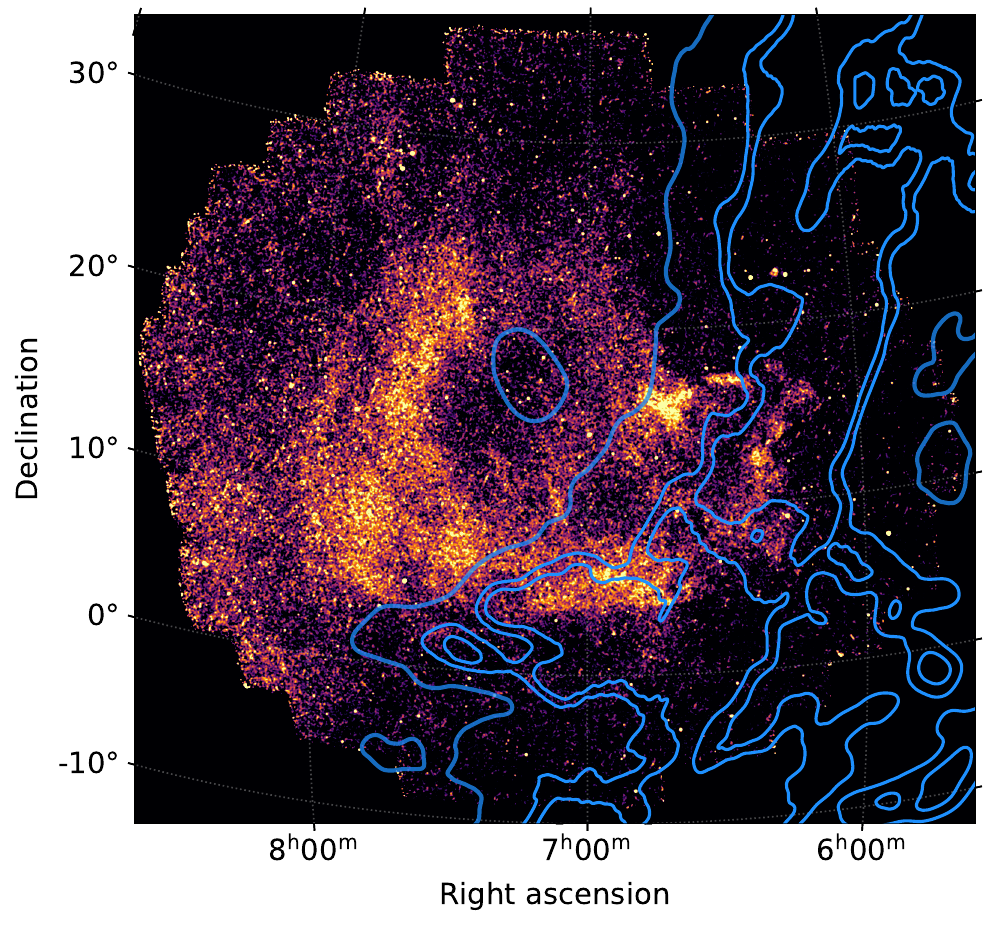}
		\caption{\label{fig:soft_gal_dust_900-1120pc}}
\end{subfigure}
\caption{Soft X-ray image in the energy range 0.2-0.4 keV, overlaid with distance-integrated slices of the $A_v$ absorption derived by \citet{dust_lallement}.
The images show $A_v$ contours for 300-600\,pc (a), 600-900\,pc (b), and 900-1120\,pc (c). The contours correspond to the 65\%-95\% percentiles of the $A_v$ data, linearly spaced.\label{fig:soft_xrays_dust_3d}}
\end{figure*}

\paragraph{H$\alpha$:}
An overlay between X-ray image and H$\alpha$ contours are shown in \autoref{fig:xray_rgb_mosaic_halpha}. Toward the Galactic Plane (right), we observe strong H$\alpha$ emission with a complex morphology. 
The Gemini H$\alpha$ ring is located toward the north in the H$\alpha$ emission \citep{kim_uv}. This structure appears to be almost perfectly anticorrelated with the soft X-ray emission. The Gemini H$\alpha$ ring (white dashed ring, \autoref{fig:xray_rgb_mosaic_halpha}) is believed to be interacting with the Monogem Ring SNR and caused by massive stars, which existed at similar distances as the Monogem Ring SNR, as shown in our previous studies \citep{mg_paper_2018, Knies2022} and the UV study by \citet{kim_uv}. The H$\alpha$ emission toward the south and west appears to be slightly weaker while also not showing any clear (anti)correlation with the X-rays. Therefore, these parts of the  H$\alpha$ emission are most likely unrelated with the diffuse X-ray emission.
\paragraph{Dust:}
From the $A_v$ 3D cubes by \citet{dust_lallement} we created contours, which we overlay on the soft X-ray emission in \autoref{fig:soft_xrays_dust_3d}. We integrated the cube along the line of sight for three distance slices ranging from 300 to 1120\,pc. We also integrated below 300\,pc but observed no interesting feature.

In the distance slice 300-600\,pc we observe a small overlap in the south without any clear anticorrelation with X-rays. In the next slice $600-900$\,pc we find an interesting absorption feature right in the center of the remnant, which almost perfectly matches a depression in X-ray intensity (edge of green emission, \autoref{fig:finding_chart}). Also in the south toward the Galactic plane, there  appears to be a significant absorption; however, here we do not observe any anticorrelation with X-rays. In the last slice $900-1120$\,pc, we observe no (anti)correlation whatsoever, despite significant $A_v$ absorption in the lower half of the shown frame. An in-depth discussion of the implications is given in \autoref{sec:distance_constrain}.

%

%
%
\section{Study of the astrophysical X-ray background}
\label{sec:background}
\subsection{Aim}
Estimating an appropriate X-ray background for diffuse sources can be challenging. This is especially true for large extended objects and regions in the sky, like the Gemini Monoceros X-ray enhancement, subject of this study. The background may vary spatially to a great extent over the entire object. Therefore, in preparation of the X-ray spectral analysis, we performed an extensive study of the background emission. Usually, a region close to the object with little to no emission from the actual object itself is an appropriate choice to extract the background emission. However, since the diffuse emission of our region extends over several degrees, 
it makes this approach unfeasible. 

\begin{figure}
	\centering
		\includegraphics[width=0.49\textwidth]{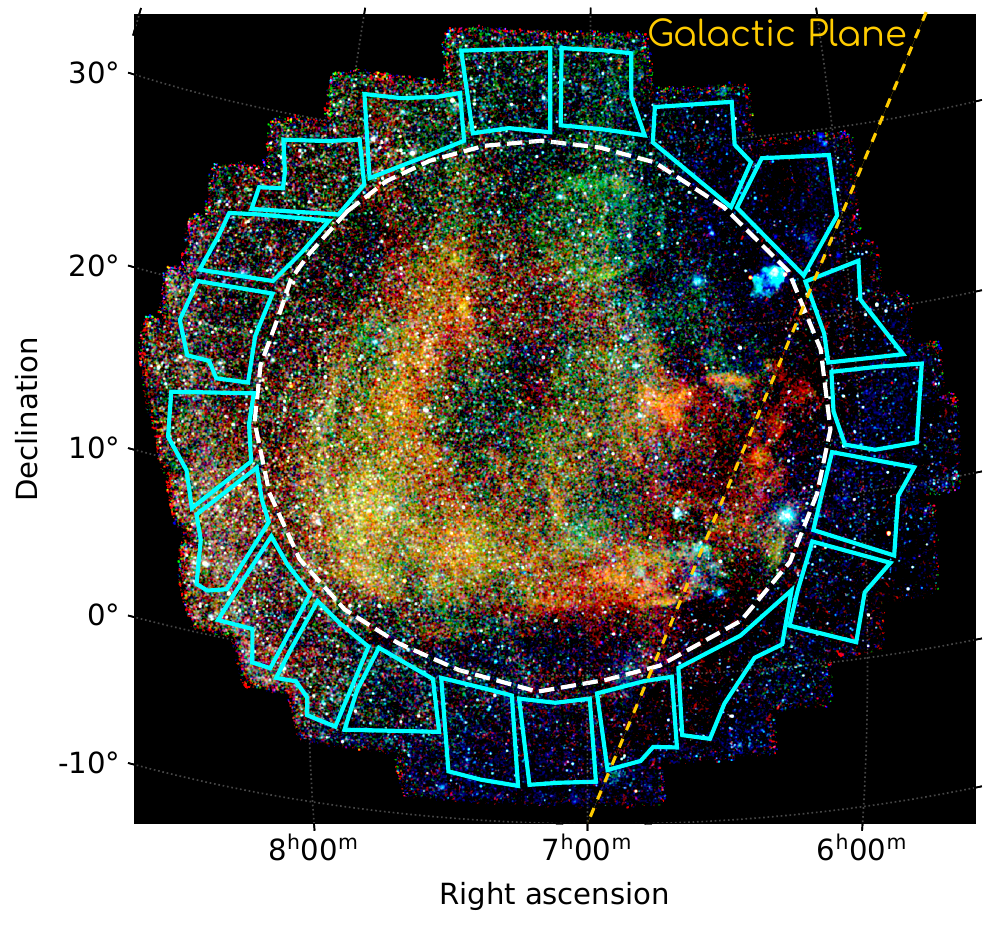}
		\caption{eROSITA X-ray image (\autoref{fig:xray_rgb_mosaic}) with the regions defined for the background analysis shown in cyan. The Galactic plane is shown in yellow for orientation.\label{fig:bkg_regions}}
\end{figure}

Instead, we defined regions surrounding the entire X-ray enhancement, where no emission from the source is visible anymore. The regions are shown in \autoref{fig:bkg_regions} and were optimized by hand over several iterations. Too large regions resulted in unreliable average values due to strong variations within the extent of one region, while too small regions did not have sufficient statistics for a precise spectral analysis. Additionally, we avoided regions in the sky that clearly showed diffuse emission from fore- and background sources. Since the spatial distance between the source regions and the background can be quite large, we performed a more in-depth study of the background. The aim was to create an accurate X-ray model of the Galactic- and instrumental background for each given coordinate inside the large extended source.

\subsection{Spectral analysis}
\subsubsection{Model}
For our spectral analysis of the background, we used the individual spectra of detectors TM$1-4$ and TM$6$. We account for the following Galactic components in our model: the local hot bubble (LHB), the circumgalactic medium (CGM) and the hot Galactic corona. 

The local hot bubble is a component visible over the entire sky, which can be described as emission from a thermal plasma with a temperature $kT \sim 0.1$\,keV \citep{fwc_model_yeung}. As shown by \citet{lhb_liu} there are slight variations in the temperature and emission strength for different parts of the sky, distributed narrowly around $kT=0.1$\,keV. We used the \textsc{apec} model \citep{apec_model} to describe the LHB, constraining the temperature to $kT = 0.09-0.12$\,keV with solar abundances, with the range constrained to the results shown in Fig. 6 in \citet{lhb_liu}. The normalization was let free to fit and later compared to the emission measure (EM) reported in their work.

Next, we introduced an absorption component (\textsc{tbabs}) to account for Galactic absorption along the line of sight. We fixed the value to the column density of the HI4PI survey value \citep{hi4pi} at the respective positions of the extracted spectra and multiplied this component with the following Galactic components.

The second Galactic component describes the circumgalactic medium, now believed to be an intrinsic property of galaxies \citep{cgm_theoretical}. We also used the thermal \textsc{apec} model for this component. The temperature of this component is believed to be close to the virial temperature $kT = 0.15-0.25$\,keV \citep{cgm_temp_virial, ponti_efeds}, while the exact structure and distribution is still not fully understood. Previous studies have shown that the chemical abundances of this component are significantly subsolar in the range from $\sim 0.05-0.3$ Z$_{\odot}$ \citep[e.g.,][]{cgm_theoretical, cgm_abundance, ponti_efeds}. We initially let this parameter free to fit and then fixed it to the median of the best-fit value over all regions, as we proceeded with the analysis. This was done to account for the relatively low statistics, which led to the parameter not being well constrained, as our uncertainty calculations showed. We obtained a median of $Z \sim 0.11$ Z$_{\odot}$ for the abundance, which appears consistent with the low values obtained by \citet{ponti_efeds}.

The third Galactic component accounts for emission from a hotter Galactic component, which might also be part of the CGM and could be caused by unresolved sources in the Galactic disk. Previous studies have suggested a temperature $kT > 0.6$\,keV, and therefore significantly hotter than the CGM, with a thermal nature \citep[e.g.,][]{hot_corona_temp_suzaku,cgm_temp_virial, ponti_efeds, halo_locatelli}. Therefore, we again used the \textsc{apec} model. We initially let the temperature free to fit, but similar to the CGM abundance, the uncertainties on the temperature was large. Therefore, we fixed the value to the median best-fit value of $kT \sim 0.7$\,keV, which is in good agreement with the previously mentioned studies. The abundances were fixed to solar values. 

To describe the extra-galactic background (CXB), we used a double-broken power law model. The CXB was also multiplied with the \textsc{tbabs} Galactic absorption component. The indices were defined as $\Gamma_1 = 1.96$ for $E < 0.6$\,keV, $\Gamma_2 = 1.75$ for $0.6  < E < 1.2$\,keV, $\Gamma_3 = 1.45$ for $E > 1.2$\,keV. This model was shown to yield an accurate description of the background also down to lower energies \citep{CXB_index}. The model includes the small modification for eROSITA used by \citet{ponti_efeds} to account for residuals at lower energies that were present without the modification \citep["CXBs" in][]{ponti_efeds}. This model performed slightly better in our case compared to the canonical model ("CXB") in their study. We set the initial normalization to $7.7\cdot 10^{-7}$\,photons/keV/cm$^2$/s which is equal to $\approx 8.3$\,photons\,s$^{-1}$\,cm$^{-2}$\,sr$^{-1}$. After the initial fit, the normalization is set free to fit. 

We also tried an alternative description of the background by decoupling the CGM and hot halo component absorption into individual components. The results are discussed in \autoref{sec:alternative_bkg}. However, this alternative model was not chosen for the following spectral analysis due to inconsistencies of the description of the SWCX contamination. Also, the increased complexity of the model did not yield better spectral fits of the background regions. While the coupled absorption of the Galactic and extragalactic background may not be physically correct, we achieve a good description of the background for our goal of performing a detailed spectral analysis on our sources. For a full physically motivated background model we would need more statistics, since using three individual absorption components would increase the complexity of the model too much.

The extraction areas $A$ of the respective spectra are taken into account with a normalization constant, set to the exact area in arcmin$^2$. This also has the benefit of making the normalization between different regions and detectors comparable.

We also considered possible contamination by solar wind charge exchange (SWCX). For this, we introduced an unabsorbed ``AtomDB Charge Exchange Model v2.0'' \citep[ACX2, ][]{swcx_temp, ACX2}\footnote{\url{https://github.com/AtomDB/ACX2}} (v1.1.0) component to our spectral model described in \autoref{eq:model_expr}. This model takes into account the complex charge exchange reactions based on atomic data. We used the model by fixing the temperature to the typical temperature of $kT = 0.1$\,keV in the initial fit \citep{swcx_temp} and then constraining the temperature to $kT = 0.07-0.13$\,keV in the second iteration. The study by \citet{ponti_efeds} yielded a slightly higher temperature of $kT = 0.136\pm 0.007$\,keV; however, if we left the temperature completely free to fit, most best-fit values centered around $kT \approx 0.1$\,keV. A possible cause might be the different ecliptic latitude of our background regions as well as different times and therefore solar activities when the data were taken. The collision parameter, equal to the wind velocity, was fixed to $450$ km/s, typical for the slow solar wind \citep{swcx_cumbee}. This velocity appears to also be appropriate as a mean solar wind velocity \citep{swcx_speed}. We also tried to let this parameter free; however, it was poorly constrained while not improving the fit. We only considered a single recombination for the calculation of the spectrum, as expected for SWCX.

In addition, we account for inter-detector calibration differences with a constant free to fit close to unity in the range $0.75-1.25$ (CAL). 
 In summary, we used the following model expression in \texttt{xspec}:

\begin{equation}
\begin{split}
\text{CONSTANT}_{\text{CAL}}\times\text{CONSTANT}_{\text{A}}\times(\text{ACX2}_{\text{SWCX}} + \text{APEC}_{\text{LHB}} \\ +  \text{ TBABS}\times (\text{APEC}_{\text{CGM}} +   \text{APEC}_{\text{Corona}} + \text{BKN2POW}_{\text{CXB}})\,\,.
\end{split}
\label{eq:model_expr}
\end{equation}

The second part of our model accounts for the non-X-ray background which can be studied using the data taken with the filter-wheel closed (FWC). We adopted the model based on a detailed study of the FWC data by \citet{fwc_model_yeung}. The FWC data were taken by eROSITA during calibration and service intervals. The phenomenological model describes the FWC background for each individual detector with high precision. Since this background does not conform to the regular photon paths, this model was only folded with the Redistribution Matrix File (RMF), but not the Auxiliary Response File (ARF). We added a normalization constant to the FWC model to account for differently sized true detector extraction region compared to the original FWC data. A detailed definition of the model is given in \textit{Appendix B} in \citet{fwc_model_yeung}. We refer to the entire model as ``Model B'' hereafter. An alternative model (``Model A'') is also described in \autoref{sec:alternative_bkg}. For all models and spectral fits in this work, we used the elemental abundances of \textsc{wilm} \citep{abund_wilm}.
\subsubsection{Spectral fit}
In all our spectral fits presented in this paper, we used the cash statistic \citep[cstat, ][]{cash} on unbinned spectra to optimize the use of the sparse X-ray data. We use the metric cstat/d.o.f. to asses our spectral fits in the following, since it leads to more accessible numbers than the absolute cstat value. While strictly speaking the cstat/d.o.f. is not directly comparable to the typical $\chi^2/\mathrm{d.o.f}$, since we use roughly always the same d.o.f. for our background fits, and within the spectral analysis later, this quantity is still very useful for comparing the quality of the fits. For most contour bins, background regions, and manually defined regions, we have enough counts to safely use cstat/d.o.f as a goodness-of-fit quantity \citep{cstat_goodness}.

For the spectral fit of the individual regions, we first estimated the normalization of the FWC model. We only used the data in the hard energy tail of the spectra in the $E=3.0-9.0$\,keV range, where the FWC emission dominates the spectrum. The normalizations for the different TM FWC models were independent from each other, and completely free to fit. We than performed a fit and calculated the uncertainties at the $90\%$ confidence interval, using the \textsc{error} command. The normalization constants were then constrained to the calculated uncertainty range and set to the best-fit value.

Next, to model the X-ray background, we considered the data in the energy range of $E=0.2-3.0$\,keV for TM1, TM2, and TM6, and $E=0.25-3.0$\,keV for TM2 and TM4. We fixed the detector normalization constant (CAL) of TM1 to unity, while we constrained the other constants individually to the aforementioned range. The second constant in the model was fixed to the extraction sky-area $A$ in arcmin$^2$ and linked between different TMs. All other parameters of the X-ray model were linked between detectors to reduce the number of free parameters, under the assumption that any calibration difference is accounted for with the first constant. 
%
%
\subsection{Results}
\begin{figure*}
	\centering
	\begin{subfigure}[t]{0.49\textwidth}
		\includegraphics[width=1.0\textwidth]{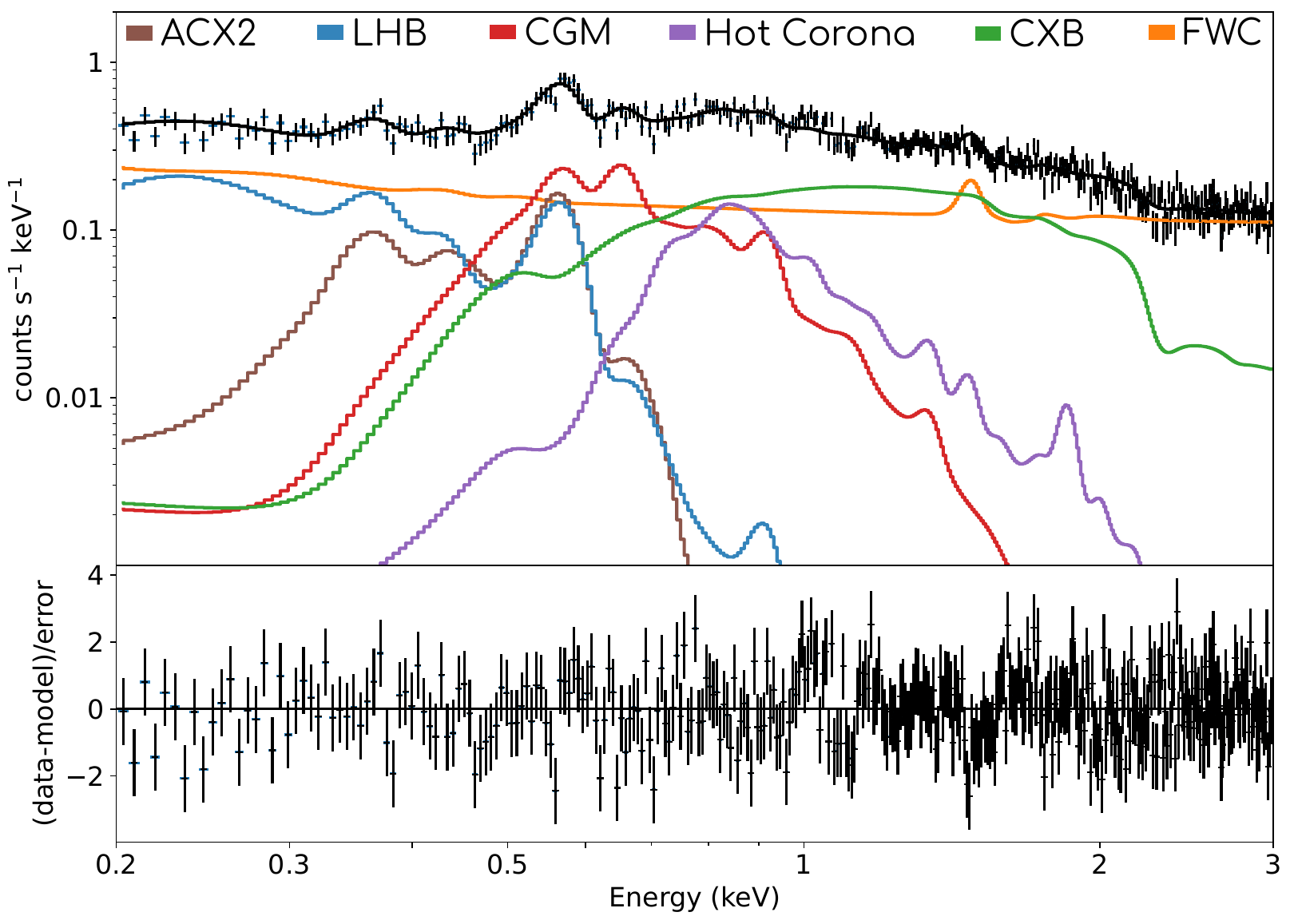}
		\caption{\label{fig:bkg_inplane_spectrum}}
\end{subfigure}
\hfill
\begin{subfigure}[t]{0.49\textwidth}
	\centering
		\includegraphics[width=1.0\textwidth]{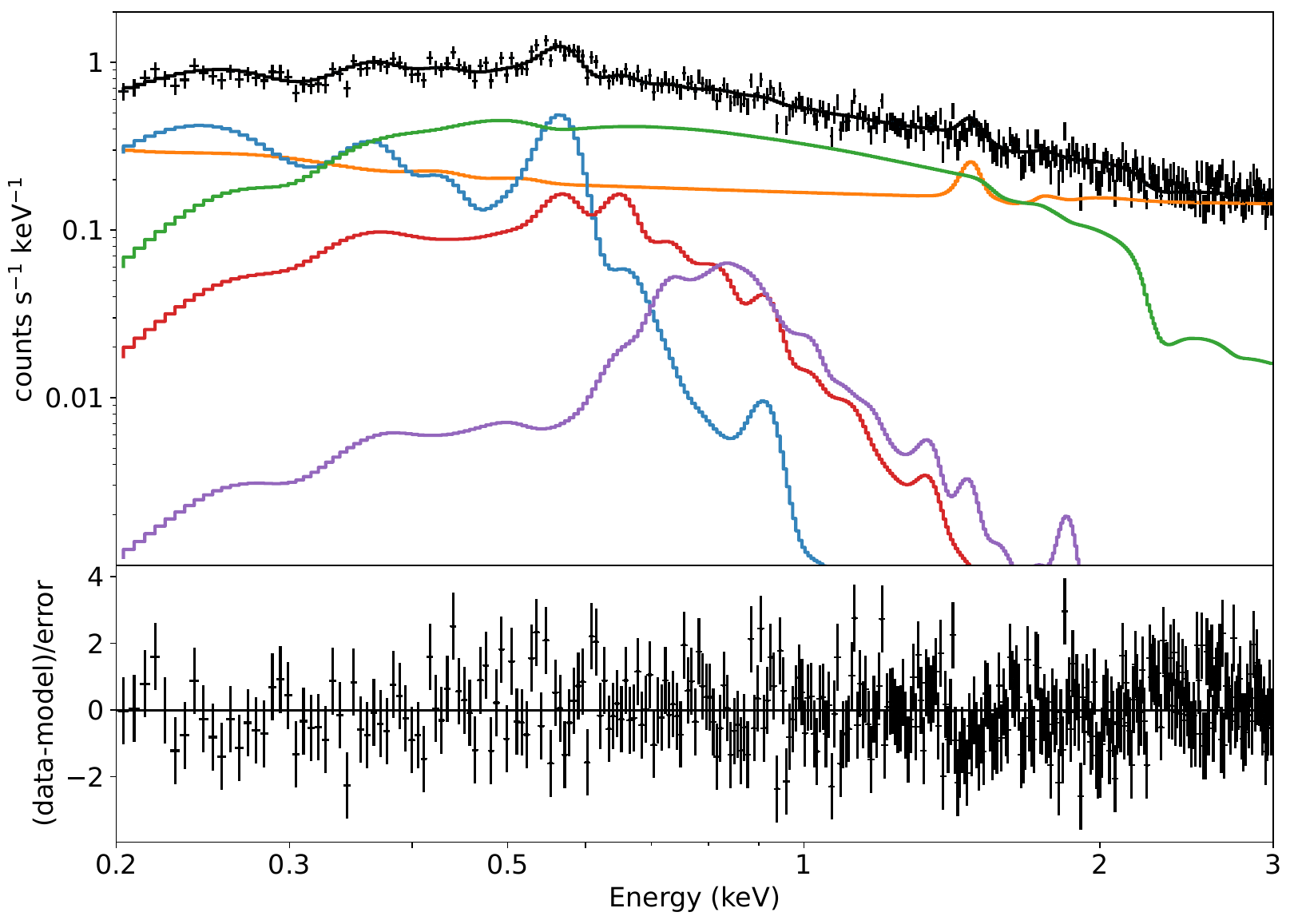}
		\caption{\label{fig:bkg_outplane_spectrum}}
		 \end{subfigure}
\caption{Background spectrum and fit model near the Galactic plane at $b\sim 3.6^{\circ}$ (a) and out of the plane at $b\sim 22.1^{\circ}$ (b). The data and total best-fit model are shown in black and the different model components are shown with the colors  according to the legend in (a). For visual purposes only, the spectra were binned with a minimum $5\sigma$ or $50$ counts per bin and for better visibility only the spectrum and model for TM1 is shown.}
\end{figure*}

\begin{figure*}
	\centering
\begin{subfigure}[t]{0.49\textwidth}
	\centering
		\includegraphics[width=1.0\textwidth]{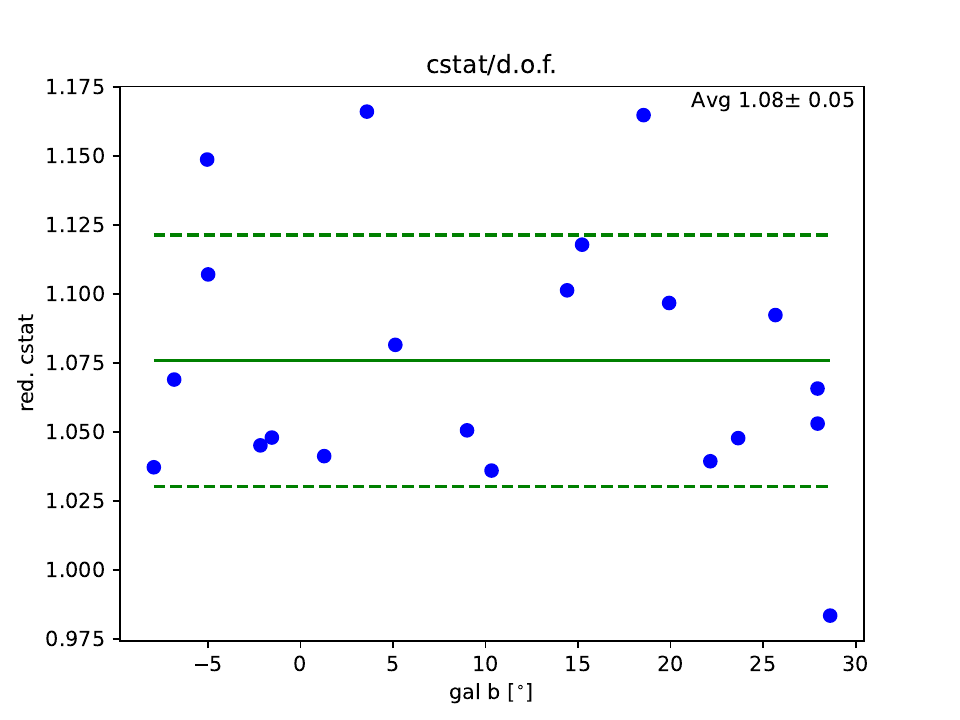}
		\caption{\label{fig:cstat_bkg}}
\end{subfigure}	
\hfill
\begin{subfigure}[t]{0.49\textwidth}
	\centering
		\includegraphics[width=1.0\textwidth]{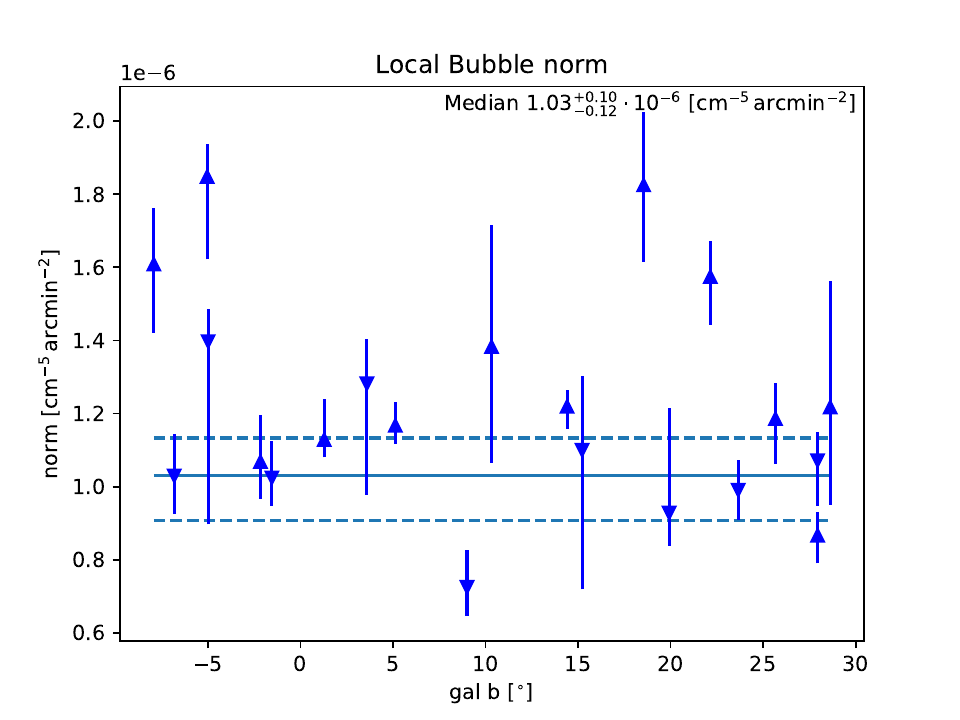}
		\caption{\label{fig:lhb_norm_bkg}}
		 \end{subfigure}
\\
	\begin{subfigure}[t]{0.49\textwidth}
		\includegraphics[width=1.0\textwidth]{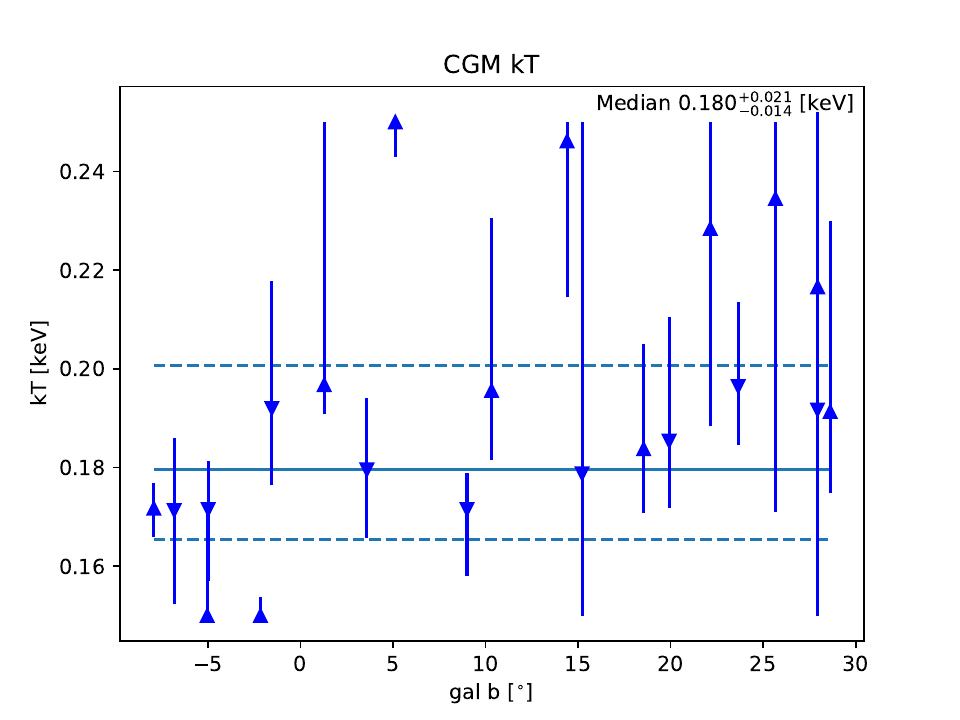}
		\caption{\label{fig:cgm_kT}}
\end{subfigure}
\hfill
\begin{subfigure}[t]{0.49\textwidth}
	\centering
		\includegraphics[width=1.0\textwidth]{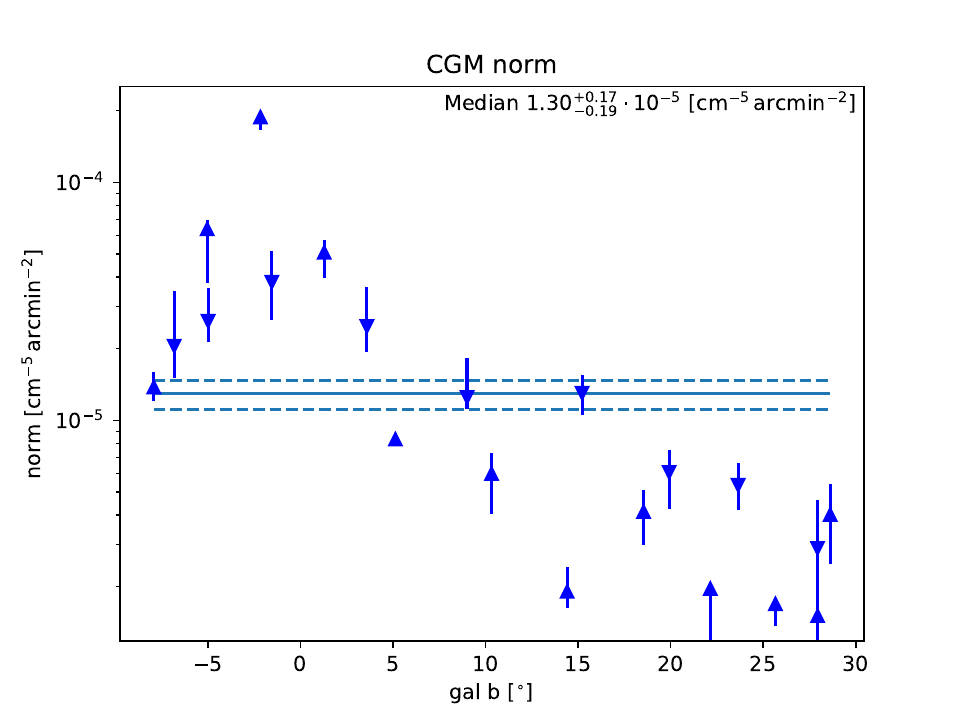}
		\caption{\label{fig:cgm_norm}}
		 \end{subfigure}
\\
	\begin{subfigure}[t]{0.49\textwidth}
		\includegraphics[width=1.0\textwidth]{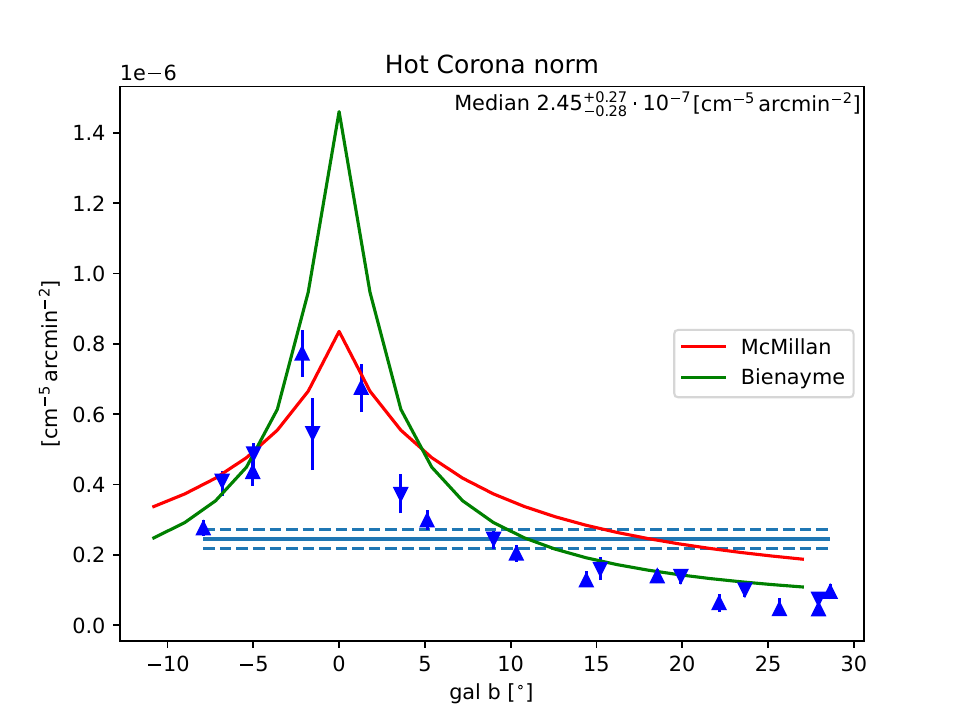}
		\caption{\label{fig:corona_norm}}
\end{subfigure}
\hfill
\begin{subfigure}[t]{0.49\textwidth}
		\includegraphics[width=1.0\textwidth]{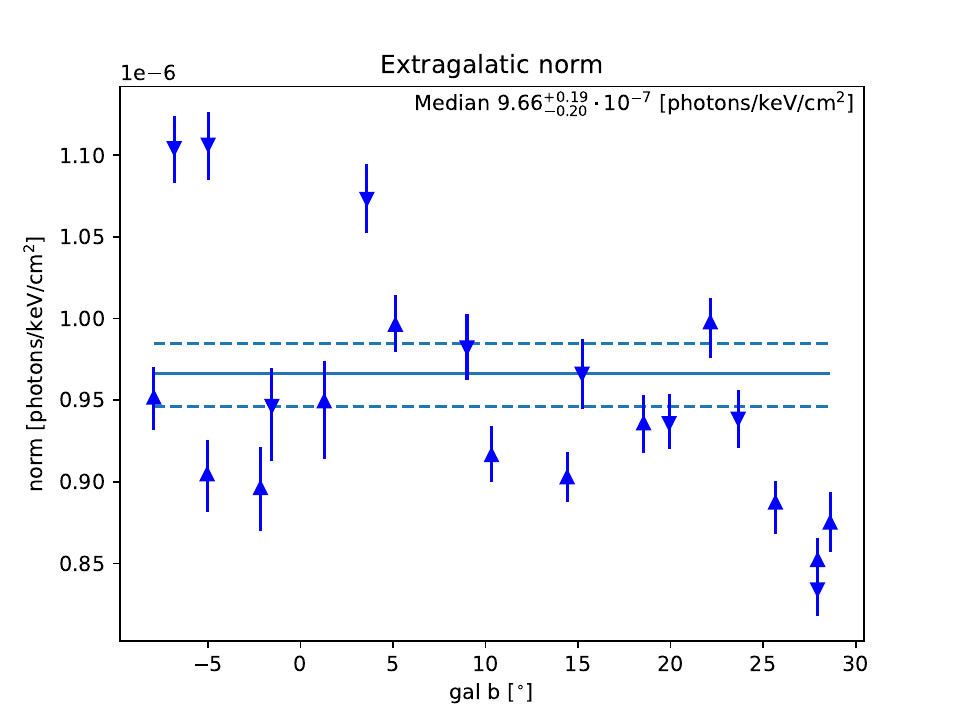}
		\caption{\label{fig:cxb_norm}}
		 \end{subfigure}
\caption{\label{fig:bkg_components_results}Variation of the X-ray background model component best-fit parameters over galactic latitude (gal b). (a) Best-fit reduced fit statistic (cstat/d.o.f.). The solid line is the average, while the dashed lines show the $1\sigma$ standard deviation. (b) the LHB normalization; (c) the CGM plasma temperature (\textsc{apec}); (d) the CGM normalization;
(e) the hot corona component normalization ($kT = 0.7$\,keV). We also overlaid two Galactic mass profiles, based on \citet{mass_profile_01} and \citet{mass_profile_02}. The profiles were kindly provided by N. Locatelli, averaged over the projected profiles at $l = 175^{\circ}-230^{\circ}$. The scale of the profiles is arbitrary, chosen for the best visual representation; (f) The extra-galactic component (double-broken power law) normalization. For (b)-(f) the triangles facing down indicate gal $l < 200^{\circ}$ while facing up triangles are used for $l > 200^{\circ}$.}
\end{figure*}

With ``Model B'' as described above, we obtained very good fit statistics for all regions, as shown in \autoref{fig:cstat_bkg}. The average reduced fit statistic and standard deviation is $\text{cstat/d.o.f.} = 1.08 \pm 0.05$. An example of two spectra, one close to the Galactic plane at $b\sim 3.6^{\circ}$ and another far from the Galactic plane at $b\sim 22.1^{\circ}$, are shown in \autoref{fig:bkg_inplane_spectrum} and \autoref{fig:bkg_outplane_spectrum}, respectively, and the different model components are indicated.

In the following we discuss the individual components. The most important fit parameters of the individual components are shown in \autoref{fig:bkg_components_results}. The respective median values were calculated from the best-fit values, while the lower and upper limits shown in the diagrams are equal to the median lower and upper uncertainties. The respective values are shown along the Galactic latitude of the associated background regions.
The normalization of the LHB component (\autoref{fig:lhb_norm_bkg}) appears to be very consistent across all background regions, yielding a median of $\eta_\text{LHB} = 1.03^{+0.10}_{-0.12}\cdot 10^{-6}$ cm$^{-5}$\,arcmin$^{-2}$. We see a small increase in the normalization for some regions at higher longitudes (triangle up), but considering the uncertainties, they are only slightly outside the expected statistical scatter. We compared our median normalization $\eta$ with the LHB study by \citet{lhb_liu} by converting it to an emission measurement (EM). The \textsc{apec} normalization is defined as 
\begin{equation}
\eta = \frac{10^{-14}}{4\pi D^2}   \int n_\mathrm{e} n_\mathrm{H} dV \,[\mathrm{cm}^{-5}] \,\mathrm{ ,}
\label{eq:norm_apec1}
\end{equation}
where $n_\mathrm{e}$ and $n_\mathrm{H}$ are the electron and hydrogen densities, respectively, and $D$ the distance to the source.
If we only consider the area in the line of sight, this can be expressed as 
\begin{equation}
\eta = 10^{-14}\frac{\theta}{4\pi}   \int n_\mathrm{e} n_\mathrm{H} dl\,\mathrm{ ,}
\label{eq:norm_apec2}
\end{equation}
where $\theta$ is the emission area in the sky in units sr \citep{ponti_efeds}. Since we normalized the models to arcmin$^2$, this yields $\theta =  (1/60 \cdot \pi / 180)^2$.
With this, we obtain 
\begin{equation}
\text{EM}  = \frac{\eta}{2.081\cdot 10^{-4}} \, [\mathrm{cm}^{-6} \mathrm{pc}] \,\mathrm{ .}
\label{eq:norm_apec3}
\end{equation}
Using \autoref{eq:norm_apec3} we obtain an emission measure of $\text{EM} = (4.95_{-0.58}^{+0.48})\cdot 10^{-3} \mathrm{cm}^{-6}\, \mathrm{pc}$ which is consistent with the measurements by \citet{lhb_liu} at the position of the Monogem Ring, within uncertainties.

The CGM results are show in \autoref{fig:cgm_kT} and \autoref{fig:cgm_norm}. We obtained relatively uniform temperatures around the median value of $kT \sim 0.18$\,keV for the majority of the regions, where some show higher temperatures up to $0.25$\,keV, albeit with relatively large uncertainties in most cases. The normalization of the CGM appears strongly enhanced for background spectra obtained near the Galactic disk ($\pm 5^{\circ}$). If we exclude the outlier near $\sim 0^{\circ}$, the increase is up to a factor of $\sim 5$ around the disk, compared to the median value $\eta_\text{CGM} = 1.30^{+0.17}_{-0.19}\cdot 10^{-5}$ cm$^{-5}$\,arcmin$^{-2}$.

For the hot corona, we also obtained a strong dependence on the Galactic latitude for the normalization, as shown in \autoref{fig:corona_norm}. The median of the normalization is $\eta_\text{Corona} = 2.45^{+0.27}_{-0.28}\cdot 10^{-7}$ cm$^{-5}$\,arcmin$^{-2}$. In comparison to the CGM, this component's normalization is smaller by two orders of magnitude. One possible origin of the hot corona component are stars, and especially dwarf M star spectra show a remarkably similar temperature to this component \citep{stars_coronae, dm_dwarfs}. Since the population should roughly correlate with the mass profile of the Milky way, we compared model predictions to our results (solid lines, \autoref{fig:corona_norm}). The models are based on the theoretical predictions by \citet{mass_profile_01} and \citet{mass_profile_02}, reprojected onto the average galactic longitude of the Gemini-Monoceros X-ray enhancement. While the normalizations of the profiles are arbitrary, we obtain a surprisingly good agreement between the profile shape of the predicted mass profile denoted "Bienayme" and the hot corona normalization. This result indicates that indeed this component could be caused, at least partially, by the stellar population. In addition, our results for the hot corona show, that we are indeed fitting a physical component of the Milky way.

For the extragalactic background, we obtain a median normalization of $\eta_\text{CXB} = 9.66^{+0.19}_{-0.20}\cdot 10^{-7}$ photons/keV/cm$^2$\,arcmin$^2$. We obtain a few outliers outside the expected statistical scattering with values lower or higher by about $\sim 10\%$, which might be caused by local enhancements in the background. Compared to previous studies, our values are slightly higher by about $\sim 10\%$ \citep{cxb_study_01, cxb_study_02}. This might be caused by our very high number of point-source candidates removed from the spectra. Since removing point sources effectively changes the slope of the CXB, the higher normalization compensates the fixed power law indices in our model.
\subsection{Application to the spectral analysis}
\label{sec:bkg_model_application}
From our results it is clear that there is a strong variation over Galactic latitude of the background, and to a lesser degree also over the Galactic longitude. Therefore, in order to apply the results, we need to take into account both the Galactic latitude, as well as the absolute distance between background and source region. 

Based on this, we extrapolated the background of our source regions by averaging the best-fit parameters with uncertainties of the most appropriate background regions distributed around the source.
 The background regions were selected by calculating the smallest angular separation in Galactic latitude compared to the source region. After some initial tests, we obtained the best results with limiting our calculations to the four regions with the least angular separation. Adding additional regions did not improve the background estimate. Then, we extrapolated the background parameters in the source region with the weighted average given by 
 \begin{equation}
    \bar{x} = \frac {\sum \limits_{i=1}^{n}w_{i}x_{i}}{\sum \limits_{i=1}^{n}w_{i}} \,\, \text{ ,}
    \label{eq:weighted_mean}
    \end{equation}
    where the weights are given by 
     \begin{equation}
    1/w_i = 2 d_{\mathrm{gal,b},i}+d_{\mathrm{abs},i} \,\, \text{ ,}
    \label{eq:bkg_weights}
    \end{equation}
    with $d_{\mathrm{gal,b},i}$ and $d_{\mathrm{abs},i}$ being the distance in Galactic latitude and the angular distance between the source and background region centroid positions, respectively. Because the greatest varation in background was found across Galactic latitude, we weigh this angular separation double, compared to the absolute angular separation.
	From the resulting weighted lower and upper limits (90\% confidence interval, CI) of the parameters, we construct the respective parameter ranges that are used to approximate the background in the source regions.

Applying this background estimate, we obtained very good fits for most of the source regions. For a handful of isolated regions, we were able to further improve the fit significantly by increasing the allowed parameter fitting range by a modest $10\%$ of what we obtained from the weighted average. This can be explained by the very large angular separation of the source regions from the background regions used in the calculaton of the weighted averages. Additionally, since the background regions are relatively large, there might be gaps which lead to the over- or under-estimation of the background in certain cases. Increasing the parameter range beyond $10\%$ did not improve the spectral fits further, that is, for some regions our background-estimate is ``off'' by that amount.
We therefore adopted this small modification and increased the uncertainty/fit range if we obtained a suboptimal fit ($\text{cstat}/\text{d.o.f.} > 2000/1752 = 1.14$) which solved the remaining issues for all source regions. 
%
\section{X-ray spectral analysis}
\label{sec:spectral_analysis}
\subsection{Spectral analysis regions}
\subsubsection{Manually defined regions}
\begin{figure*}
	\centering
		\includegraphics[width=0.8\textwidth]{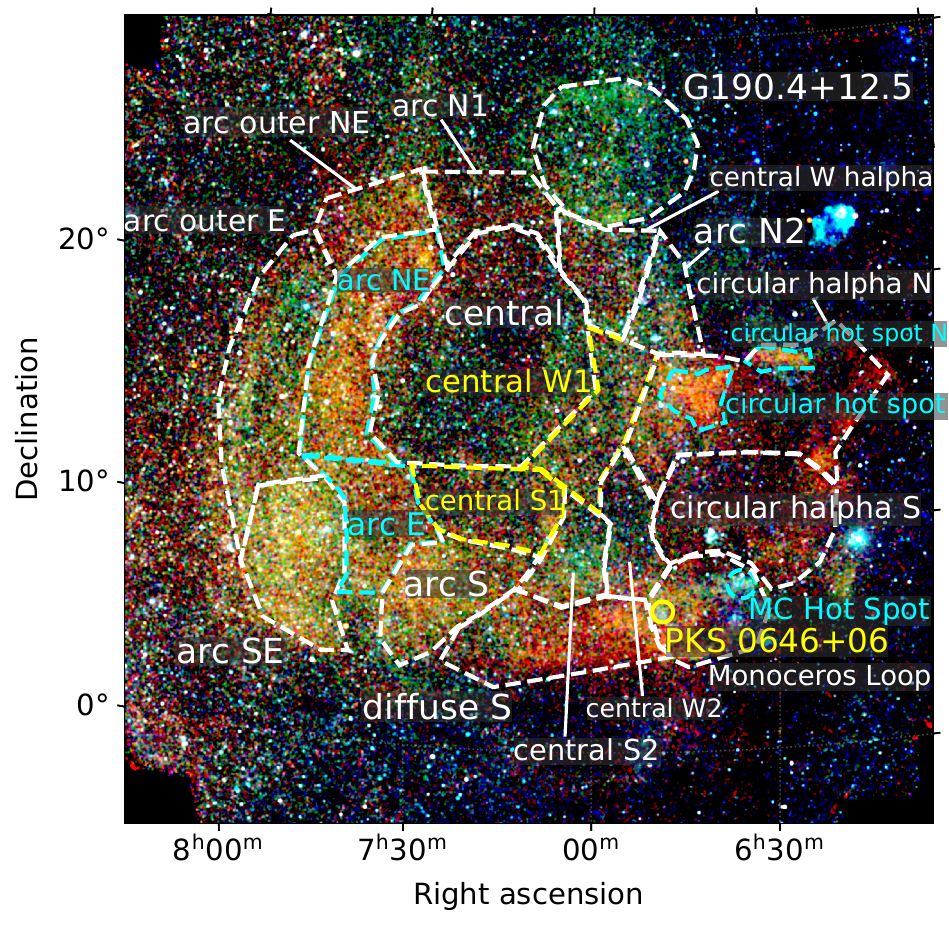}
		\caption{\label{fig:manual_regions}eROSITA X-ray mosaic as in \autoref{fig:xray_rgb_mosaic} with the manually defined regions. The regions are shown in white, cyan, and yellow.}
\end{figure*}

We manually defined spectral analysis regions based on three criteria: the X-ray three-color image in \autoref{fig:xray_rgb_mosaic}, the H$\alpha$ emission in \autoref{fig:xray_rgb_mosaic_halpha}, and finally the dust optical depth $A_v$ in the line of sight as shown in \autoref{fig:soft_xrays_dust_3d}. The regions are shown in \autoref{fig:manual_regions}.
We first defined the regions based on the color in the X-ray three-color image, which corresponds to the hardness, to obtain regions with apparently similar spectral properties. In a second step, we optimized those regions by comparing them to the H$\alpha$ emission. Regions which had vastly different H$\alpha$ emission gradients were adjusted to better reflect the H$\alpha$ morphology. In a last step, we compared the regions to the dust map in the range of $600-900$\,pc, where we observe a possible (anti)correlation. Again, we modified the regions to align with strong gradients, here in the $A_v$ dust absorption.
\subsubsection{Contour binning}
\begin{figure}
	\centering
	\includegraphics[width=0.49\textwidth]{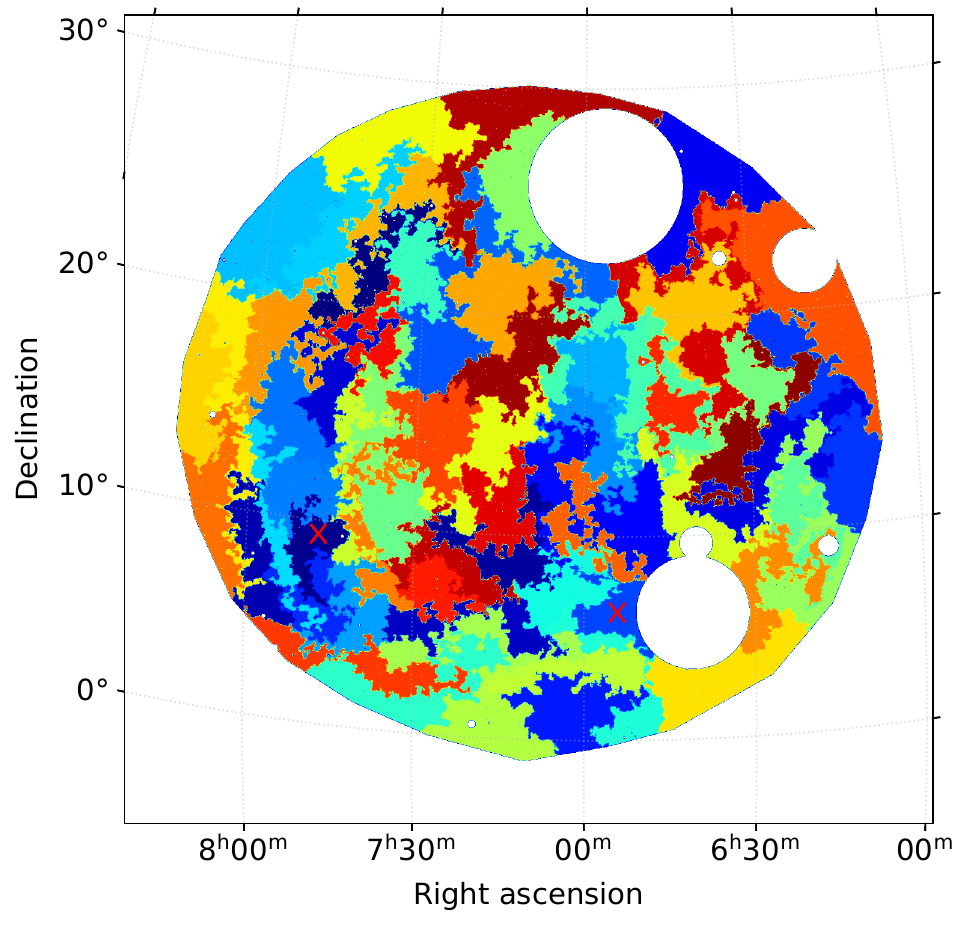}
	\caption{\label{fig:contbin_radec_plot}}
\caption{\label{fig:contbin_map_plot}Map of the resulting contour bins based on a minimum S/N of 250. Different colors correspond to different bins. For bins marked with a red ``x'', we show example spectra in \autoref{fig:contbin_example_spectra}.}
\end{figure}

In addition to the manually defined spectral analysis regions, we also created regions purely based on the statistics of the data. This allows us to systematically cover the entire region while minimizing any observer bias. We implemented the contour binning method by \citet{contbin_sanders} which allows us to specify a certain S/N while also taking into account the shape of the diffuse emission. In our previous works \citep[e.g.,][]{spur_paper, sasaki_lmc_erosita} we implemented the Voronoi binning method to achieve similar goals. However, the Voronoi binning often leads to bins that include strongly different emission (based on X-ray spectral color) and therefore averaging over different physical conditions is almost unavoidable. The contour binning method circumvents this problem, with the regions smoothly tracing along any strong gradients in emission. We required a minimum S/N of 250 for each bin, based on the count statistics of the X-ray data. The resulting contour bin map is shown in \autoref{fig:contbin_map_plot}. In total, we obtained 66 contour bin regions using this method with an average S/N of 287. For each contour bin we extracted point-source subtracted spectra and calculated response matrices, similar to what we described in \autoref{sec:xray_data_reduction}. The response matrices were sampled with a size of 2\arcmin, which is accurate enough for the average bin size on the order of a degree or more.
\subsection{Model}
\label{sec:source_model}
Similar to the background analysis, we only used the individual spectra of detectors TM$1-4$ and TM$6$. Again, we used the elemental abundances of \textsc{wilm} \citep{abund_wilm}.

For both the contour bins and manually defined regions, we adopted the background model, with the parameters estimated based on the respective centroid positions, as described in \autoref{sec:bkg_model_application}.
We first introduced an absorbed (\textsc{tbabs}) thermal \textsc{apec} component for the source emission; however, the fit quality was bad for some of the regions. Therefore, we instead introduced an absorbed nonequilibrium ionization (NEI) model \textsc{nei} for the source emission for the contour bins.
This significantly improved the fit in the aforementioned regions, indicating that the plasma is in NEI there. For the manually defined regions, due to the higher statistics, we instead used the \textsc{vnei} model that allows to vary individual elemental abundances.
In summary, the model described in \autoref{eq:model_expr} was modified to:
\begin{equation}
\begin{split}
\text{CONSTANT}_{\text{CAL}}\times\text{CONSTANT}_{\text{A}}\times(\text{ACX2}_{\text{SWCX}} + \text{APEC}_{\text{LHB}} \\ 
+ \text{ TBABS}_{\text{source}}\times \text{ (V)NEI}_{\text{source}}+ \\
\text{ TBABS}\times (\text{APEC}_{\text{CGM}} +   \text{APEC}_{\text{Corona}} + \text{BKN2POW}_{\text{CXB}}))\,\,.
\end{split}
\label{eq:model_expr_source}
\end{equation}
As described above, we constrained the X-ray background parameters to the weighted averages of the four most appropriate background regions, which were determined for each source region or contour bin individually. The four background regions were chosen based on the lowest Galactic latitude difference to the respective source region. The source component parameters were linked between the different TMs. The detailed spectral fit routine for the contour bins and manually defined regions is described in \autoref{sec:spectral_fitting_routine_source}.

Another issue in the low-count-rate regime is the foreground absorbing column, which is often poorly constrained. Therefore, we utilized the distance integrated $A_v$ dust maps shown in \autoref{fig:soft_xrays_dust_3d} to estimate the $N_{\text{H,X}}$. We made use of the gas-to-dust ratio derived by \citet{gas_to_dust_xmm}, which is based on \textit{XMM-Newton} data. The gas-to-dust ratio is given as
\begin{equation}
N_{\text{H,X}} = A_j \cdot 7.2 (\pm 0.5) \cdot 10^{21} \text{ cm}^{-2}
\label{eq:gas-to-dust}
\end{equation}
where we used the conversion factor of $A_j = A_v \cdot 0.34^{+0.02}_{-0.04}$ to convert the $A_v$ dust absorption to $A_j$ \citep{gas_to_dust_xmm}. We mainly applied this estimation to the Monoceros Loop and PKS 0646+06 SNR, but also made consistency-checks with regions of the Monogem Ring, where we observed indications for absorption. The final results of the spectral analyses are discussed in detail below. In case the source was beyond our map range of $> 1.1$\,kpc, we cross-checked the extinction via the ``EXPLORE: G-Tomo'' app\footnote{\url{https://explore-platform.eu/sda/g-tomo}} which provides profiles up to several kpc. The G-Tomo data is also based on the cube by \citet{dust_lallement}, as well as \citet{dust_vergely}. However, for most sources the changes in $A_v$ beyond $\approx 1.1\,$kpc were small and below our sensitivity. The small increase is most likely due to the increasing distance from the Galactic plane and therefore low ISM density.
\subsection{Results}
\subsubsection{Contour bins}
\begin{figure*}
	\centering
	\begin{subfigure}[t]{0.49\textwidth}
		\includegraphics[width=1.0\textwidth]{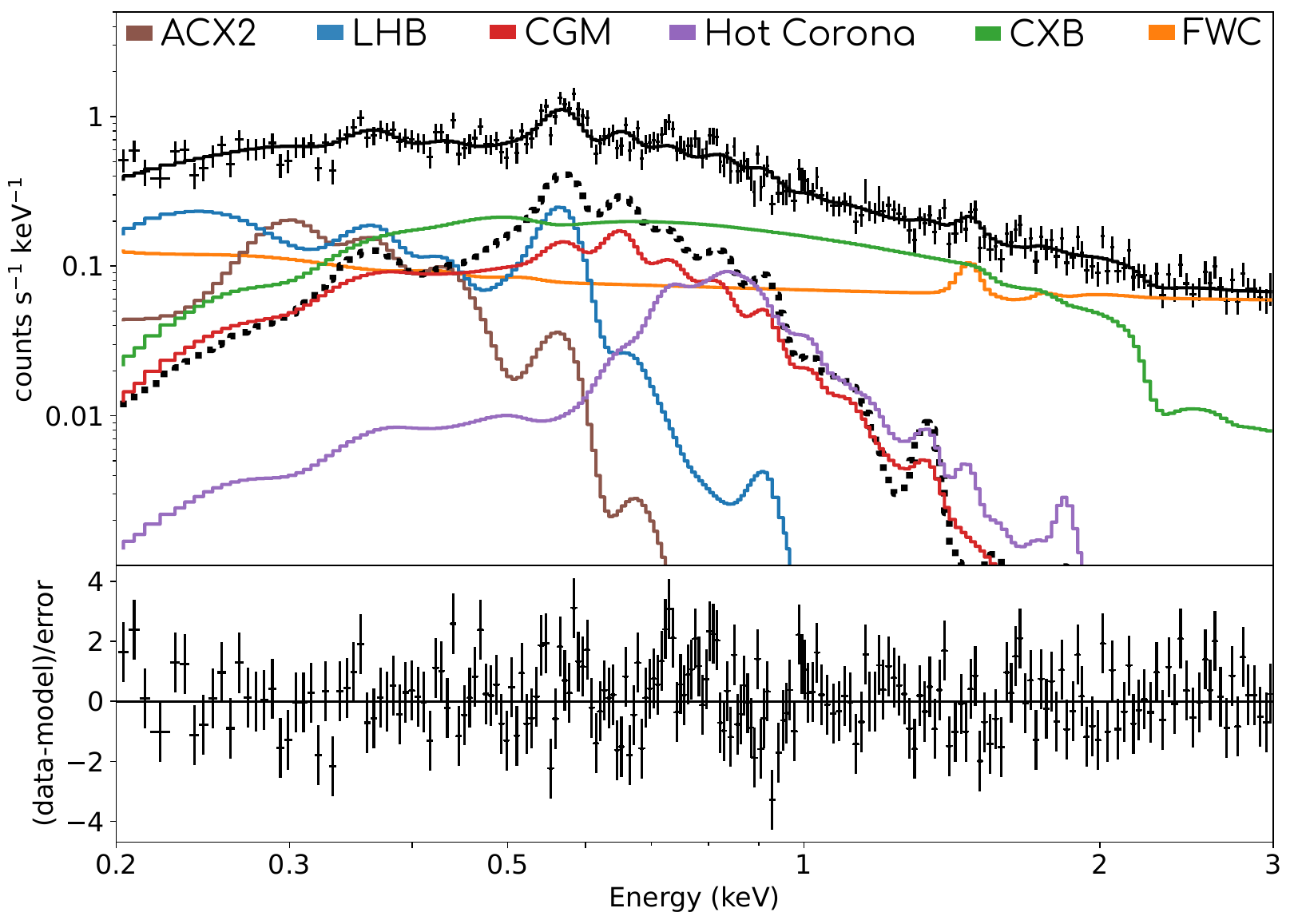}
		\caption{\label{fig:example_contbin_near_arc_SE}}
\end{subfigure}
\hfill
\begin{subfigure}[t]{0.49\textwidth}
	\centering
		\includegraphics[width=1.0\textwidth]{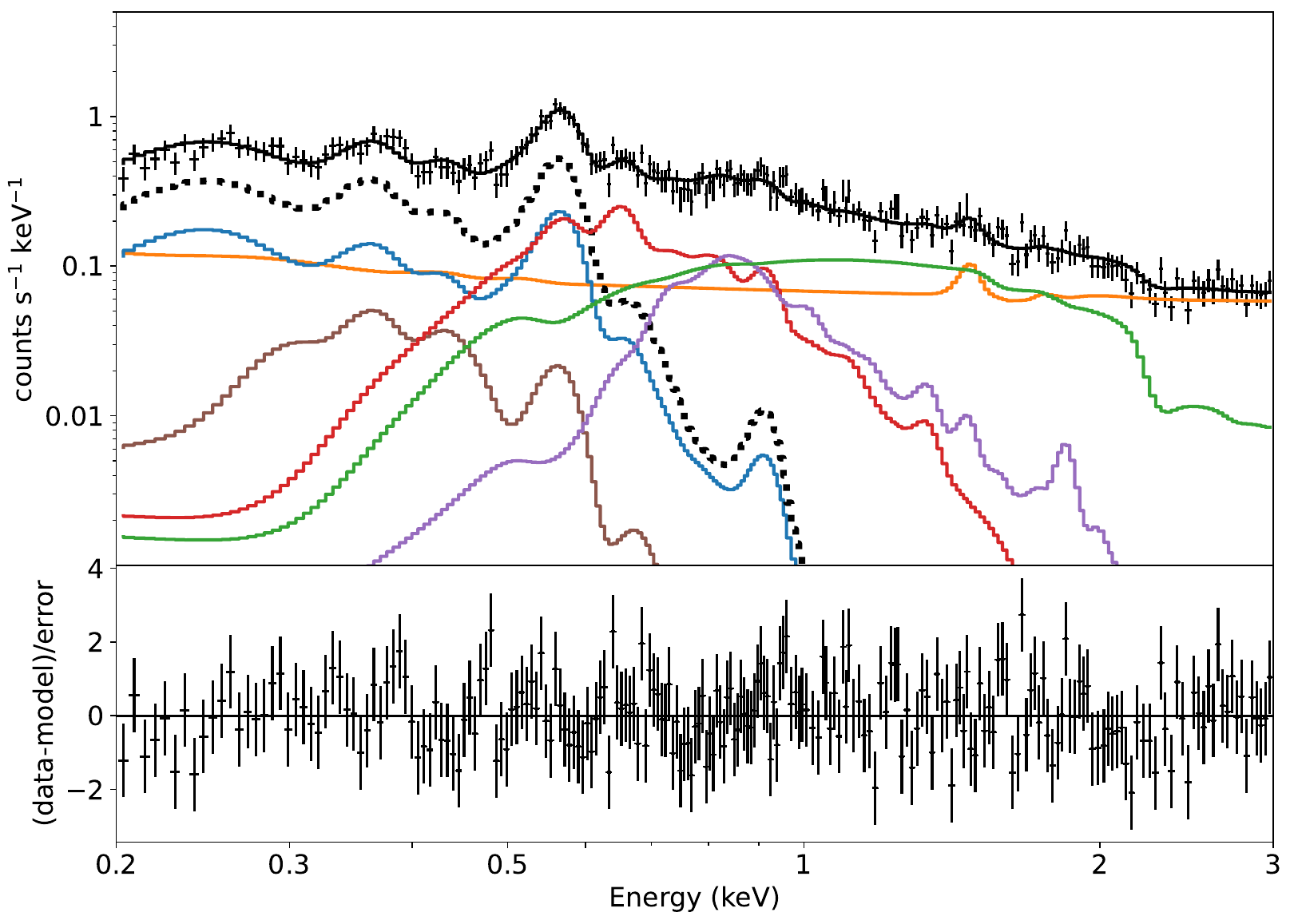}
		\caption{\label{fig:example_contbin_near_diffuse_S}}
		 \end{subfigure}
\caption{\label{fig:contbin_example_spectra}Two example spectra of contour bins with the best-fit model. We used the single \textsc{nei} component model for the source. The background is approximated as described above, with the same colors for the various model components as shown in \autoref{fig:bkg_inplane_spectrum}. The source component is shown with a thick, dotted black line. In (a) we show the spectrum for a bin near the ``arc\_SE'' region and in (b) near the ``diffuse\_S'' region. The corresponding bins are also marked in \autoref{fig:contbin_map_plot}. For visual purposes only, the spectra were binned with a minimum $5\sigma$ or $50$ counts per bin and for better visibility only the spectrum and model for TM1 is shown.}
\end{figure*}
\begin{figure*}
	\centering
	\begin{subfigure}[t]{0.49\textwidth}
		\includegraphics[width=1.0\textwidth]{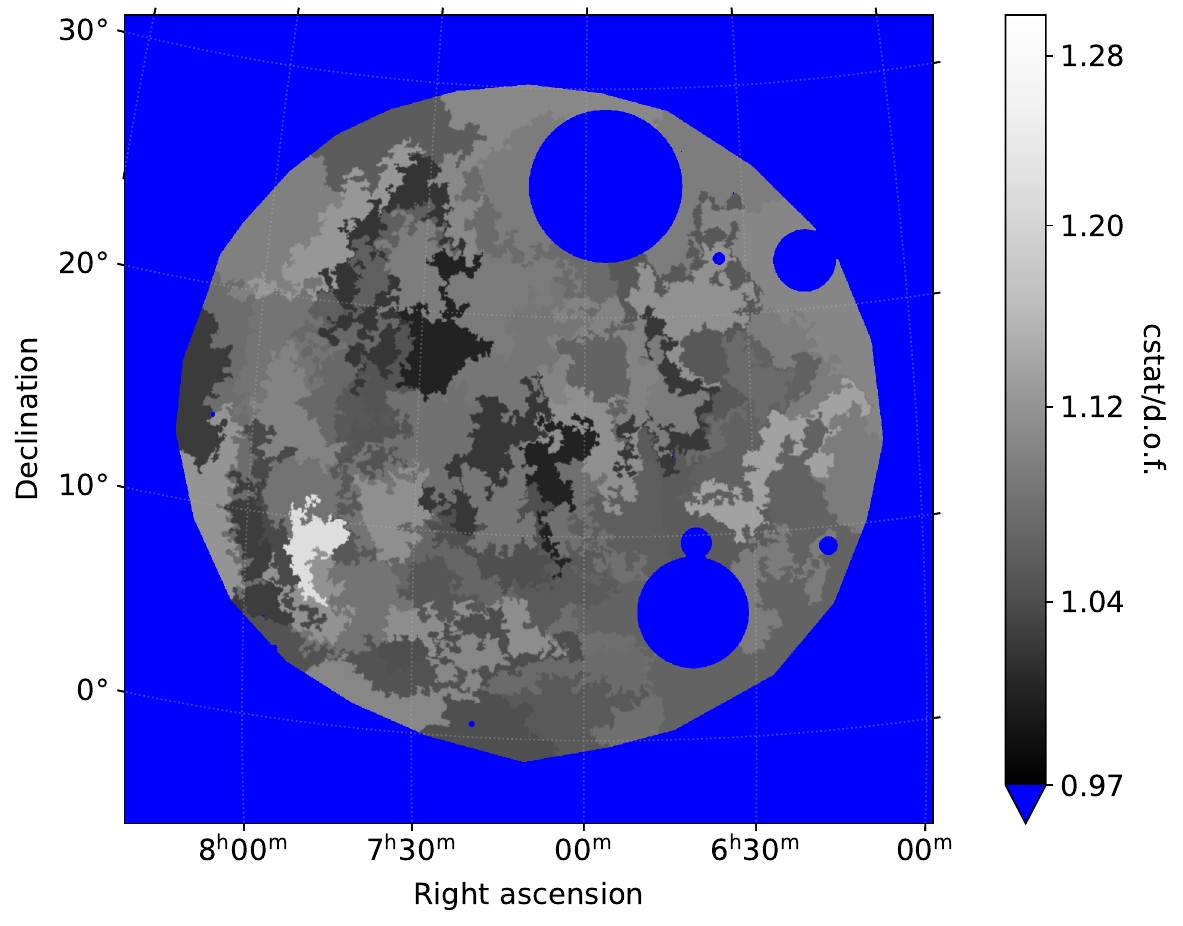}
		\caption{\label{fig:contbin_cstat}}
\end{subfigure}
\hfill
\begin{subfigure}[t]{0.49\textwidth}
	\centering
		\includegraphics[width=1.0\textwidth]{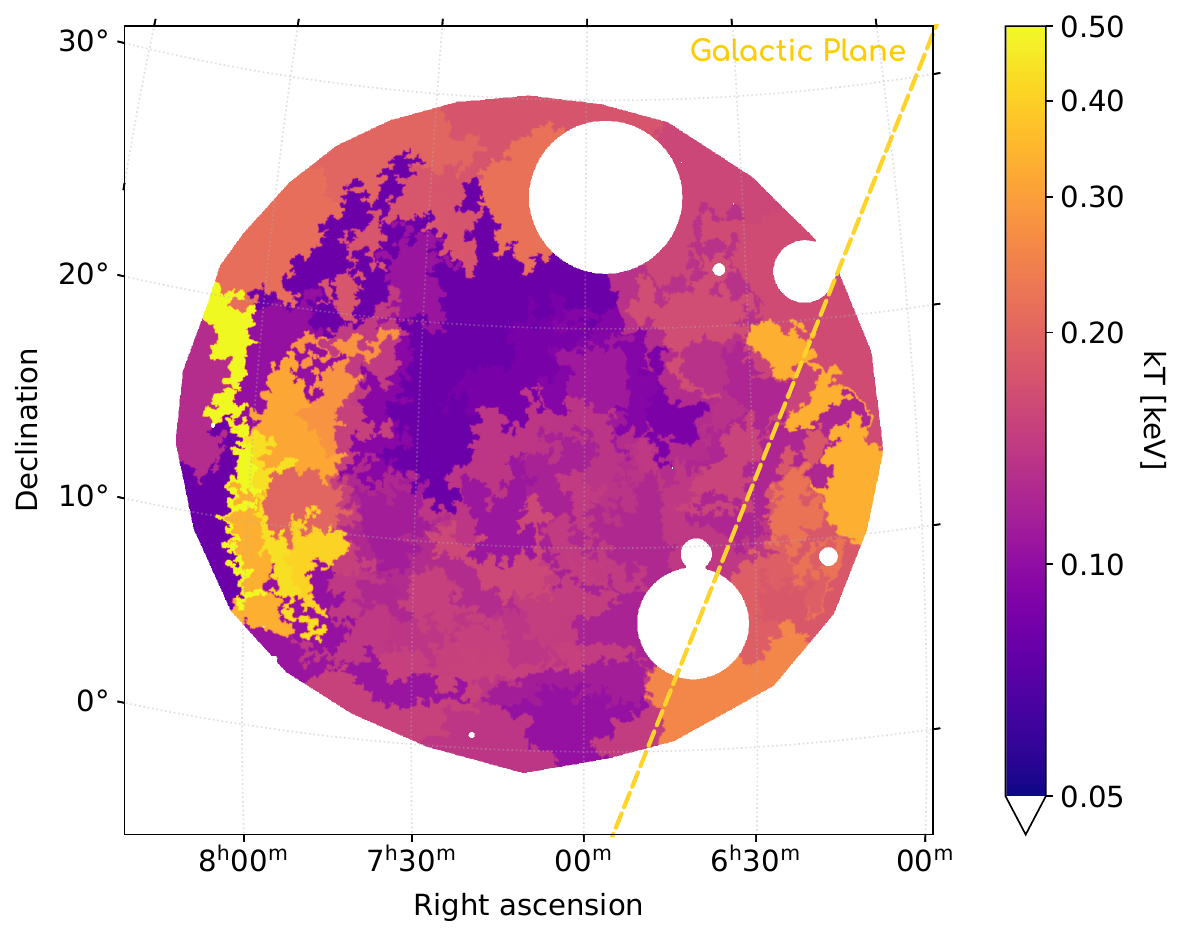}
		\caption{\label{fig:contbin_kT}}
		 \end{subfigure}
		 \\
		\begin{subfigure}[t]{0.49\textwidth}
		\includegraphics[width=1.0\textwidth]{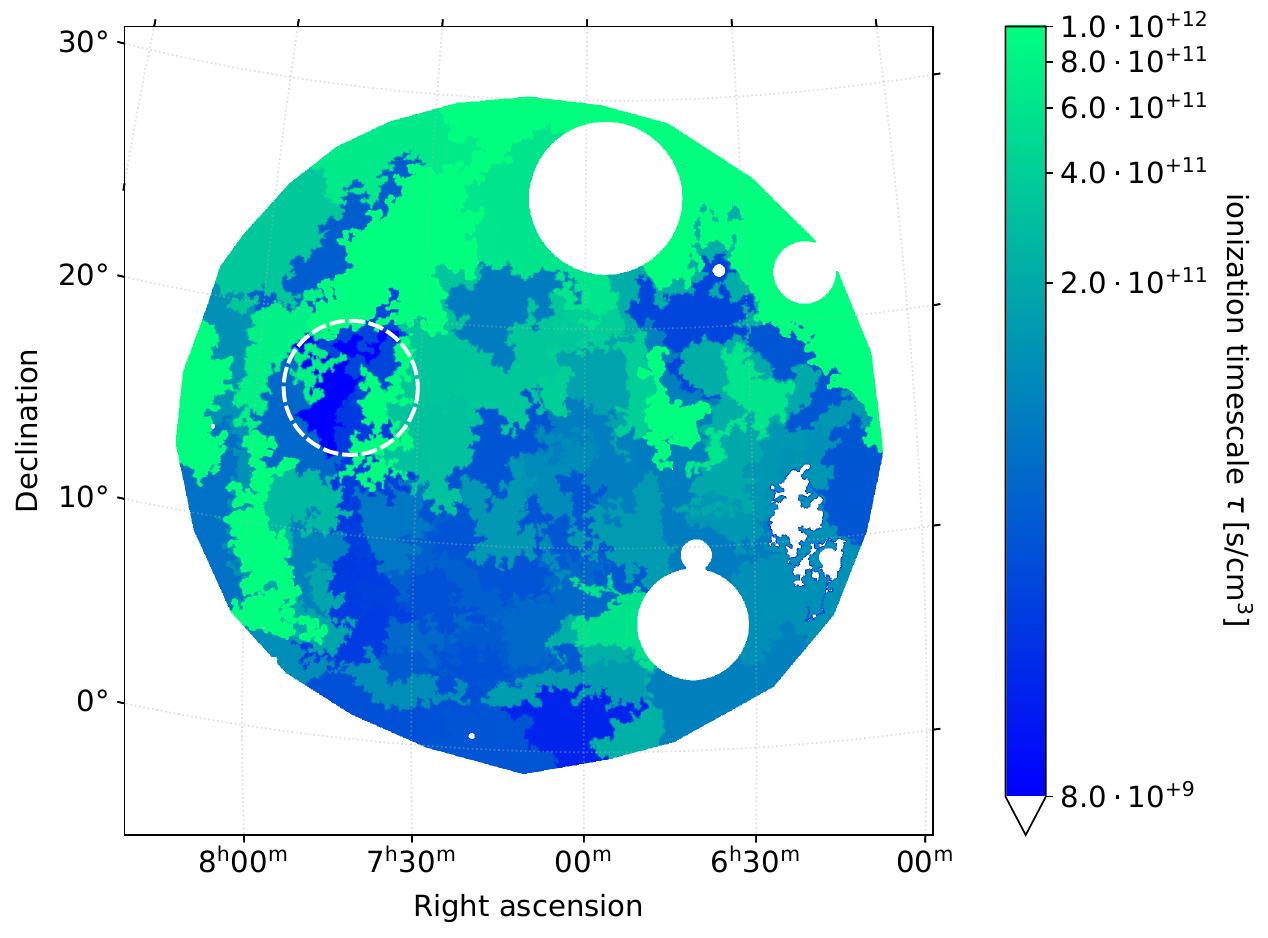}
		\caption{\label{fig:contbin_tau}}
\end{subfigure}
\hfill
\begin{subfigure}[t]{0.49\textwidth}
	\centering
		\includegraphics[width=1.0\textwidth]{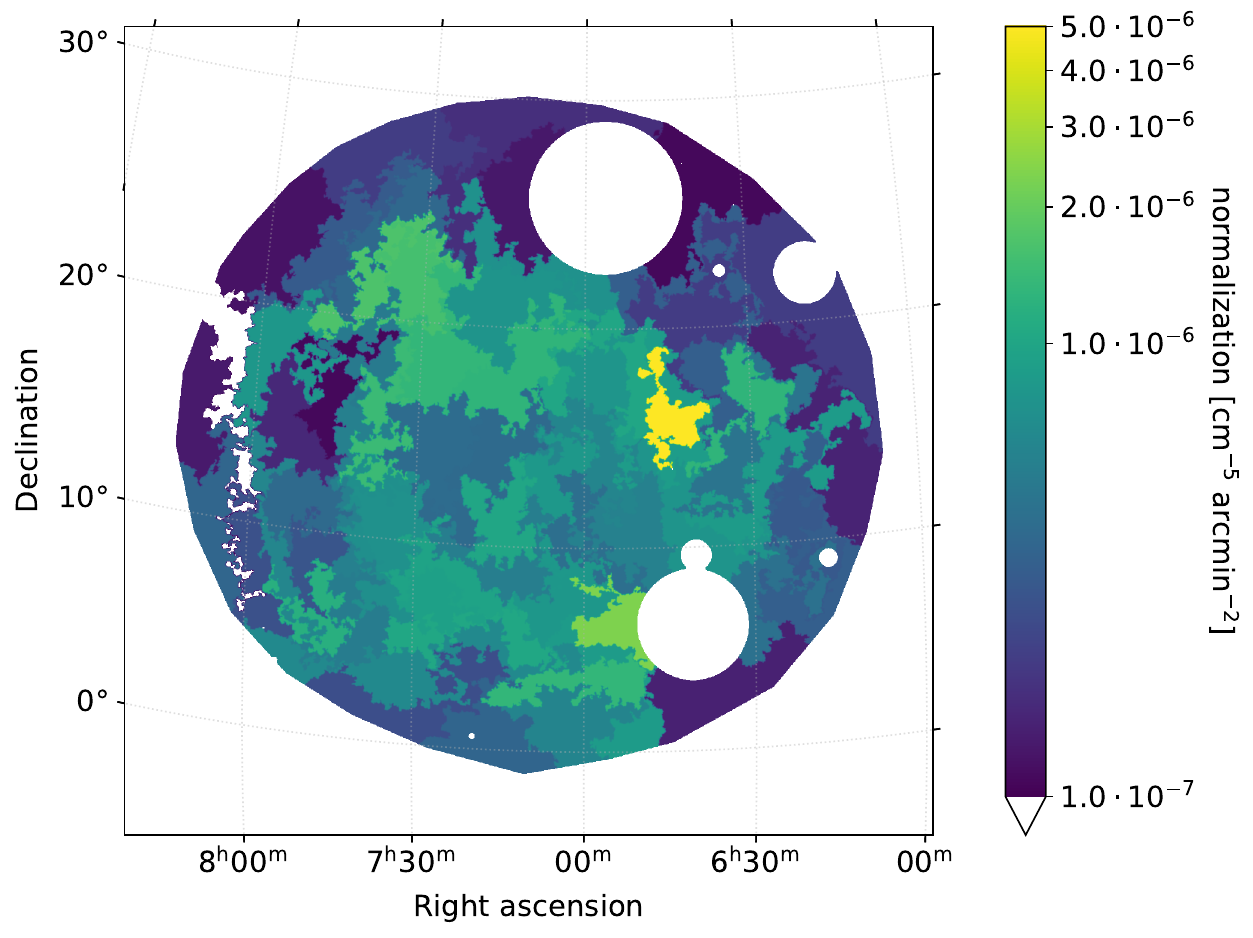}
		\caption{\label{fig:contbin_norm}}
		 \end{subfigure}
  \caption{\label{fig:contour_bin_results_modelB_v2}Spectral analysis results with the contour binning method. We used the \textsc{tbabs~$\times$~nei} model to account for the source emission. In (a) we show the red. fit statistic, in (b) the plasma temperature $kT$, in (c) the ionization timescale $\tau$, and in (d) the normalization of the plasma component, normalized to the respective area of the bins for comparability.}
\end{figure*}

With our model that uses a simple \textsc{vnei} with solar abundances, we achieved very good fits for the entire region, with an average cstat/d.o.f. of $1.08$ in a narrow range of $1.00-1.22$.  With our model and unbinned data, $\mathrm{d.o.f.} = 1749$ follows for the contour bins. An example of two contour bin spectra and best-fit models is shown in \autoref{fig:contbin_example_spectra}. The distribution of cstat/d.o.f. is shown in \autoref{fig:contbin_cstat}.

Thanks to the relatively fine spatial resolution of the contour bin method, we obtained detailed model parameter maps, as shown in \autoref{fig:contbin_kT}-\autoref{fig:contbin_norm}. We derived the median value for the best-fit, as well as lower and upper confidence limit values to approximate the overall statistical uncertainty on the median best-fit value.
The temperature appears to be relatively uniform, with an average $kT = 0.14 \pm 0.03$\,keV for all bins. However, there are several regions that deviate from this mean value. Toward the center of the remnant, coinciding with a lower intensity, we obtain the lowest temperatures $kT \sim 0.09$\,keV. In the east (left-handed side) we find a higher plasma temperature, with $kT = 0.18_{-0.03}^{+0.02}$\,keV on average, with a significant enhancement toward the southeast with $kT = 0.25_{-0.05}^{+0.04}$\,keV (see also \autoref{fig:xray_rgb_mosaic}, green cross). In addition, we notice an enhancement close to the disk, toward the west. This is most likely caused by the hot plasma of (unresolved) background sources in the Galactic disk. 

The diffuse emission of the Monogem Ring coincides with a low $\tau < 10^{12}$\,s/cm$^3$ for most bins, while the regions outside of the structure to the north show a high $\tau$ typical for CIE.
The ionization timescale (\autoref{fig:contbin_tau}) shows a small gradient approximately from south to north. In the bins to the southeast where we find an enhanced plasma temperature, we also obtained a relatively low $\tau$, indicating NEI. This is especially true for the low-$\tau$ region marked in \autoref{fig:contbin_tau} (dashed circle). This region also shows a low normalization, which coincides with a depression in X-ray brightness (see \autoref{fig:xray_rgb_mosaic}). A study by \citet{kim_uv} found bright C IV emission here, which suggests that the shock interacted with denser ISM region more recently, which would explain the low $\tau$ values. We also investigated if the ionization timescale might be degenerate with the temperature by performing simultaneous 2D \textsc{steppar} calculations of $kT$ and $\tau$. We only found degeneracies for the bins on the edge of our coverage - just outside the X-ray enhancement - or in the center where we have a significant reduction in X-ray surface brightness, leading to a low amount of source counts and the fit parameters not being well contrained.

In general, the normalization shown in \autoref{fig:contbin_norm} corresponds very well to the X-ray morphology shown in \autoref{fig:xray_rgb_mosaic}. 
The bright spot at RA,~Dec~$=$~$6^\mathrm{h}43^\mathrm{m}16^\mathrm{s}$, $16^{\circ}$00$\arcmin$10$\arcsec$ also appears with the highest normalization of all bins. This region was also observed by Suzaku in our previous study of the Monogem Ring \citep[P2,][]{mg_paper_2018}, where we also obtained the highest normalization with an \textsc{nei} model. Regions outside the diffuse emission or at the boundary are fit with a very low normalization, as expected.
\subsubsection{A possible second plasma component}
Our results in \autoref{fig:contour_bin_results_modelB_v2} show an interesting temperature enhancement in the southeast, while some residuals still remain in the $0.5-1.0\,$keV energy range, as shown in \autoref{fig:example_contbin_near_arc_SE}. Therefore, we tried a more complex model consisting of two plasma components for the source emission instead of one, that is, \textsc{tbabs~$\times$~apec~+~apec}. The background model remains the same as before. The unabsorbed \textsc{apec} component (second component hereafter) represents the emission from the Monogem Ring, while the absorbed \textsc{tbabs~$\times$~apec} component accounts for additional emission in the line of sight, possibly located at higher distances. We applied this new model to the all regions, similar to before.

\begin{figure*}
	\centering
	\begin{subfigure}[t]{0.49\textwidth}
		\includegraphics[width=1.0\textwidth]{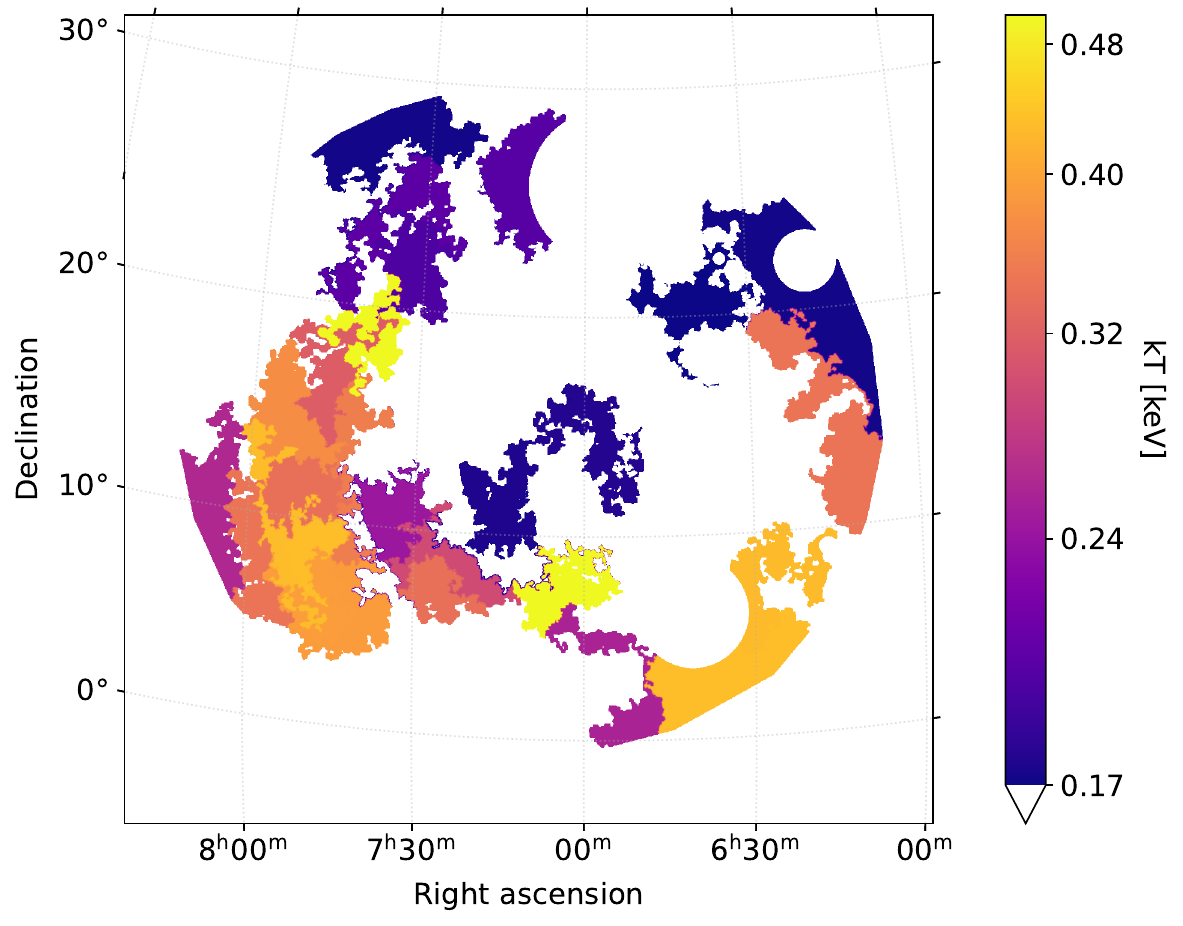}
		\caption{}
\end{subfigure}
\hfill
\begin{subfigure}[t]{0.49\textwidth}
	\centering
		\includegraphics[width=1.0\textwidth]{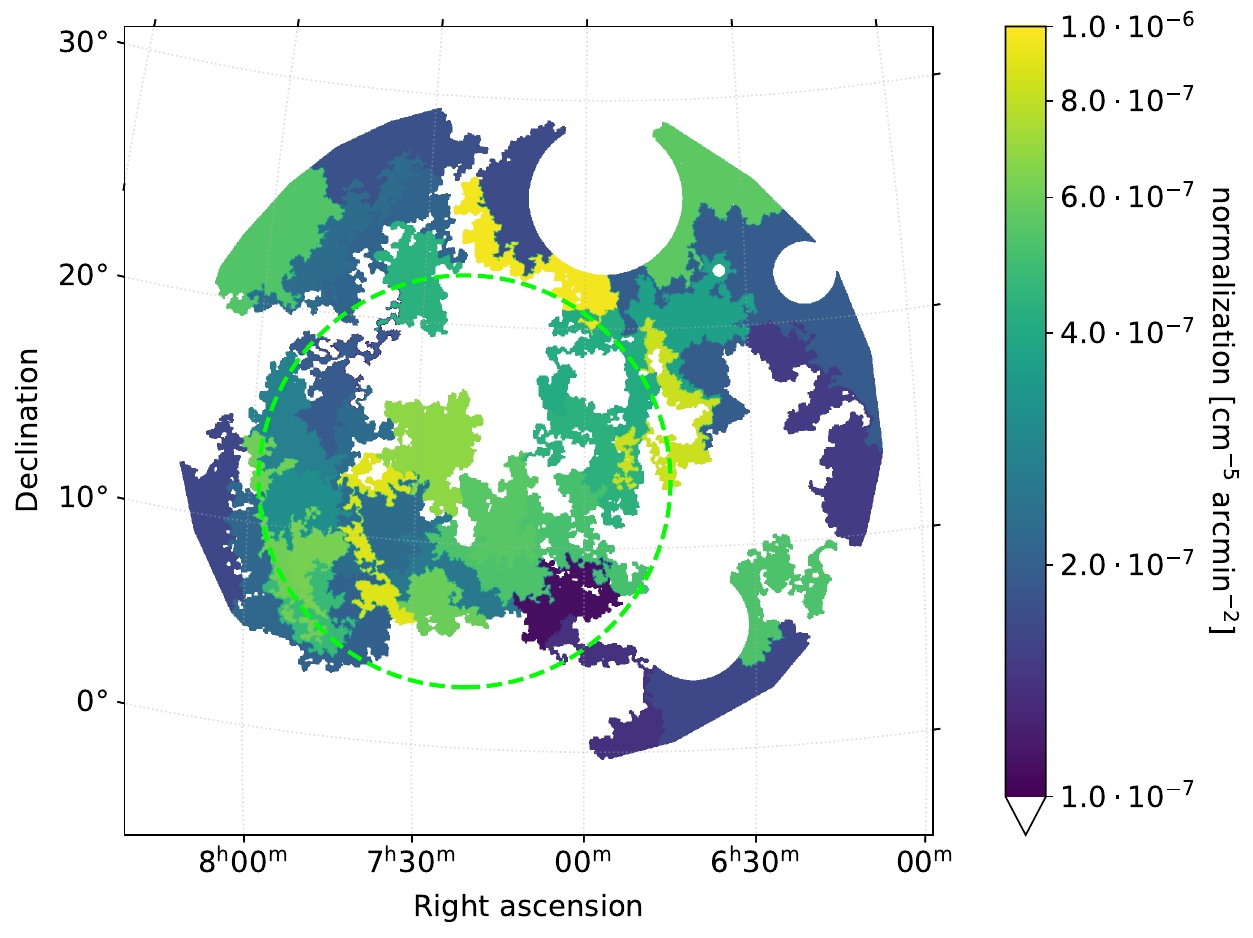}
		\caption{}
		 \end{subfigure}
\caption{\label{fig:contour_bin_results_two_apec}Spectral analysis results with the contour binning method. We used the \textsc{tbabs~$\times$~apec~+~apec} model to account for the source emission. (a) the temperature $kT$ and (b) the normalization of the first (absorbed) plasma component. The green dashed circle highlights the bins where we obtain significant contribution of the hotter plasma component. Regions below the threshold as indicated in the colorbar were made blank.}
\end{figure*}

Using this model, we were able to significantly improve the fit statistic in the southeast. The results are shown in \autoref{fig:contour_bin_results_two_apec} for the first (absorbed) plasma component. The temperature is fit with much higher values compared to the mean temperature for the Mongem Ring ($kT = 0.14 \pm 0.03$\,keV) while being consistent with the previously found temperature enhancement. In addition, the normalization shows that we obtain significant contribution of this component localized to the circular marked area (G205.6+12.4 in \autoref{fig:new_geometry}), excluding some outliers outside the larger diffuse structure and in the Galactic plane. Outside of this region, most bins show a very low or consistent with zero normalization (blank regions). For all regions inside the diffuse emission, we obtained an absorption consistent with zero, suggesting that the additional plasma component is also located rather close.
Moreover, for the second (unabsorbed) component we obtained very similar results compared to the previous \textsc{tbabs~$\times$~nei} model discussed above, sans the temperature enhancement. 
These results suggest the presence of a second diffuse plasma overlapping in projection with the emission of the Monogem Ring. Since the statistics with the smaller counter-bin regions is limited, we investigated the highlighted regions in more detail below.
\subsubsection{Manually defined regions}
\label{sec:spectral_analysis_mdr}
\begin{figure*}
	\centering
	\begin{subfigure}[t]{0.49\textwidth}
		\includegraphics[width=1.0\textwidth]{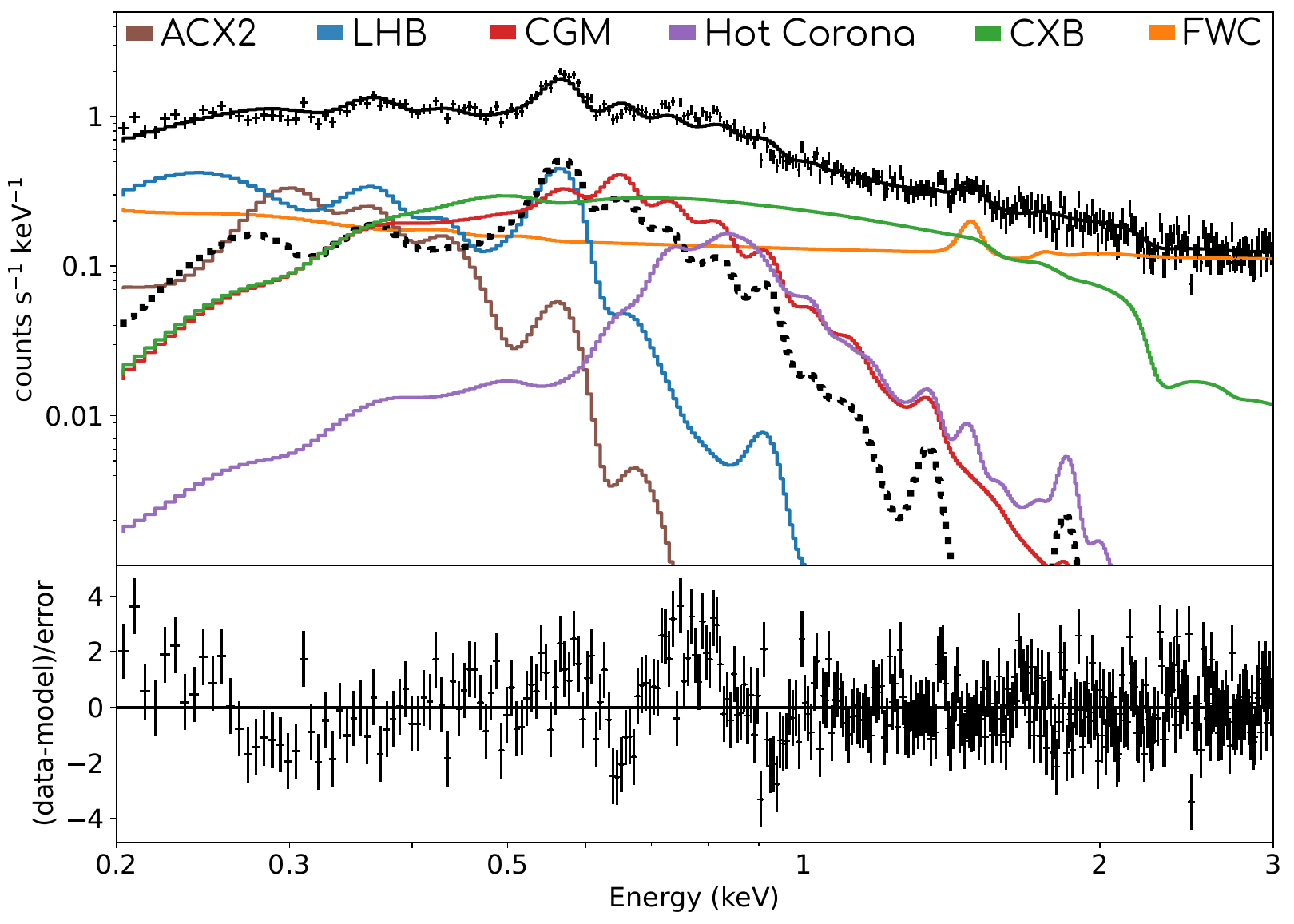}
		\caption{\label{fig:example_arc_SE_one_apec}}
\end{subfigure}
\hfill
\begin{subfigure}[t]{0.49\textwidth}
	\centering
		\includegraphics[width=1.0\textwidth]{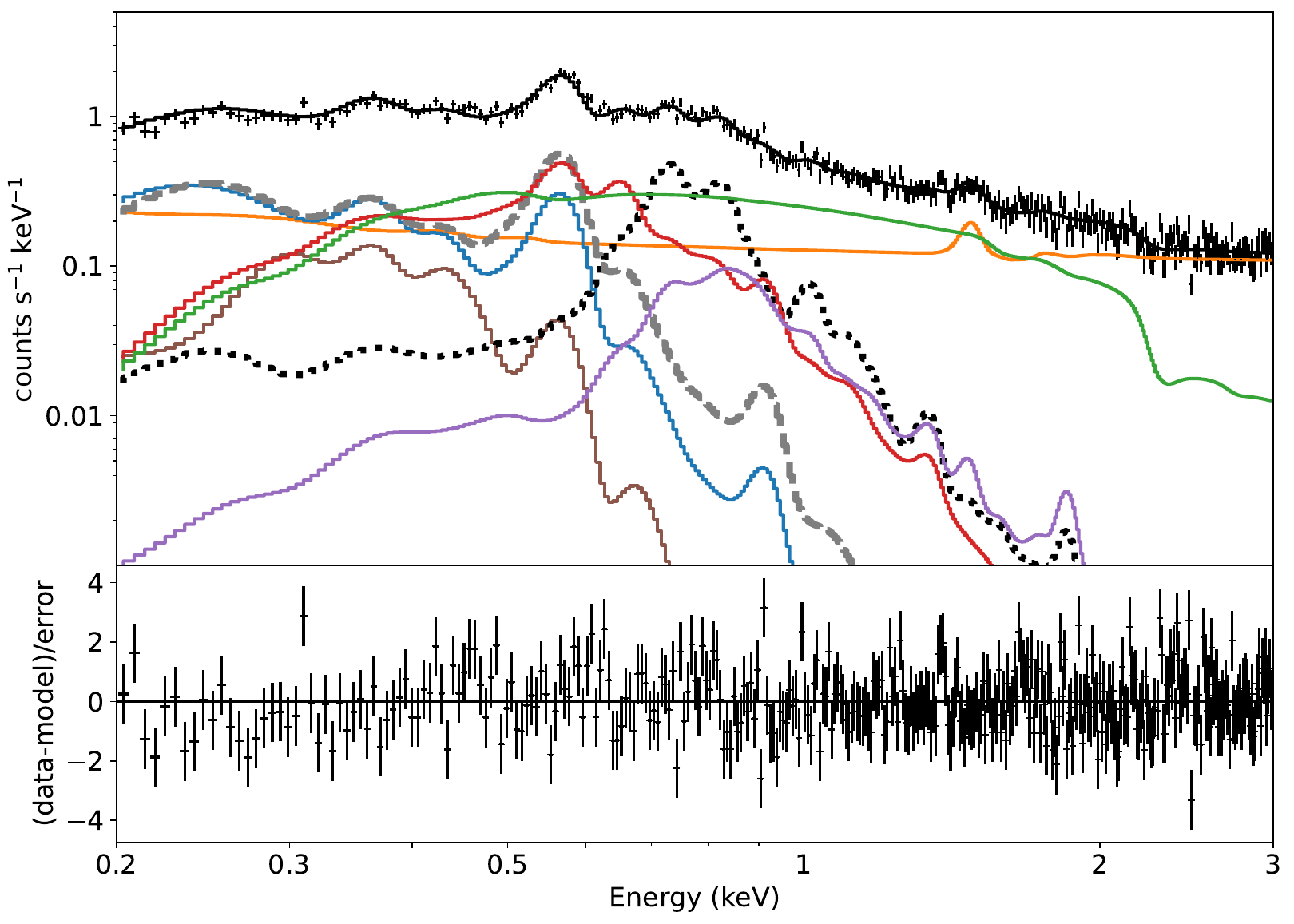}
		\caption{\label{fig:example_arc_SE_two_apec}}
		 \end{subfigure}
\caption{\label{fig:manual_example_spectra}Spectra and best-fit models for some of the manually defined regions. (a) ``arc\_SE" with a single \textsc{tbabs}~$\times$~\textsc{vapec} component and in (b) using \textsc{tbabs}~$\times$~\textsc{vapec+apec} for the source. }
\end{figure*}
Following up on the interesting results of the contour bin analysis that suggest a second plasma component, we performed a similar analysis with the higher statistic spectra of the manually defined regions.
We slightly modified the source model expression to also allow nonsolar abundances for one of the components: \textsc{tbabs}~$\times$~\textsc{vapec}~+\textsc{apec}. The general fitting routine is described in detail in \autoref{sec:spectral_fitting_routine_source}.
We focused on the regions in the southeastern part of the (previously assumed) Monogem Ring, where we expect the highest intensity from a possible second component in addition to the Monogem Ring, based on the X-ray morphology. Additionally, here we obtained significant contribution of a hotter plasma, as shown in \autoref{fig:contour_bin_results_two_apec}. The regions for which we tried the model are listed in \autoref{tab:abundance_fit_two_apec_components_results}. Additionally, we limited the abundances with possible nonsolar values to O, Ne, Si and Fe to reduce the number of free parameters, considering the limited statistics. It is important to note that we also analyzed the regions with a simple \textsc{tbabs}~$\times$~\textsc{vapec} model described in detail in \autoref{sec:manually_defined_analysis} for consistency checks, which shows that the results of the manually defined regions agree very well with the contour bin results when using the same model.

The fit results with this approach are shown in \autoref{tab:abundance_fit_two_apec_components_results}. The spectrum and best-fit model for ``arc\_SE'' is shown in \autoref{fig:manual_example_spectra}, as well as a one-component source model fit to illustrate the necessity of a second source component here. With this model, the spectral fits improved significantly for the "arc\_SE" and "arc\_outer\_E" regions, where we also expect the strongest contribution from the possible second, more distant SNR. The temperatures of the hotter component $kT \approx 0.3-0.5$\,keV are roughly consistent with the previous results, and for the "arc\_SE" region, we also obtain enhanced abundances for O and Fe, where the ratio $\mathrm{Fe/O} > 1$ is typical for a Type Ia supernova (SN). We obtain significant absorption in the upper limits $N_\mathrm{H} < 0.4 \cdot 10^{20}$\,cm$^{-2}$ which coincides with strong UV emission observed by \citet[][Fig. 1, R1]{kim_uv} which could indicate an interaction with denser material here, but due to the large uncertainties this is not conclusive. We also tried to couple the absorption component with both source components; however, the spectral fits remained the same with the absorption component being consistent with zero. This indicates, that the assumption of the unabsorbed \textsc{apec} component describings a plasma located at a close distance holds.

In the central regions, the second component temperature $kT_2$ is slightly higher compared to the average $kT$ of the Monogem Ring (using a single component) but still lower compared to the "arc\_SE" and "arc\_outer\_E" region $kT_1$. If we indeed have two sources overlapping in projection, we most likely measure an average between a hotter plasma and the soft emission of the Monogem Ring plasma. Due to the drop in brightness toward the center, the lower statistics prevent us from disentangling the different plasma. For both the "arc\_SE" and "arc\_outer\_E" regions we obtain a relatively high upper limit of the foreground absorption of the hot thermal source component. However, from the $A_v$ maps, we expect very little absorption of $\approx 0.01 \cdot 10^{22}$\,cm$^{-2}$ at 900\,pc for both regions. Therefore, the absorption is not a good indicator of the distance for those regions, far from the Galactic plane.

In order to probe the possible absorption feature we observe in for example \autoref{fig:soft_gal_dust_600-900pc}, we also extracted a spectrum from the central part of the X-ray enhancement, matching where the $A_v$ contours are the strongest (i.e., second and third level in \autoref{fig:soft_gal_dust_600-900pc}). For this region, we obtain the following expected foreground absorption column, using \autoref{eq:gas-to-dust}: $0.007 \cdot 10^{22}$\,cm$^{-2}$ ($D < 300\,$pc), $0.012 \cdot 10^{22}$\,cm$^{-2}$ ($D < 600\,$pc), and $0.126 \cdot 10^{22}$\,cm$^{-2}$ ($D < 900\,$pc). We used a model with two \textsc{apec} components, where the hotter model was multiplied with a \textsc{tbabs} absorption component. We obtained a foreground absorption best-fit of $N_H < 0.03 \cdot 10^{22}$\,cm$^{-2}$ (90\% CI). While the upper-limit would be consistent with a distance of $\approx 600$\,pc, a distance of 900\,pc is rejected by our spectral fits. We note, that the statistics for this X-ray dark central region are low. Therefore, the result could also indicate a sensitivity limit. Based on the results at hand, the most likely distance to the hotter plasma is $D < 600\,$pc, and possibly in line with the distance of the Monogem Ring.
\begin{table*}
\caption{Results from selected regions in the southeast, fit with a two-plasma spectral model.\label{tab:abundance_fit_two_apec_components_results}}
\renewcommand{\arraystretch}{1.25}
        \begin{center}
        \begin{adjustbox}{width=\textwidth,center}
\begin{tabular}{r r r r r r r r r}
                        Region & $N_\mathrm{H}$ & $kT_1$ & O$_1$ & Fe$_1$ & norm$_1$ & $kT_2$ & norm$_2$ & cstat/dof \\
                        & [$10^{22}$ cm$^{-2}$] & [keV] & [solar] &  [solar] & [cm$^{-5}$\,arcmin$^{-2}$] &[keV] &  [cm$^{-5}$\,arcmin$^{-2}$] &  \\ \hline
                        arc\_E & $0.000_{}^{+0.314}$ & $0.130_{-0.130}^{+0.021}$ & $2.100_{-2.050}^{+7.900}$ & $0.098_{-0.048}^{+9.900}$ & $(0.53_{-0.51}^{+0.32})\cdot 10^{-6}$ & $0.089_{-0.018}^{+0.029}$ & $(1.18_{-0.66}^{+0.45})\cdot 10^{-6}$ & 1.08 \\
                        arc\_S & $0.009_{-0.003}^{+0.008}$ & $0.130_{-0.130}^{+0.009}$ & $1.320_{-0.028}^{+0.251}$ & $0.098_{-0.098}^{+2.390}$ & $(1.46_{-0.62}^{+0.16})\cdot 10^{-6}$ & $0.086_{-0.018}^{+0.037}$ & $(0.49_{-0.44}^{+0.50})\cdot 10^{-6}$ & 1.16 \\
                        arc\_SE & $0.000_{}^{+0.412}$ & $0.426_{-0.043}^{+0.021}$ & $2.080_{-1.030}^{+3.270}$ & $3.490_{-1.320}^{+4.680}$ & $(0.14_{-0.08}^{+0.07})\cdot 10^{-6}$ & $0.125_{-0.004}^{+0.004}$ & $(1.27_{-0.09}^{+0.07})\cdot 10^{-6}$ & 1.07 \\
                        arc\_outer\_E & $0.000_{}^{+0.231}$ & $0.339_{-0.061}^{+0.058}$ & $1.190_{-0.384}^{+0.440}$ & $1.340_{-0.375}^{+0.530}$ & $(0.37_{-0.11}^{+0.16})\cdot 10^{-6}$ & $0.115_{-0.006}^{+0.005}$ & $(0.70_{-0.08}^{+0.06})\cdot 10^{-6}$ & 1.05 \\
                        central & $0.000_{}^{+0.005}$ & $0.187_{-0.006}^{+0.004}$ & $1.320_{-0.205}^{+0.324}$ & $0.098_{-0.098}^{+0.239}$ & $(0.30_{-0.09}^{+0.06})\cdot 10^{-6}$ & $0.082_{-0.007}^{+0.011}$ & $(1.06_{-0.22}^{+0.24})\cdot 10^{-6}$ & 1.09 \\
                        central\_S2 & $0.013_{-0.013}^{+0.055}$ & $0.197_{-0.050}^{+0.058}$ & $0.098_{-0.098}^{+1.290}$ & $2.920_{-1.850}^{+3.900}$ & $(0.20_{-0.10}^{+1.04})\cdot 10^{-6}$ & $0.152_{-0.120}^{+0.003}$ & $(0.79_{-0.39}^{+0.03})\cdot 10^{-6}$ & 1.06 \\
                        central\_W1 & $0.000_{}^{+0.020}$ & $0.164_{-0.021}^{+0.016}$ & $1.140_{-0.320}^{+0.581}$ & $0.098_{-0.098}^{+0.679}$ & $(0.37_{-0.18}^{+0.47})\cdot 10^{-6}$ & $0.107_{-0.030}^{+0.013}$ & $(0.84_{-0.54}^{+0.26})\cdot 10^{-6}$ & 1.09 \\
                \end{tabular}
                \end{adjustbox}
				\tablefoot{
					The source model is \textsc{tbabs~$\times$~vapec$_1$+apec$_2$}, where the absorbed \textsc{vapec} component represents the hotter diffuse emission in the back, and the \textsc{apec} the unabsorbed soft emission in the front. 
				}
        \end{center}
\end{table*}
%
\subsubsection{G190.4+12.5}
\label{sec:UB_spec_ana}
\begin{figure}
	\centering
		\includegraphics[width=0.49\textwidth]{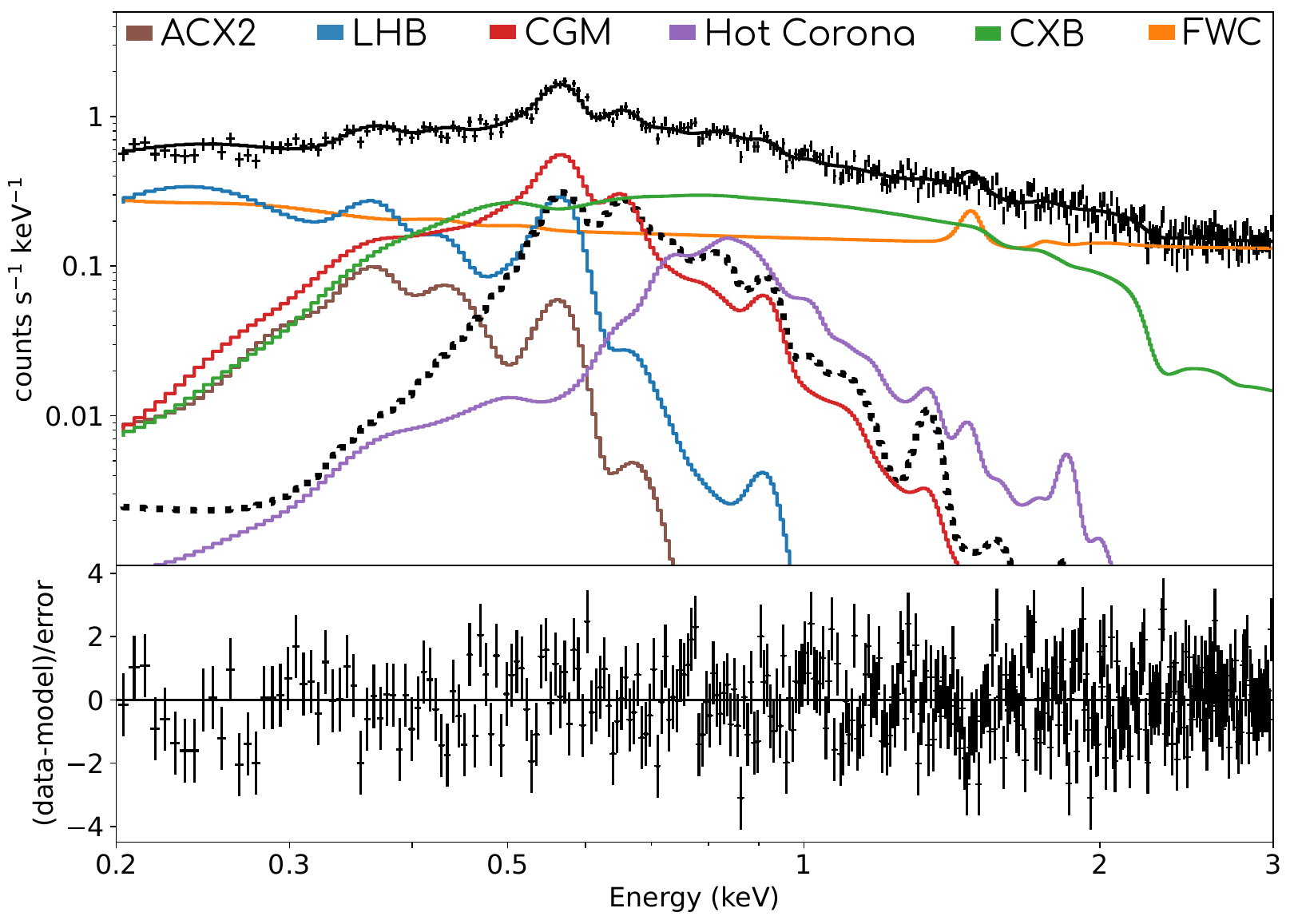}
		\caption{\label{fig:spectrum_UB}Spectrum and best-fit model for G190.4+12.5. The source model is \textsc{tbabs~$\times$~vnei}. The visual representation is similar to \autoref{fig:contbin_example_spectra}.}
\end{figure}
The fit results for G190.4+12.5 are listed in \autoref{tab:northern_SNR_candidate_fit_results} and the spectrum and best-fit model is shown in \autoref{fig:spectrum_UB}. Here, we again used the simpler \textsc{tbabs}~$\times$~\textsc{vnei} model to account for the source emission. With our model, we obtain a good fit with cstat/d.o.f. of 1.08. 
We find a significant foreground absorption to the plasma, with the lower limit of $N_H > 0.17 \cdot 10^{22}$\,cm$^{-2}$ (90\% CI). In comparison, the absorption calculated from the $A_v$ data yields only $N_H \approx 0.12 \cdot 10^{22}$\,cm$^{-2}$ at the maximum distance of the data at 1.9\,kpc \citep{dust_vergely}.
The temperature appears to be higher compared to the Monogem Ring, with $kT = 0.42_{-0.20}^{+0.43}$; however, the uncertainties are large. We also obtain indications for an enhanced O abundance, while the Fe abundance appears to be consistent with solar. The ratio $\mathrm{Fe/O} < 1$ appears to be typical for a core-collapse (CC) SN, but we lack the statistics to decide which type of SN is favored. The ionization timescale $\tau$ is relatively low with $\sim 10^{10}$s\,cm$^{-3}$, which indicates that either the timescale since the ISM was shocked is short, or the plasma density is low.

\begin{table*}
\caption{Spectral fit results for G190.4+12.5.\label{tab:northern_SNR_candidate_fit_results}}
\renewcommand{\arraystretch}{1.25}
        \begin{center}
                \begin{tabular}{r r r r r r r}
                        $N_\mathrm{H}$  & $kT$ & O & Fe & $\tau$ & norm & cstat/dof \\ 
		 \multicolumn{1}{r}{[$10^{22}\,\mathrm{cm}^{-2}$]} & [keV] & [solar] &  [solar] & [s/cm$^3$] &  [cm$^{-5}$\,arcmin$^{-2}$] &  \\
                        \hline
					  $0.29_{-0.12}^{+0.32}$ & $0.42_{-0.20}^{+0.43}$ & $1.84_{-0.42}^{+0.43}$ & $0.73_{-0.21}^{+0.32}$ & $(2.53_{-1.75}^{+11.4})\cdot 10^{10}$ & $1.91\cdot 10^{-7}$ & $1.09$\\
                \end{tabular}
				\tablefoot{
					The source model is \textsc{tbabs~$\times$~vnei}. For the norm, we were unable to determine reliable uncertainties. 
				}
        \end{center}
\end{table*}
%
\subsubsection{Monoceros SNRs}
\begin{figure*}
	\centering
	\begin{subfigure}[t]{0.49\textwidth}
		\includegraphics[width=1.0\textwidth]{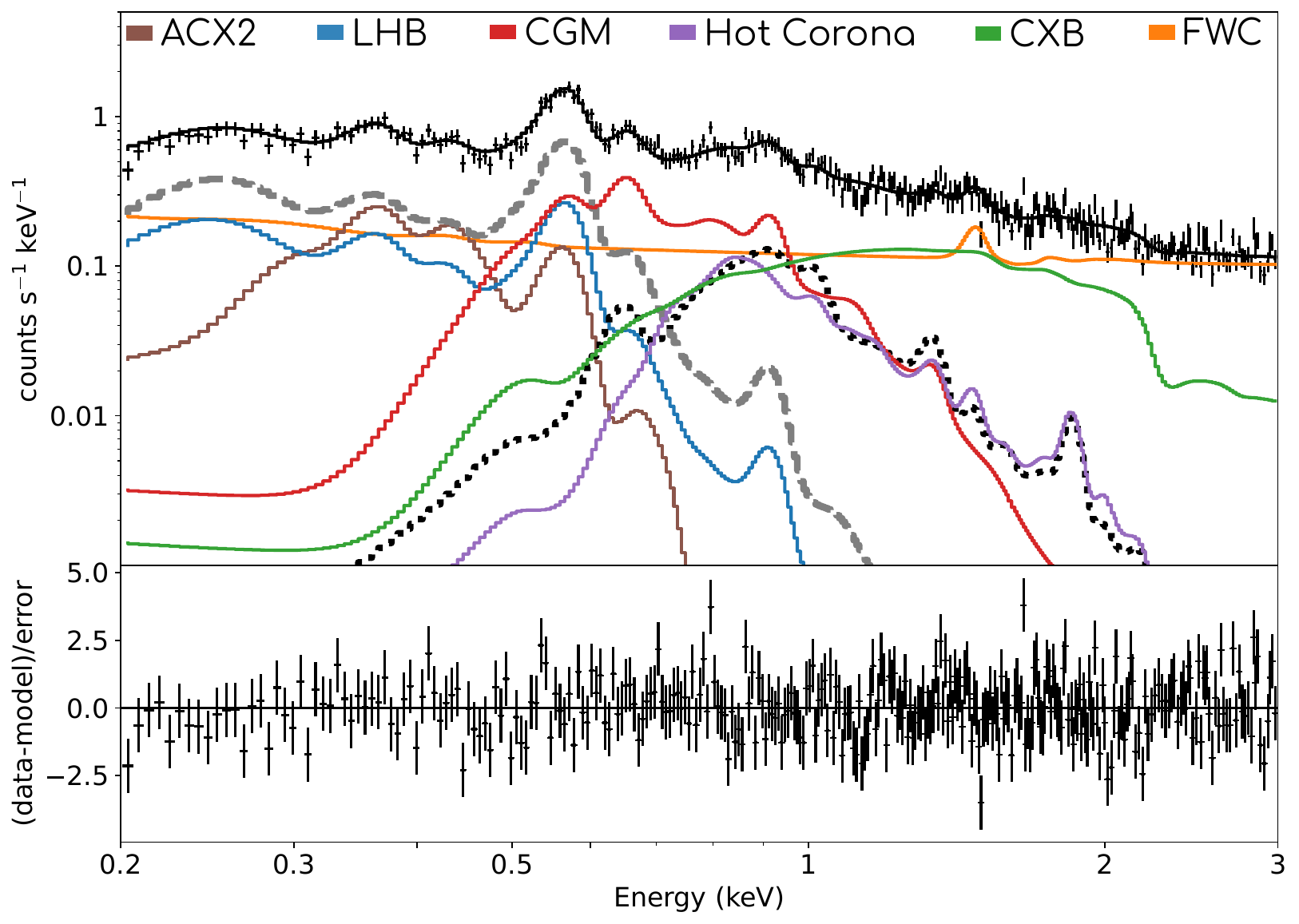}
		\caption{\label{fig:example_MC_two_apec}}
\end{subfigure}
\hfill
\begin{subfigure}[t]{0.49\textwidth}
	\centering
		\includegraphics[width=1.0\textwidth]{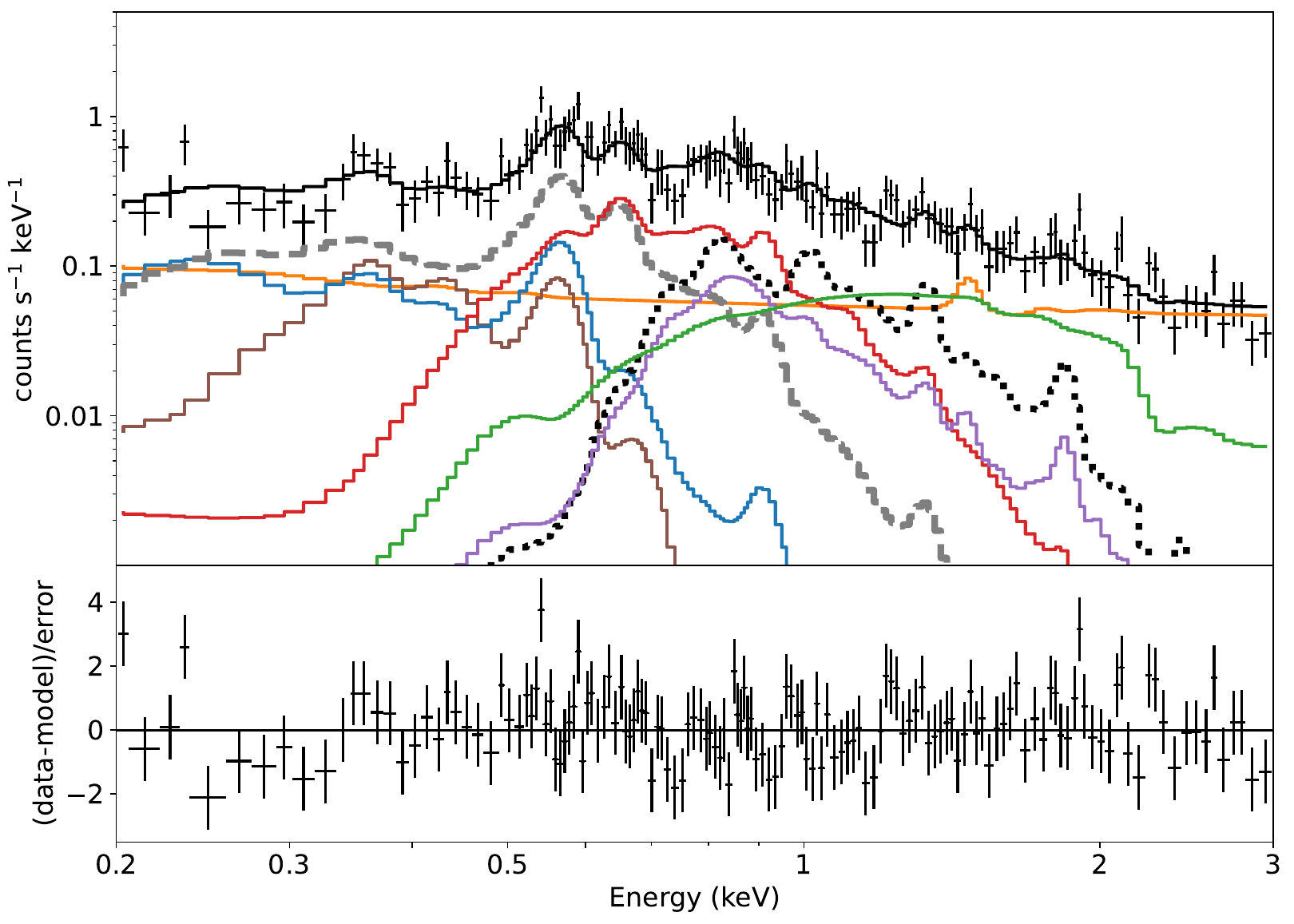}
		\caption{\label{fig:example_MC_hot_spot_two_apec}}
		 \end{subfigure}
\caption{\label{fig:MC_regions_spectra}Spectra and best-fit models for (a) the Monoceros Loop SNR and (b) the ``Hot Spot'' region inside the Monoceros Loop. The spectra were rebinned visually with $3\sigma$ or 30 counts. The source model is \textsc{tbabs~$\times$~vnei+apec} for both. The visual representation is similar to \autoref{fig:contbin_example_spectra}. The second source component (\textsc{apec}) is shown with a thick, dashed gray line.} 
\end{figure*}
\begin{figure}
	\centering
		\includegraphics[width=0.49\textwidth]{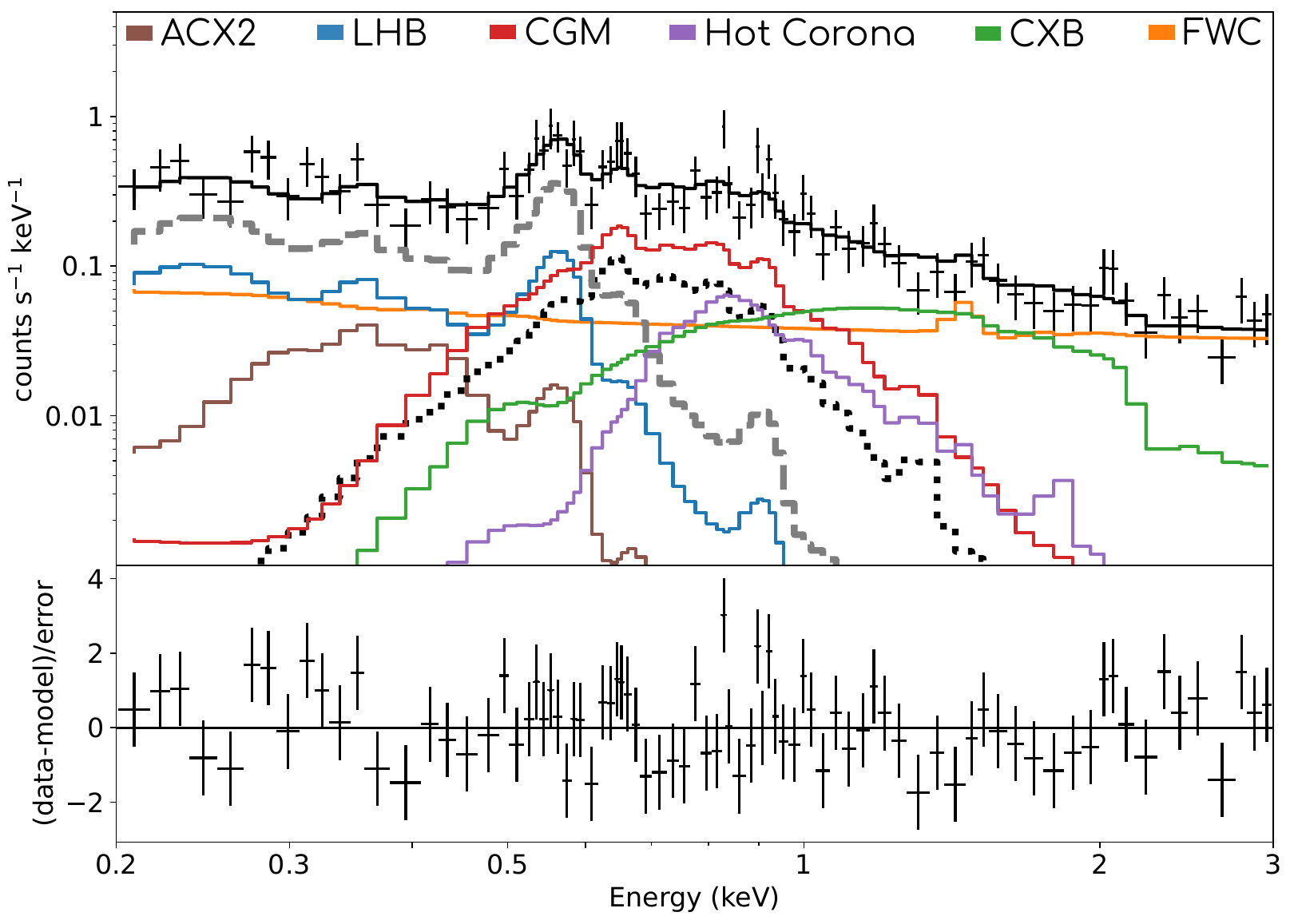}
		\caption{\label{fig:spectrum_PKS}Spectrum and best-fit model for PKS 0646+06. The spectrum was rebinned visually with $3\sigma$ or 30 counts. The source model is \textsc{tbabs~$\times$~vnei+apec}. The visual representation is similar to \autoref{fig:contbin_example_spectra}. The second source component (\textsc{apec}) is shown with a thick, dashed gray line.}
\end{figure}
For the SNRs Monoceros Loop and PKS 0646+06 we were unable to obtain reasonable fits using the simple \textsc{tbabs~$\times$~vnei} model. From the X-ray morphology it is clear that the soft (red) emission of the Monogem Ring overlaps in projection with the SNRs in the Monoceros vicinity. Introducing a second, unabsorbed additional \textsc{apec}, which accounts for the soft Monogem Ring foreground emission, significantly improved the fit quality for these regions. Therefore, we adopted the model \textsc{tbabs~$\times$~vnei$_1$+apec$_2$} in the following.
 
The results are shown in \autoref{tab:monoceros_loop_fit_results}. For all regions, we used the $N_\mathrm{H}$ estimate obtained from $A_v$ (see \autoref{eq:gas-to-dust}) as lower limit for the foreground absorption of the \textsc{vnei} component. We obtain good fits with cstat/d.o.f. $1.03-1.11$ using this model. The temperature of the unabsorbed \textsc{apec$_2$} appears to be consistent between the Monoceros Loop and PKS 0646+06 and the mean temperature for the Monogem Ring. Again, we tried to couple the absorption component with both source components, but this made the fits significantly worse. These results indicate, that we indeed fit the soft foreground emission with the second, unabsorbed plasma component. We tested for significant elemental abundancies as described in \autoref{sec:spectral_fitting_routine_source}, but found the abundancies consistent with solar. This is most likely a result of the lower statistics compared to the Monogem Ring. For the Monoceros Loop "Hot Spot" the unabsorbed \textsc{apec$_2$} temperature appears to be enhanced in comparison. This is most likely caused by the hotter emission of the Monoceros Loop, overlapping in addition to the soft emission of the Monogem Ring, and disentangling the emission is even more complicated. The normalization also appears to be similar and typical to the values derived from the contour bins, where some local variation beyond the statistical error is expected.
 
 The Monoceros Loop appears to have a high temperature of $kT_1 = 0.968_{-0.102}^{+0.165}$\,keV with a low $\tau_1 = (0.94_{-0.29}^{+0.41})\cdot 10^{11}$s\,cm$^{-3}$. The foreground absorbing column is fit with the lower limit, but appears to be constrained toward higher values. A previous study found a much lower temperature of only $\sim 0.2$\,keV \citep{einstein_monoceros_loop}. Most likely, the authors - due to a lack of spatial and spectral resolution - were fitting primarily the foreground emission of the Monogem Ring overlapping with the emission from Monoceros Loop, which lead to a much lower temperature, effectively averaging all plasma components in the line of sight. Interestingly, the "Hot Spot" of the Monoceros Loop, visible as a bright knot in the X-ray images, appears to have a lower temperature. At the same time, the normalization appears to be higher by a factor $\sim 20$ and the $\tau$ indicates that the plasma is in CIE, or has a high density. Possibly, this emission originates from a previously cold, dense cloud that was shocked by the SNR blast wave. We show the spectra and best-fit models for both regions in \autoref{fig:MC_regions_spectra}.
 
For the more distant PKS 0646+06  we also obtained a good fit with the two-component model, with cstat/d.o.f. of 1.06. The temperature also appears to be relatively high with $kT_1 \sim 0.6-1.0$\,keV, while the ionization timescale is low with $\tau_1 < 10^{11}$\,s\,cm$^{-3}$. Here, we had to freeze the foreground absorption to the value derived from $A_v$ due to low statistics, resulting in the parameter not being well constrained otherwise. A previous study by \citet{pks_einstein} found a much lower temperature of $kT \sim 0.1-0.2$\,keV which again leads us to believe that they instead were mainly sensitive to the foreground emission of the Monogem Ring. The spectrum and best-fit model is shown in \autoref{fig:spectrum_PKS}.
\begin{table*}
\renewcommand{\arraystretch}{1.25}
\caption{\label{tab:monoceros_loop_fit_results}Spectral fit results for the Monoceros SNRs.}
        \begin{center}
                \begin{tabular}{r r r r r r r r}       
                        Region & $N_\mathrm{H}$ & $kT_1$ & $\tau$ & norm$_1$ & $kT_2$ & norm$_2$ & cstat/dof \\
                        & [$10^{22}$ cm$^{-2}$] & [keV] & [s/cm$^3$] & [cm$^{-5}$\,arcmin$^{-2}$] &[keV] &  [cm$^{-5}$\,arcmin$^{-2}$] &  \\ \hline
                         MC\tablefootmark{1} & $0.282^{+0.067}$ & $0.968_{-0.102}^{+0.165}$ & $(0.94_{-0.29}^{+0.41})\cdot 10^{11}$ & $(0.23_{-0.04}^{+0.04})\cdot 10^{-6}$ & $0.133_{-0.004}^{+0.005}$ & $(1.50_{-0.11}^{+0.07})\cdot 10^{-6}$ & 1.11 \\
                        MC HS\tablefootmark{2} & $1.040_{-0.208}^{+0.156}$ & $0.467_{-0.107}^{+0.138}$ & $> 1.00\cdot 10^{11}$ & $(5.83_{-2.43}^{+4.62})\cdot 10^{-6}$ & $0.194_{-0.014}^{+0.012}$ & $(1.27_{-0.10}^{+0.10})\cdot 10^{-6}$ & 1.07 \\
                        PKS\tablefootmark{3} & $0.245$ & $0.773_{-0.134}^{+0.279}$ & $(0.66_{-0.40}^{+0.40})\cdot 10^{11}$ & $(0.74_{-0.21}^{+0.17})\cdot 10^{-6}$ & $0.147_{-0.015}^{+0.016}$ & $(2.35_{-0.24}^{+0.34})\cdot 10^{-6}$ & 1.06 \\

                \end{tabular}
				\tablefoot{
					The source model is \textsc{tbabs~$\times$~vnei$_1$+apec$_2$}, where the absorbed \textsc{vnei} component represents the hotter diffuse emission in the back, and the \textsc{apec} the unabsorbed Monogem Ring emission. No significant elemental abundances different from solar were found for the regions.	\\
					\tablefoottext{1}{Monoceros Loop}
					\tablefoottext{2}{Monoceros Loop Hot Spot}
					\tablefoottext{3}{PKS 0646+06}
				}
        \end{center}
\end{table*}
\section{Properties of the SNRs}
\label{sec:properties}
\subsection{Determining the center}
\paragraph{Monogem Ring:}
Determining the SN's point of origin is central to estimating its characteristic properties such as the age and explosion energy. In our previous study, based on ROSAT and Suzaku observations, we determined a center close to PSR B0656+14 \citep{mg_paper_2018}. However, proper motion measurements have indicated that the pulsar seems to be moving toward - rather than away from - the center of the Gemini-Monoceros X-ray enhancement \citep{1994ApJ...421L..13T,pulsar_age}.  With the new data, we obtained a much clearer picture of the diffuse emission in the vicinity of the Gemini-Monoceros X-ray enhancement. Instead of a manual approach based on the morphology, we decided to use the BANANA tool to determine the center of the ring structure, originally developed for bubble detection, based on Minkowski tensors \citep{banana}. This tool provides a relatively unbiased way to determine the center of the diffuse structure, based on the emission. As we have discussed in \autoref{sec:morphology}, the different energy bands show different features, and therefore morphology. We determined the center in two different bands: 0.2-0.4 and 0.4-0.8 keV.

Since BANANA - simply speaking - calculates lines perpendicular to edges, we first processed the images using a Gaussian gradient magnitude (GGM) filtering \citep{GGM_sanders}. This highlights the features we are most interested in: the shell-like filaments in the diffuse emission. We applied the GGM filter with a filter size of 4 pixels to binned images (factor 2) in the respective bands, where we removed the bottom 0.1\% and top 1\% percentiles of the brightness distribution to reduce outliers. Additionally, we excluded point-source candidates with the list already used for the spectral analysis. Holes were filled with surrounding pixel values using the \texttt{astropy:convolve} function. For the 0.4-0.8 keV band we also masked the possible SNR candidate G190.4+12.5.

\begin{figure*}
	\centering
	\begin{subfigure}[t]{0.49\textwidth}
		\includegraphics[width=1.0\textwidth]{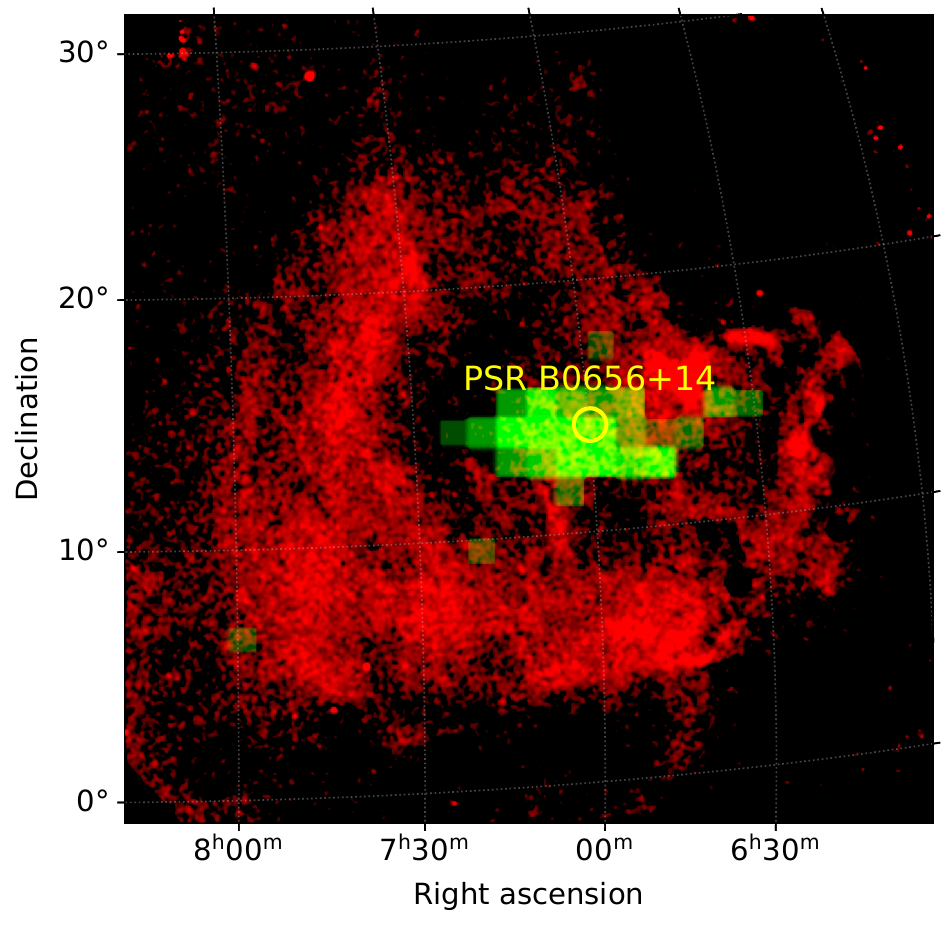}
		\caption{\label{fig:xray_soft_ggm_linedens_s=200}}
\end{subfigure}
\hfill
\begin{subfigure}[t]{0.49\textwidth}
	\centering
		\includegraphics[width=1.0\textwidth]{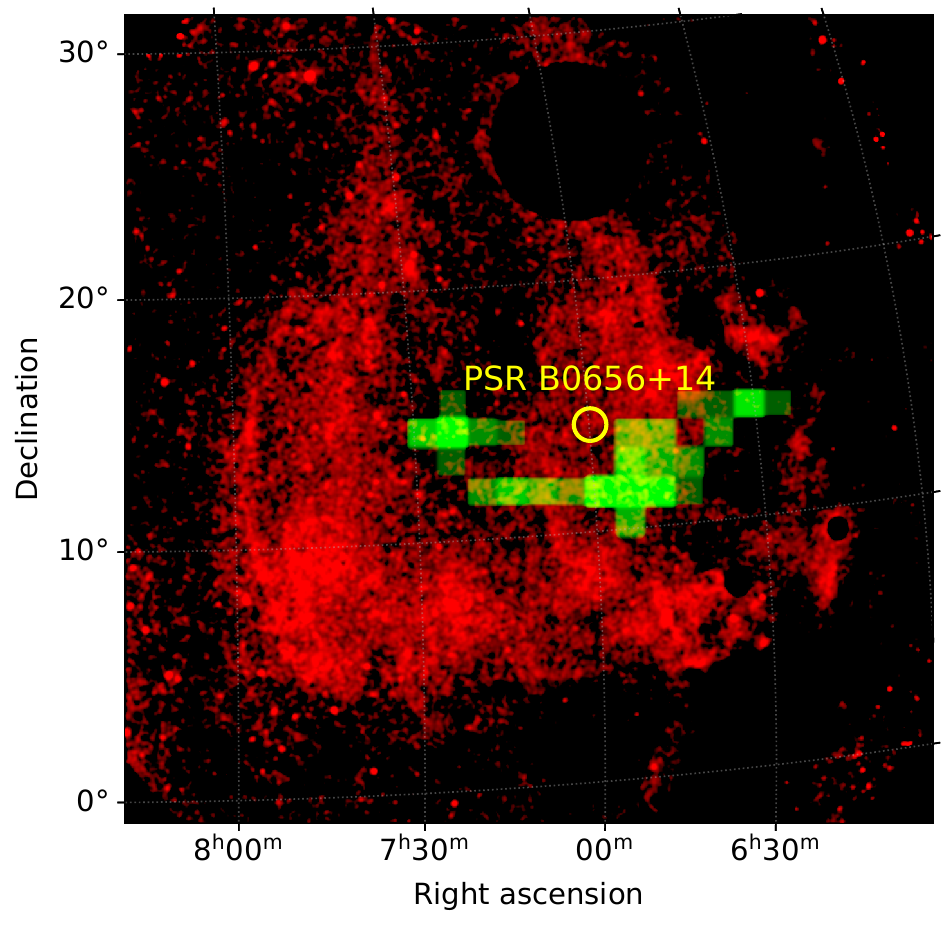}
		\caption{\label{fig:xray_medium_ggm_linedens_s=200}}
		 \end{subfigure}
\caption{\label{fig:ggm_banana}eROSITA X-ray images (red) overlaid with the BANANA results. In green, the line-density maps derived from the GGM convolved X-ray image and BANANA applied are shown. Higher line-densities correspond to brighter colors. In (a) we show in the X-ray image and line density in $0.2-0.4\,$keV and (b) for $0.4-0.8\,$keV. Also shown is the position of the compact object PSR B0656+14 in yellow.
}
\end{figure*}

The GGM convolved maps were then used as input for BANANA. We chose the length of the calculated lines sufficiently long to allow possible intersections over the entire structure. The thresholds for the calculations were based on the 80-99.9\% percentile range of the GGM maps to focus on the strongest features. The line-density maps were smoothed in a final step with a diameter of 200 pixels. This was necessary, since the X-ray morphology is clearly not a well-defined shell as typically seen in young supernova remnants. This leads to the lines not intersecting exactly in the same spot, since they are perpendicular to the edges that are significantly distorted from a perfect circle. Instead, we have an "area" of intersection. Using the smoothing, we can reconstruct the position where the line-density is the highest. The results are shown in \autoref{fig:xray_soft_ggm_linedens_s=200} for 0.2-0.4 keV (soft) and \autoref{fig:xray_medium_ggm_linedens_s=200} for 0.4-0.8 keV (medium).
We find a clear difference in the recovered center position, depending on the analyzed energy band. For the soft band the position with the highest line density is close to the position of the pulsar candidate, albeit with a relatively low accuracy (\autoref{fig:xray_soft_ggm_linedens_s=200}). For the medium band, we find two clear maxima, one close to 7$^h$23$^m$, 14$^{\circ}$ and the other one to the west, with some smaller enhancements to the northwest.

From this analysis, we recover two possible center positions for the Monogem Ring SNR. One is based on the soft emission and appears to be close to the previously assumed center \citep{mg_paper_2018} and associated pulsar position. The second possible center position was calculated from the medium band and appears to be significantly different from the pulsar position.  It is unclear if the diffuse emission originates from the same structure, since the shape of the diffuse emission changes significantly between the different bands. The recovered line-density maxima suggest that we are instead dealing with two distinct overlapping structures: the Monogem Ring SNR being the larger structure, with a center near 6$^h$58$^m$, 11$^{\circ}$30$\arcmin$, and another slightly smaller ring-like structure with the center near 7$^h$23$^m$, 14$^{\circ}$08$\arcmin$ (green arc and edge of green emission \autoref{fig:finding_chart}). Another explanation would be a highly asymmetric and distorted morphology due to a strong surrounding ISM density gradient, as well as the likely high age. However, the fundamentally different properties of the two plasma components in the southeast, that is, the normalization/density and abundancies (\autoref{sec:spectral_analysis_mdr}), imply two distinct sources.
\paragraph{Monoceros Loop, PKS 0646+06 and G190.4+12.5:}
For PKS 0646+06 (G206.9+2.3) we used the center position from the SNR catalog by \cite{snr_catalog_green} at RA,~Dec~$=$~$6^\mathrm{h}48^\mathrm{m}40^\mathrm{s}$, $6^{\circ}$26$\arcmin$, and a radius of $25\arcmin \pm 2.5\arcmin$, resulting in a size of $50\arcmin \times 50 \arcmin$. For the Monoceros Loop we also adopted the established extent and position reported by \citet{snr_catalog_green} in the following, which is given as RA,~Dec~$=$~$6^\mathrm{h}39^\mathrm{m}00^\mathrm{s}$, $6^{\circ}$30$\arcmin$ with $R = 110\arcmin \pm 11\arcmin$.
Since the morphology of the G190.4+12.5 candidate is well defined shell-like, we directly determined the size and center positions from the X-ray images by hand. We defined the circular region so that the shell-like filaments overlapped with the circle. 
For G190.4+12.5, we obtain the center at RA,~Dec~$=$~$6^\mathrm{h}57^\mathrm{m}48^\mathrm{s}$, $25^{\circ}$30$\arcmin$14$\arcsec$ with a radius of $3.3^{\circ} \pm 0.3^{\circ}$. We assume an uncertainty of $10\%$ on the radius for each SNR.

\subsection{Constraining the distance}
\label{sec:distance_constrain}
Without a secure compact object counter-part of an SNR, distance estimates are challenging. One way to indirectly determine the distance of an SNR is by determining the foreground absorption. The absorption by dust is a good tracer for colder material, which  absorbs X-rays on the way from the source to us. The 3D dust absorption map by \cite{dust_lallement} is perfectly suited for constraining the distance to any object that shows anticorrelation in the emission.

To make use of the 3D information in the map by \cite{dust_lallement}, we integrated the dust absorption $A_v$ along the line of sight in distance slices: 0-300 pc, 300-600 pc, 600-900 pc, and 900-1120 pc. Since soft X-rays are more sensitive to absorption \citep{abund_wilm}, we overlayed contours of the dust slices with the soft X-ray maps, as shown in \autoref{fig:soft_gal_dust_300-600pc}-\autoref{fig:soft_gal_dust_900-1120pc}. We do not show the map up to 300 pc since it is relatively feature-less in the vicinity of the Monogem Ring (i.e., no significant absorption). We note, however, that the dust contours trace the outer\_Edges of the X-ray emission to some degree (i.e., anticorrelation). 

In the slice 300-600pc (\autoref{fig:soft_gal_dust_300-600pc}) we already see features that overlap with the X-ray emission. Toward the south the soft X-ray emission shows no anticorrelation with the dust contours. This indicates that this part of the diffuse extended X-ray emission must be located at $\sim 300$\,pc or closer. 
At 600-900\,pc we note several interesting features. 
There appears to be an absorption feature located near the center of the diffuse X-ray emission. This region also shows a low surface brightness in X-rays. Interestingly, the dust contours directly south of the center show no correlation whatsoever with the X-ray emission. However, the $N_\mathrm{H}$ we obtained from the spectral analysis of the region overlapping with the $A_v$ feature limits the distance to $< 600\,$pc.
There are several possible implications from this:
\begin{itemize}
\item If we assume, as in previous studies, that the whole diffuse structure (sans Monoceros Loop and PKS 0646+06) is only one object and located at a distance of $\sim 300$ pc, the apparent anticorrelation in the center is purely coincidental and the absorbing feature instead located behind the diffuse plasma. In this case, the lower background emission due to the absorption feature, together with a lower surface brightness in the center of the remnant might mimic an anticorrelation. The low upper limits of the absorption column from the spectral fits supports the chance coincidence. This would also explain why we see no anticorrelation in other parts of the remnant. 
\item If we instead assume the superposition of two objects at different distances, the absorption feature would put the Monogem Ring SNR at the previously determined $D=300$\,pc, since the absorption features for $D > 300$\,pc overlap with the soft diffuse emission without any anticorrelation. Furthermore, another ring-like diffuse structure would be located at larger distances $> 600$ pc which is visible with green emission in \autoref{fig:xray_rgb_mosaic}. This ring-like structure then would explain the perfect anticorrelation in the center of the remnant between X-rays and the $A_v$ data for the distance slice of $600-900$\,pc. However, our spectral analysis puts an upper limit of $D < 600\,$pc on the plasma distance, whether or not it belongs to the Monogem Ring or another, second SNR. This leaves the possibility of a second SNR at roughly the same distance than the Monogem Ring. As we discuss below, this scenario best fits our findings.
\end{itemize} 
At higher distances $> 900$ pc we see no anticorrelation between the contours and X-rays from the Monogem Ring SNR anymore. Therefore it is safe to assume that most of the diffuse emission must be located $< 900$ pc (except the Monoceros SNRs and the new SNR candidate).

\subsection{Discussion}
\begin{figure*}
	\centering
	\begin{subfigure}[height=3cm]{0.49\textwidth}
		\includegraphics[width=1.0\textwidth]{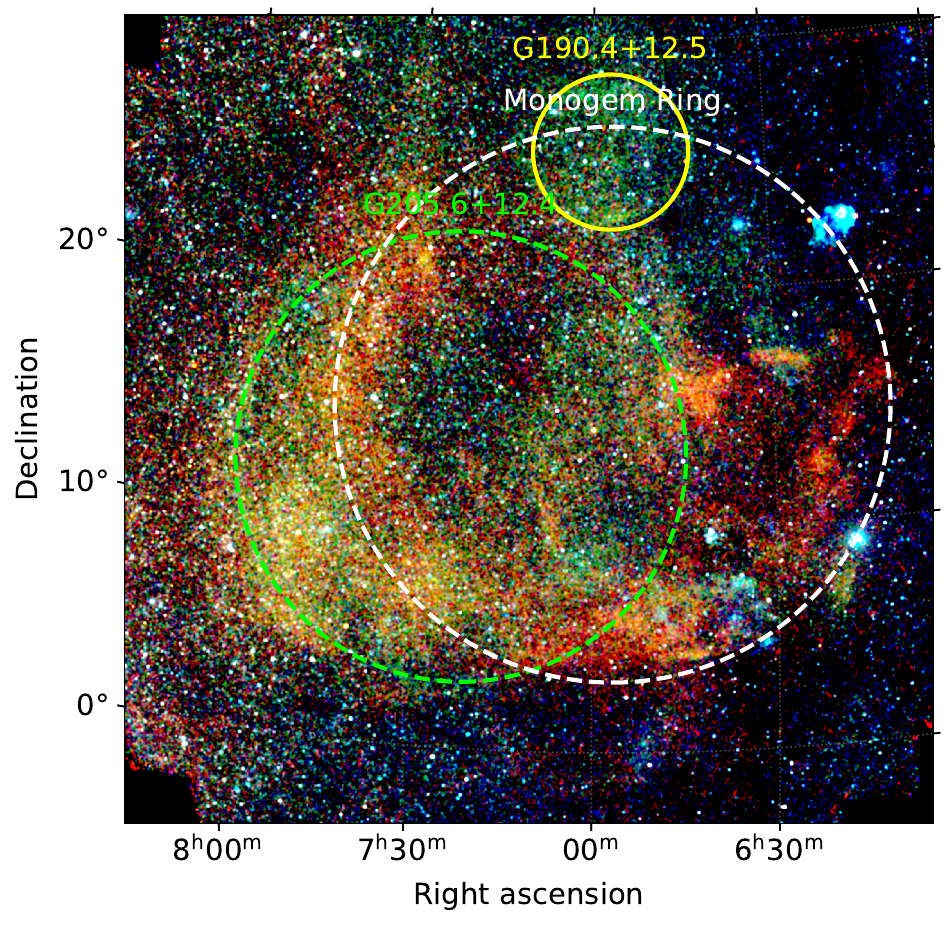}
		\caption{}
\end{subfigure}
\hfill
\begin{subfigure}[height=3cm]{0.49\textwidth}
	\centering
		\includegraphics[width=1.0\textwidth]{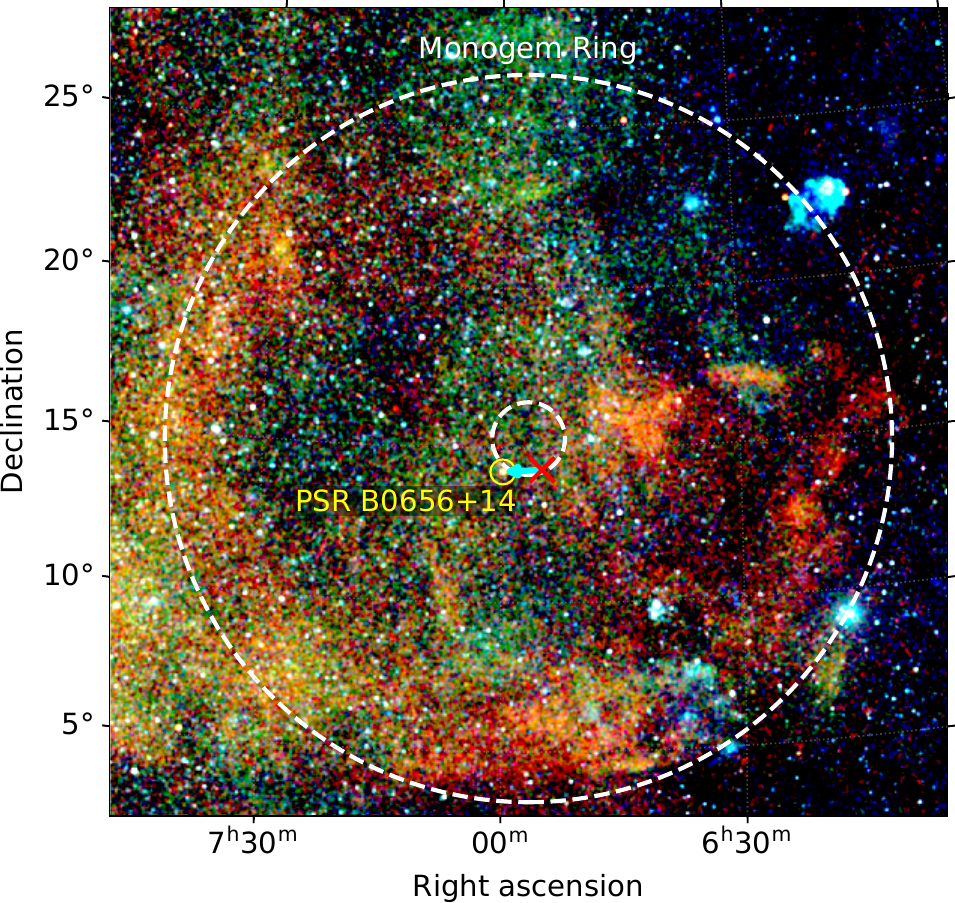}
		\caption{}
		 \end{subfigure}
\caption{\label{fig:new_geometry}Overview of the SNR and pulsar geometries, and their relation. (a) eROSITA X-ray mosaic as in \autoref{fig:xray_rgb_mosaic}, with the estimated geometry of the SNRs based on the X-ray morphology. The Monogem Ring is shown with dashed white, G205.6+12.4 in dashed green, and G190.4+12.5 in yellow. (b) eROSITA X-ray mosaic as in \autoref{fig:xray_rgb_mosaic}, with the newly found geometric center of the Monogem Ring as a white dashed circle, where the radius indicates the uncertainty ($R=1.1^{\circ}$). The current position of PSR B0656+14 as yellow circle, and a cyan arrow pointing from the birthplace (red x) to the pulsar, as derived from its proper motion \citep{distance_psr} and assuming an age of $t = 10^{5}$\,yr. 
}
\end{figure*}

From the spectral analysis and multiwavelength data, we have indications that the Monogem Ring SNR could instead be composed of two individual SNRs. The completely different plasma properties near the ``green arc'' (\autoref{fig:finding_chart}) make it difficult to just explain the difference by an asymmetry in the expansion or initial explosion alone ($T$, norm, $\tau$, abundance). In addition, if we assume two distinct SNRs instead of one, this shifts the approximate geometric center of the Monogem Ring to the northwest in the frame of \autoref{fig:finding_chart} which could solve the long-standing discrepancy of the proper motion of PSR B0656+14, which appeared to be  moving toward the geometric center of the SNR \citep{1994ApJ...421L..13T}, even with newer proper-motion measurements \citep{pulsar_age}. With the geometry as shown in \autoref{fig:new_geometry} we obtain a geometric center located very close to the birthplace of the pulsar, when assuming an age of $t = 10^{5}$\,yr and taking the proper motion into account. In this new geometric configuration, the current proper motion of the pulsar roughly points away from the center, as expected. We assumed an uncertainty of $10\%$ relative of the SNR radius, that is, $\approx 1.1^{\circ}$, on the center position. Indeed, the pulsar birthplace agrees within uncertainties and lies just on the edge of our estimated center. Moreover, the center of the new candidate G205.6+12.4 is only $2.2^{\circ}$ away from the center recovered from the medium band with BANANA, as shown in \autoref{fig:xray_medium_ggm_linedens_s=200}. A small difference is expected since the true dynamic center is difficult to determine precisely for very old remnants. Therefore, assuming two remnants at roughly the same distance ($D=300-500\,$pc) elegantly explains all our findings without the need to rely on extreme asymmetries in the SNR evolution or SN explosion. In the following we consider that the previously believed monolithic structure ``Monogem Ring'' consists of two individual old SNRs. However, we stress that assuming only one remnant at $D=300\,$pc, our results are still in line with previous findings, as we discuss later, albeit with the aforementioned strong assumptions on the explosion and evolution, as well as the different plasma properties.
\subsection{3D model}
\begin{figure}
	\centering
		\includegraphics[width=0.49\textwidth]{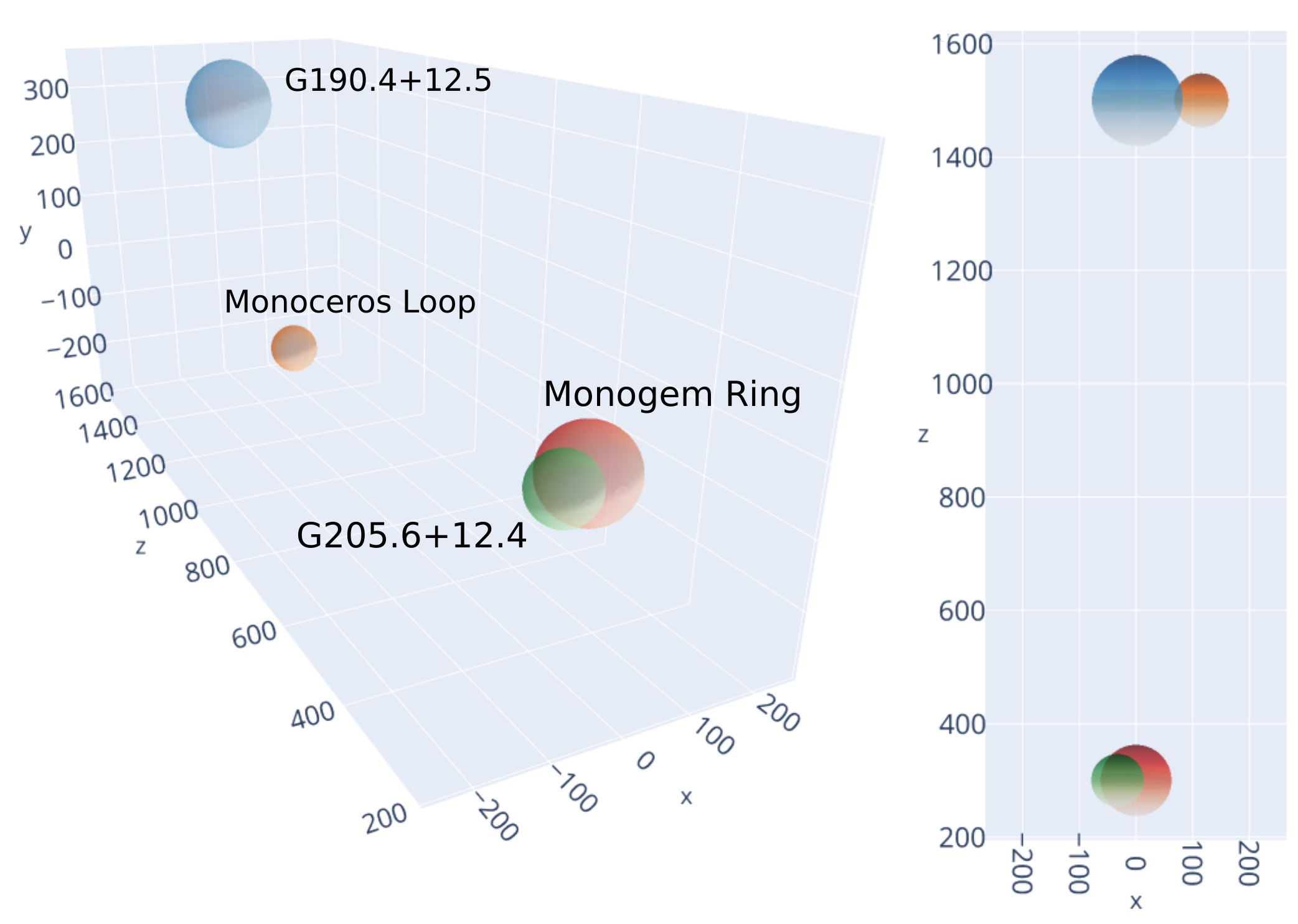}
		\caption{\label{fig:3d_geometry}3D geometrical model of the Gemini-Monoceros X-ray enhancement. The different panels show the perspective (left) and orthographic top view (right). The units of the coordinates are pc, with the axes: $x$ and $y$ being the Right ascencion and Declination distance relative to the Monogem Ring, and the distance $z$ from us. The Monogem Ring is located at $(x,y,z) = (0,0,300)$. The different colored spheres indicate the different SNR (candidates) with the correct radius at the respective assumed distance. The Monogem Ring and G205.6+12.4 are shown in red and green, respectively, at $D= 300$\,pc. The Monoceros Loop SNR at $D=1500\,$pc in orange, and G190.4+12.5 in blue at $D = 1500\,$pc (assumed distances). The radii are to scale.}
\end{figure}

For a better visualization of the geometry of the three SNRs in the proximity of the Monogem Ring, we created a 3D view shown in \autoref{fig:3d_geometry}. We assumed the distances as indicated in the figure, and projected the positions relative to the Monogem Ring to obtain $x$, $y$, and $z$ coordinates in units pc. The Monogem Ring is situated at $(x,y,z) = (0,0,300)$ in this view.
\subsection{Estimations}
We used the results from the spectral analyses to estimate the characteristics of the SNRs and SNR candidates. For this, we used a modified version of SNRpy by \citet{snrpy}. Originally, this tool provides a graphical interface to calculate SNR properties based on a set of input parameters and SNR evolution models (e.g., Sedov-Taylor).
We modified the code to provide a grid of SNR evolutionary models for a range of input parameters, such as different ambient densities, ages, and explosion energies.
We calculated grids with 15,000 distinct SNR models by varying the input parameters for two evolutionary models: Sedov-Taylor \citep[e.g.,][]{sedov_original, sedov_taylor_jk} and the cloudy ISM model \citep{cloudy_ism_reference}. Despite the name, all evolutionary phases of an SNR are calculated for both models (not limited only to the Sedov phase).

The parameter ranges for the grid were determined as follows.
The explosion energy $E_0$ of the supernova is widely unknown without additional information, therefore, we constrained the parameter range to a reasonable range of $E_0 = (0.1-2.0)\cdot 10^{51}\,$erg, based on simulations \citep{mueller_simulations_explo}. From the spectral analysis we can derive the pre-shock density, as shown below. We used our lowest derived value as a lower limit of $n_0$, and literature values as upper limits. We take the value of $n_0 \approx 0.06\,$cm$^{-3}$ from the study by \citet{ism_density_snrs} as the upper limit, which appears to be consistent with other measurements, as the authors show. For the values derived from our spectral analysis we took the respective uncertainties into account. In summary, the grids cover the ranges of $n_0 = 0.5\cdot 10^{-3}-0.06\,$cm$^{-3}$ for the ambient density, $E_0 = (0.1-2.0) \cdot 10^{51}$\,erg for the explosion energy, and $t_0 = 5000-140,000$\,yr for the age.

In the next step, to constrain the possible SNR evolutionary parameter space, we used the X-ray temperature derived from the $kT$ values of the spectral analysis results. We compared this with the predicted X-ray temperature of the respective model. We take the emission measure (EM) weighted temperature, which should be close to the measured X-ray temperature of the plasma \citep{snrpy}. The models and regions from which we take the input parameters are also referenced in \autoref{tab:parameters_snrpy}. For the Monogem Ring SNR, we also estimated the plasma properties from the annulus shown in \autoref{fig:new_geometry_single_snr}, that is only taking the brighter shell parts into account, to derive median values from the spectral fit parameters. The parameters were estimated from the contour bin regions \autoref{fig:contour_bin_results_modelB_v2}.

\begin{figure*}
	\centering
	\begin{subfigure}[t]{0.49\textwidth}
		\includegraphics[width=1.0\textwidth]{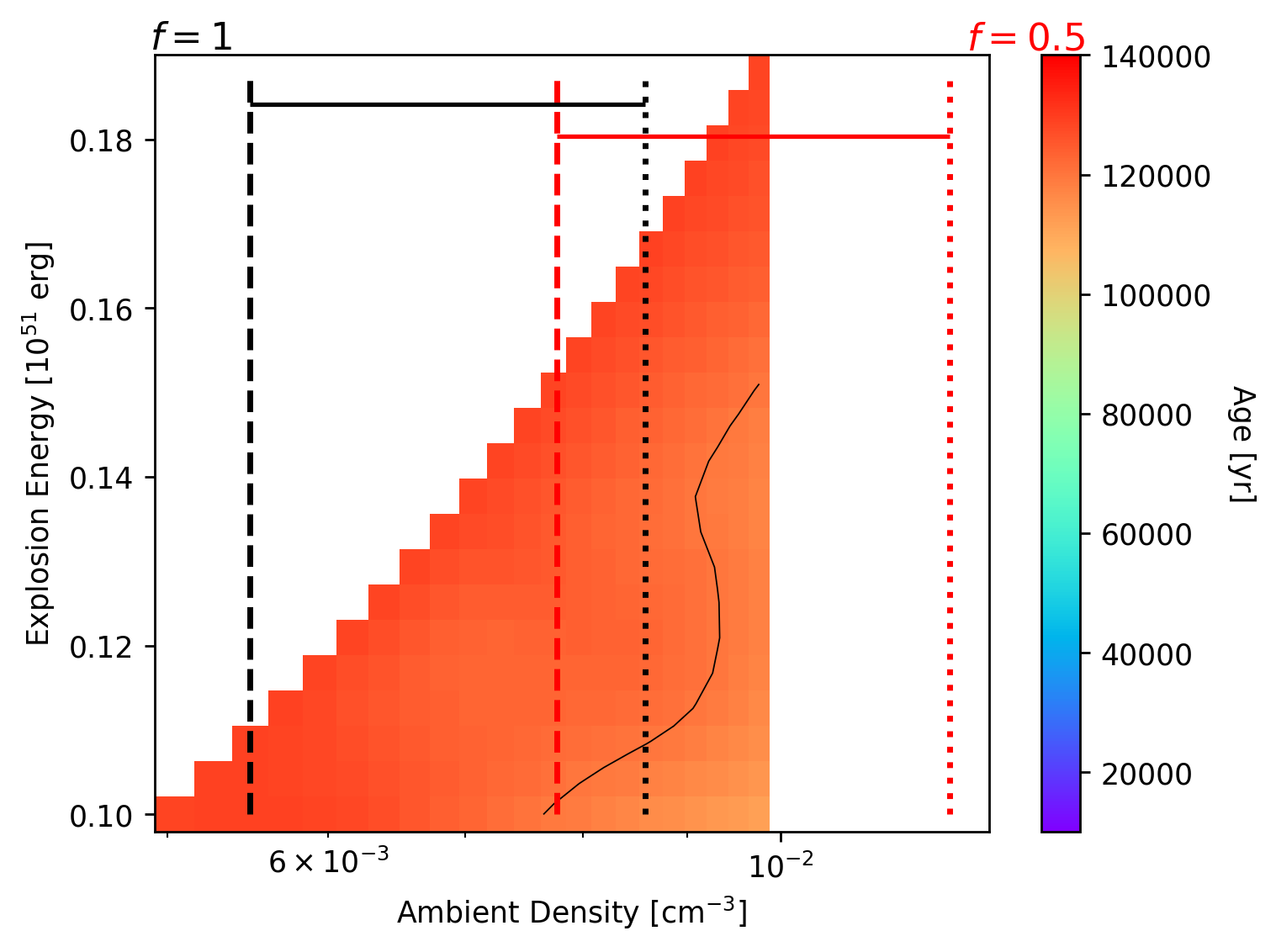}
		\caption{\label{fig:mg_sedtay}}
\end{subfigure}
\hfill
\begin{subfigure}[t]{0.49\textwidth}
	\centering
		\includegraphics[width=1.0\textwidth]{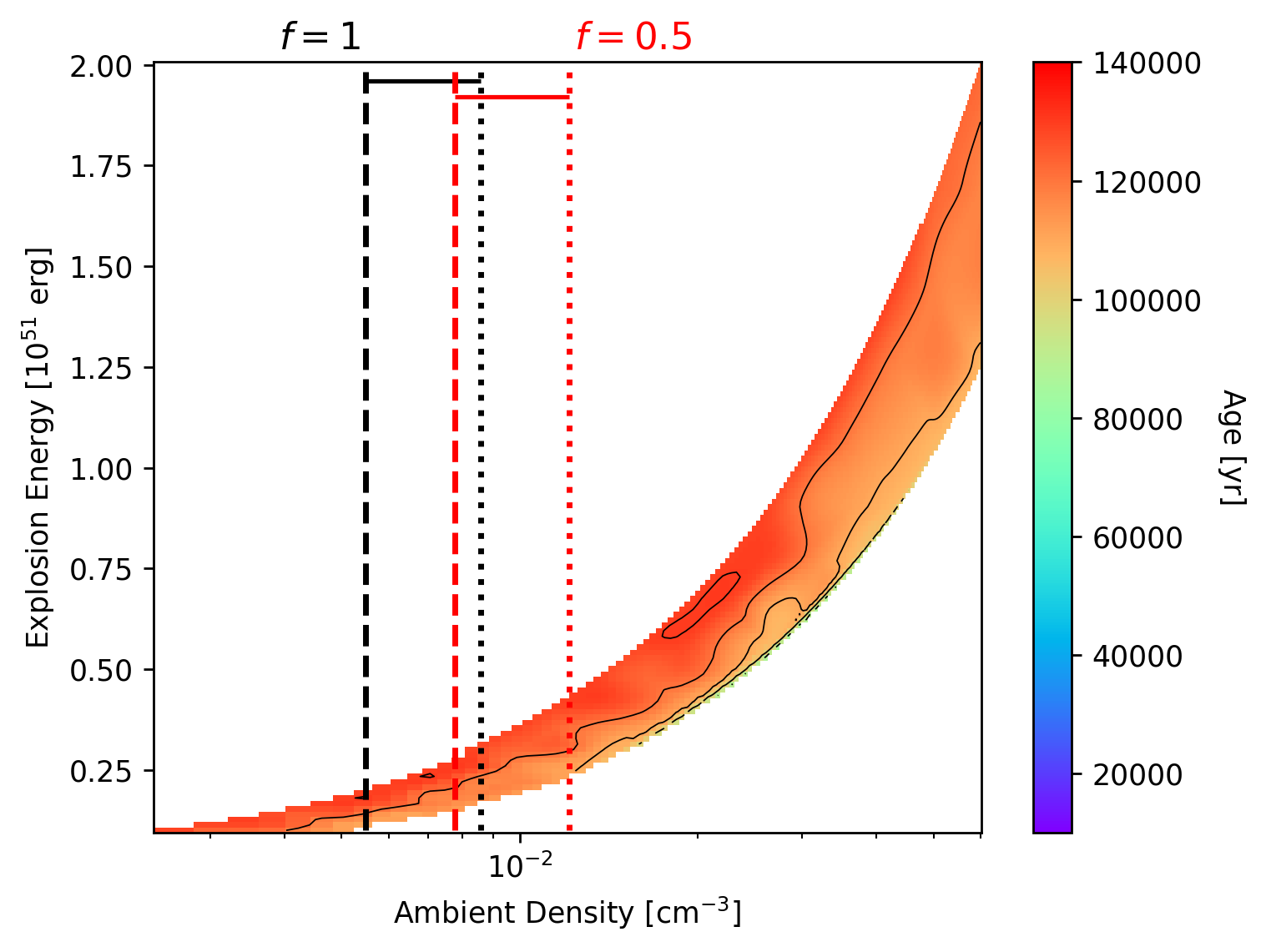}
		\caption{\label{fig:mg_cism}}
\end{subfigure}
\caption{\label{fig:mg_snr_estimates}SNR model parameters for the Monogem Ring SNR. We assumed a radius of $58$\,pc at $D=300$\,pc. The estimated uncertainty on the radius is $10\%$. The color scale indicates the calculated age of the respective model. The vertical dashed and dotted lines show the ambient ISM density $n_0$ lower and upper limits, respectively, for different filling factors. The density $n_0$ was inferred from the spectral analysis for $D=300$\,pc. The black dashed and dotted lines show $n_0$ for $f=1$, while the red lines show $n_0$ for $f=0.5$. Models with predicted radii and X-ray temperatures outside the derived confidence intervals are not shown. The SNR evolutionary models are Sedov-Taylor (a) and Cloudy ISM (b) for $\tau = 2$ \citep{snrpy}. The color scale for the age is in absolute values for better comparability with the other SNRs. The black solid contours on top of the surface plot denote different levels of the age in steps of $10,000$\,yr, similar to the color scale shown on the right.}
\end{figure*}

From the spectral analysis, we can derive the ISM density with the normalization $\eta$ of the \textsc{apec} or \textsc{nei} model via the EM, given as:
\begin{equation}
\mathrm{EM} =  \eta \cdot \left(\frac{10^{-14}}{4\pi D^2}\right)^{-1}= \int n_\mathrm{e} n_\mathrm{H} dV \mathrm{ .}
\label{eq:EM}
\end{equation}
As shown by \citet{ism_density_formula_p}, for SNRs in the Sedov phase with a radial density profile the EM is related to the electron and hydrogen densities by 
\begin{equation}
\mathrm{EM} = 2.07\left(\frac{n_e}{n_\mathrm{H}}\right)n_\mathrm{H,0}^2 V f \mathrm{ ,}
\label{eq:density_p}
\end{equation}
where V is the SNR's volume, f the filling factor and $n_e$, $n_\mathrm{H}$ the electron and hydrogen densities of the plasma, respectively.
From this, the pre-shock density $n_0$, that is the ambient ISM density, can be estimated by assuming $n_e/ n_\mathrm{H} = 1.21$ and $n_0 \approx 1.1 n_\mathrm{H,0}$ \citep{ism_density_patrick}.
To estimate the volume $V$, we assumed a sphere with the center positions and radii shown in \autoref{tab:parameters_snrpy} for the respective SNR.
The pre-shock density $n_0$ was used to further narrow down the range of possible SNR evolutionary parameters. Finally, an additional constraint was put by comparing the predicted SNR radius with the radius estimated from the geometry of the respective remnant, assuming an uncertainty of $10\%$.
\begin{table*}
\caption{\label{tab:parameters_snrpy}Parameters used for the estimation of the SNRs' properties.}
\renewcommand{\arraystretch}{1.25}
\begin{center}
\begin{tabular}{r|rrrlrrr}
Remnant/Distance & Center (RA, Dec) & Radius & Region & Model & $kT$ [keV] & $n_0 f^{-1/2}$ [$10^{-3}\,$cm$^{-3}$] & R [pc]\\ \hline
\hline & \multicolumn{6}{c}{Single SNR at $D=300$\,pc}  \\
Monogem Ring & 107.08$^{\circ}$, 13.50$^{\circ}$ & 11.97$^{\circ}$ & annulus & \autoref{fig:contour_bin_results_modelB_v2} & $0.11-0.17$  &  & \\
$D=300$\,pc & $(l, b) \sim (202.7^{\circ}, 9.8^{\circ})$ &  & (median) & \autoref{fig:new_geometry_single_snr} & & $3.1-7.2$ & 63 \\
\hline & \multicolumn{6}{c}{Two SNRs at $D=300\,$pc}  \\
Monogem Ring & 104.15$^{\circ}$, 15.30$^{\circ}$ & 11.00$^{\circ}$ & diffuse\_S & \autoref{tab:abundance_fit_results} & $0.11-0.13$  & $5.5-8.6$ & 58\\
 & $(l, b) \sim (199.8^{\circ}, 8.0^{\circ})$ &  & &  & &  &  \\
G205.6+12.4 & 110.33$^{\circ}$, 12.94$^{\circ}$ & 8.82$^{\circ}$ & arc\_outer\_E & \autoref{tab:abundance_fit_two_apec_components_results} & $0.28-0.45$ &  & 47 \\ 
 &  &  & + arc\_SE &  &  & $2.0-4.3$ &  \\ 
\hline
G190.4+12.5 & 104.24$^{\circ}$, 25.56$^{\circ}$ & 3.07$^{\circ}$ & - & \autoref{tab:northern_SNR_candidate_fit_results} & $0.22-0.85$ & & \\
$D=1000$\,pc &  &  & &  & & $1.9-3.3$ & 53 \\
$D=1500$\,pc &  &  & &  & & $1.5-2.7$ & 80 \\
$D=2000$\,pc &  &  & &  & & $1.3-2.3$ & 107 \\   
\hline & \multicolumn{6}{c}{Monoceros Loop SNRs}  \\
Monoceros Loop\tablefootmark{$\dagger$} & 99.75$^{\circ}$, 6.50$^{\circ}$ & 1.83$^{\circ}$ & - & \autoref{tab:monoceros_loop_fit_results}\tablefootmark{$\ddagger$} & $0.87-1.13$ &  \\
$D=1500$\,pc & $(l, b) \sim (205.5^{\circ}, 0.5^{\circ})$ &  & &  & & $2.3-3.7$ & 48 \\
\hline
PKS 0646+06\tablefootmark{$\dagger$} & 102.16$^{\circ}$, 6.42$^{\circ}$ & 0.42$^{\circ}$ & - & \autoref{tab:monoceros_loop_fit_results}\tablefootmark{$\ddagger$} & $0.64-1.05$ &  \\
$D=4000$\,pc & $(l, b) \sim (206.9^{\circ}, 2.3^{\circ})$  &  & &  & & $4.9-8.7$ & 29 \\ 
\end{tabular} 
\tablefoot{
	For the radius, we assume an uncertainty of $10\%$, while the temperature and density uncertainties are based on the 90\% CI of the spectral analysis results. For the density, $f$ denotes the filling factor of the plasma. Galactic coordinates are listed with $(l, b)$ where an alternative name was used.	\\
	\tablefoottext{$\dagger$}{Position and size adopted from \citet{snr_catalog_green}.}
	\tablefoottext{$\ddagger$}{Hot \textsc{vapec} plasma component used.}
}
\end{center}
\end{table*}
\paragraph{Monogem Ring:}
The SNR evolutionary models for the Monogem Ring are shown in \autoref{fig:mg_snr_estimates} assuming a distance of $D=300\,$pc and geometry as shown in \autoref{fig:new_geometry}, by assuming two remnants. The SNR models agree with our analysis results for relatively high ages $\approx 110,000$ yr, higher by about $\sim 30\%$ to previous age estimates of $\sim 70,000-86,000$\,yr \citep{mg_rosat, mg_paper_2018}. Our estimated ambient density agree with both the pure Sedov-Taylor model and the cloudy ISM model. 
Both models suggest a low explosion energy, with the Sedov-Taylor estimation lower by at least a factor of two with $E_0 \sim (0.1-0.2)\cdot 10^{51}\,$erg. With the cloudy ISM model we obtain $E_0 \sim (0.2-0.4)\cdot 10^{51}\,$erg which appears more realistic, suggesting that colder, dense ISM was situated in the vicinity prior to the explosion. 

The multiwavelength observations in the vicinity show that the Monogem Ring could be surrounded by denser ISM in several directions. Toward the Galactic disk we observe significant $H\alpha$ emission that appears to be anticorrelated with the Monogem Ring. Additionally,\citet{kim_uv} observed UV emission here, which they interpreted as possible interactions between the SNR's shock and colder, dense ISM. Toward the south (\autoref{fig:xray_rgb_mosaic} frame), close to the region ``diffuse\_S,'' they also observed strong UV emission. This indicates, that the remnant could indeed be partially surrounded by dense ISM, favoring the cloudy ISM model. With this model, the mean age consistent with the density estimates is $\sim 1.1-1.2\cdot 10^5$\,yr which agrees surprisingly well with the age determined for the pulsar PSR B0656+14 to $\sim 1.1 \cdot 10^5\,$yr \citep{pulsar_age, pulsar_new_study}. 

If we instead assume a single SNR as shown in \autoref{fig:new_geometry_single_snr} and use the parameters for a single SNR described in \autoref{tab:parameters_snrpy}, we obtain an age higher by only $\sim 5\%$ with similar explosion energies. The combination of a larger radius, lower density and larger temperature range balance each other. In that scenario, the asymmetric morphology and high-temperature region in the southeast might be explained by a blow-out after encountering a denser region \citep{reynolds_snr}. Another possibility would be a highly asymmetric initial explosion together with a very strong surrounding ISM density gradient prior to the explosion.

\begin{figure}
	\centering
		\includegraphics[width=0.49\textwidth]{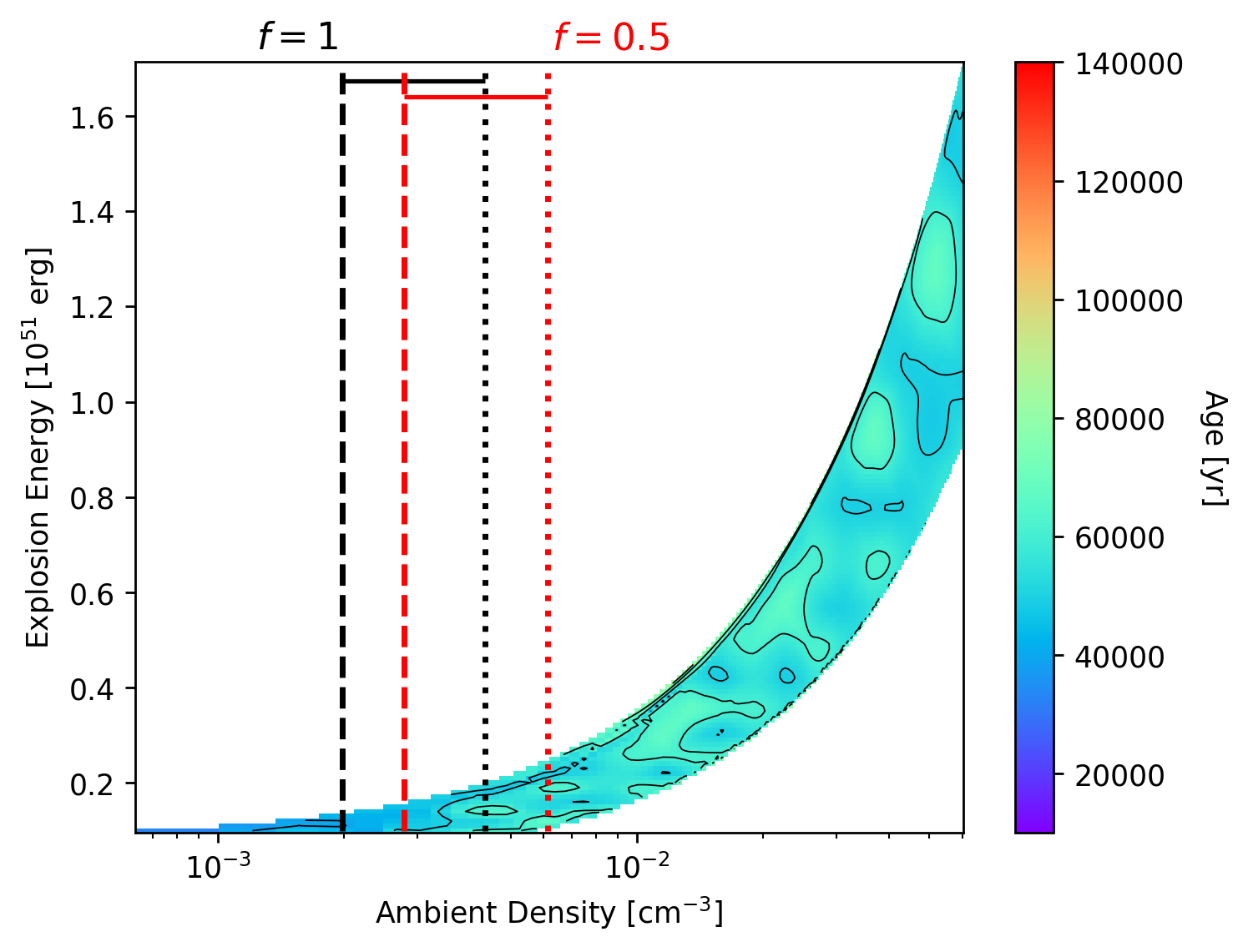}
		\caption{\label{fig:mg_green_sedtay}SNR model parameters similar to \autoref{fig:mg_snr_estimates}, but for the G205.6+12.4, using the Sedov-Taylor model. The assumed distance is $D=300$\,pc, which yields a radius of 47\,pc.}
\end{figure}

\paragraph{SNR candidate G205.6+12.4:}
If we assume the geometry as shown in \autoref{fig:new_geometry}, we obtain the parameters for the SNR model calculations as shown in \autoref{tab:parameters_snrpy}. Due to the slightly larger distance from the Galactic plane, we expect less dense or clumpy ISM. Therefore we show only the Sedov-Taylor estimation in this discussion. The estimated ambient ISM density is also consistent with a higher distance from the Galactic plane. We note, that we also obtain valid solutions with the clumpy ISM model (with a slightly lower predicted age). The possible SNR parameters are shown in \autoref{fig:mg_green_sedtay} for an assumed distance of $D=300$\,pc. Within our estimated ambient density range, we obtain ages from $40,000-55,000$\,yr. The possible explosion energies appear to be similar to the Monogem Ring, with $E_0 \sim (0.2-0.3)\cdot 10^{51}\,$erg. 

We also calculated models for $D=200$, however, here we do not find any SNR model that matches with our properties derived from the spectral analysis. For higher distances we do get SNR models  up to $500\,$pc that match our results, with the main difference being a slightly higher age and explosion energy with increasing distance.
Therefore, we can constrain the distance to the SNR candidate to $\approx 300-500$\,pc.

\begin{figure}
	\centering
		\includegraphics[width=0.49\textwidth]{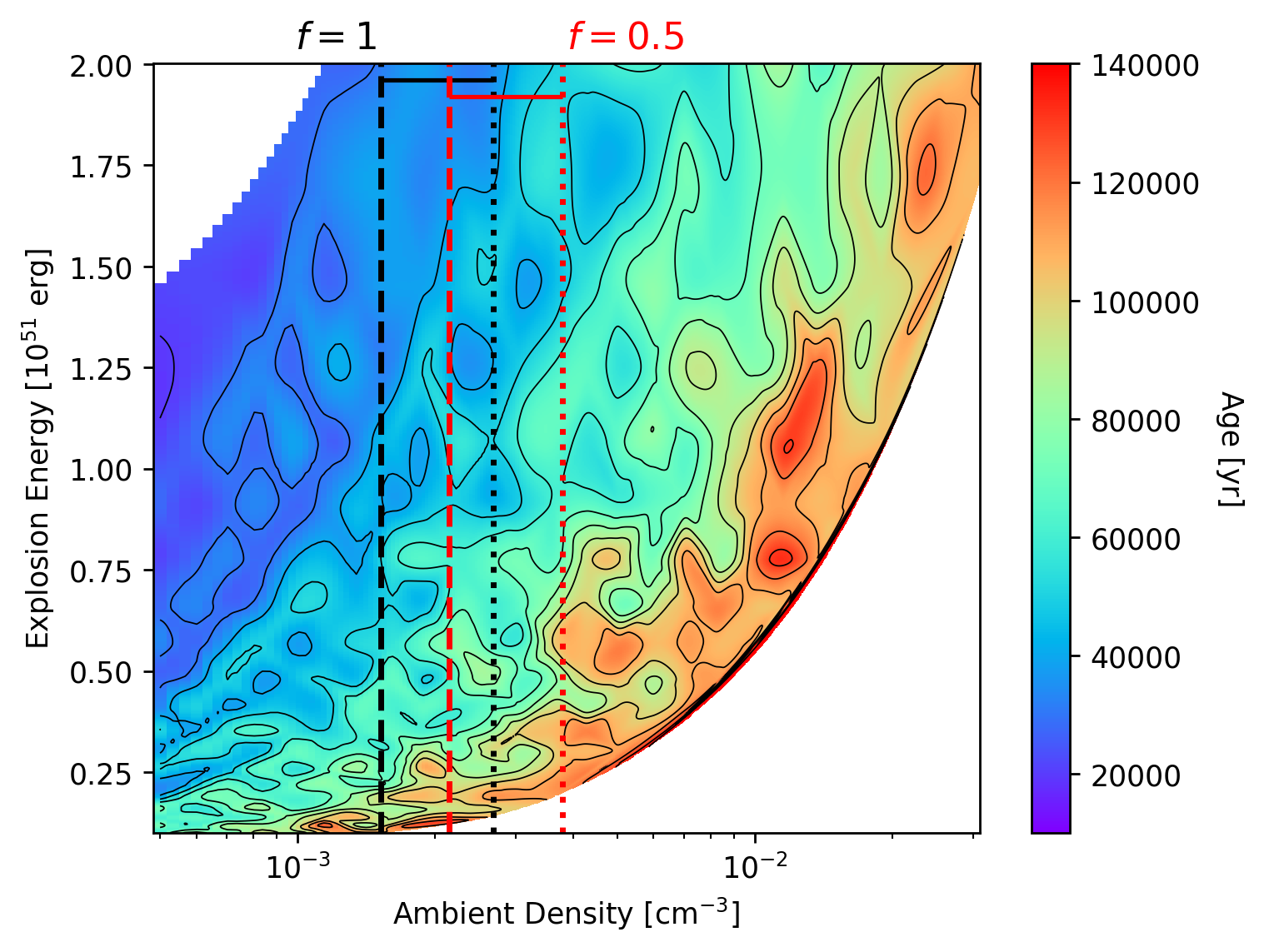}
		\caption{\label{fig:mg_UB_sedtay}SNR model parameters similar to \autoref{fig:mg_snr_estimates}, but for G190.4+12.5, using the Sedov-Taylor model. The assumed distance is $D=1500$\,pc, which yields a radius of 80 pc.}
\end{figure}

\paragraph{SNR candidate G190.4+12.5:}
For this object, we obtain low ambient densities, as shown in the SNR model parameter diagram in \autoref{fig:mg_UB_sedtay}. For the estimated possible ISM densities, we find SNR configurations in the full range of $E_0 = (0.1 - 2.0) \cdot 10^{51}$\,erg, with ages in the range 40.000 yr to 100,000 yr. If this object is indeed an SNR, it would be situated well above the Galactic plane, which explains the low ISM densities we obtained. The almost perfectly circular morphology also suggests that the expansion of the remnant appeared to be freely and relatively uniform. A younger age of $40,000-60,000$\,yr is favored, while the higher ages are found only for a very narrow parameter range at very low explosion energies. Due to the large uncertainties on the X-ray spectral fit parameters, we are only able to give a lower limit on the distance of $> 1.5\,$kpc for this SNR candidate. While the $N_\mathrm{H}$ lower limits discussed in \autoref{sec:UB_spec_ana} hint toward a distance of $> 2$\,kpc, the $A_v$ data is very limited at large distances, especially further away from the Galactic plane. Above $2-3\,$kpc, the geometric radius becomes too large while also requiring unrealistically small ISM densities.

\begin{figure}
	\centering
		\includegraphics[width=0.49\textwidth]{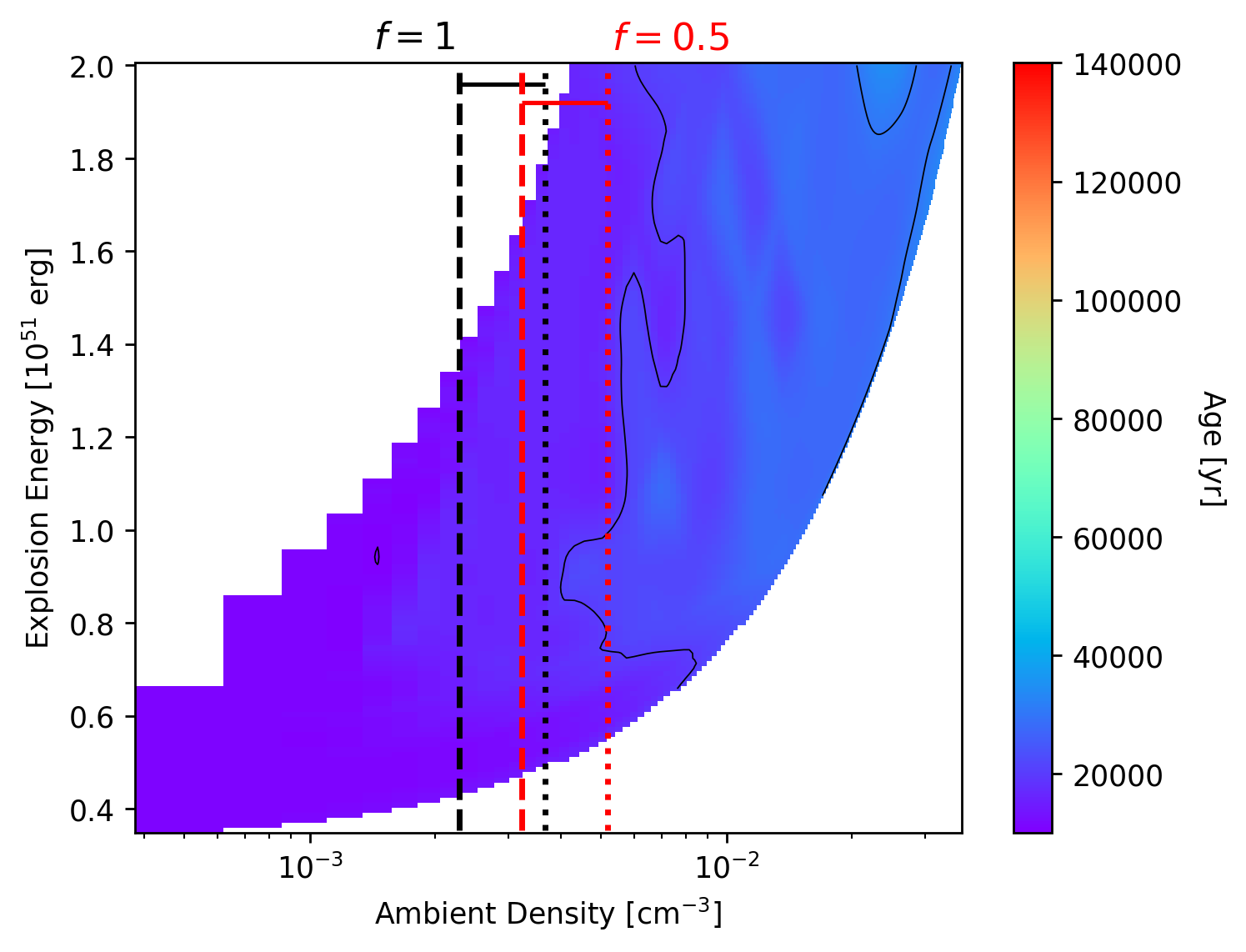}
		\caption{\label{fig:MC_sedtay}SNR model parameters similar to \autoref{fig:mg_snr_estimates}, but for the Monoceros Loop SNR cadidate, using the Sedov-Taylor model. The assumed distance is $D=1500$\,pc, which yields a radius of 48 pc.}
\end{figure}

\paragraph{Monoceros Loop:}
The possible models for the Monoceros Loop SNR are shown in \autoref{fig:MC_sedtay} for the distance $D \approx 1.5$\,kpc. We adopted this distance from the study by \citet{einstein_monoceros_loop_02}. Other studies suggest a distance of $\sim 1.6\,$kpc \citep{MC_distance, MC_radio}, however, within uncertainties this does not change the interpretation of our results. While the density appears to be low, the radio study by \citet{MC_radio} suggests, that the SN explosion might has taken place in a low-density cavity, created by several massive stars. We obtain valid models in the range $10,000-33,000$\,yr, with and age of $\sim 15,000$\,yr being most consistent with the estimated density, and explosion energies of at least $> 0.5\cdot 10^{51}\,$erg. This is roughly in agreement with previous studies that obtained an age of $\sim 30,000$\,yr \citep{einstein_monoceros_loop, MC_age_range}. We were able to disentangle the plasma emission in the line of sight into distinct components, thus getting a more accurate estimation on the properties of the X-ray emitting plasma and reducing the contamination of the soft foreground emission of the Monogem Ring. 
\paragraph{PKS 0646+06:} For this remnant, we assumed a distance of 4\,kpc for the SNR estimations. Previous studies suggest a range from $3-5\,$kpc \citep{MC_distance, monoceros_loop_radio}. We find relatively low explosion energies of $(0.2-0.4)\cdot 10^{51}$\,erg. The remnant appears to be in the Sedov phase with ages ranging $15,000-20,000\,$yr. The previous X-ray study by \citet{pks_einstein} found a much higher age of $\sim 60,000\,$yr, most likely caused by modeling primarily the contaminating foreground emission of the Monogem Ring with $kT \approx 0.14\,$keV in their analysis. A study based on optical line emission by \citet{PKS_age} estimate a lower distance of $\sim 2.2\,$kpc and an age of $6.4\cdot 10^{4}\,$yr. However, they require a very low explosion energy of $1.7\cdot 10^{49}\,$erg and derive a much lower shock velocity compared to our results from X-rays. Instead, this indicates that the remnant might interact with dense cold clouds in the vicinity. Both the low explosion energy together with a high age are rejected by our analysis.
\begin{figure}
	\centering
		\includegraphics[width=0.49\textwidth]{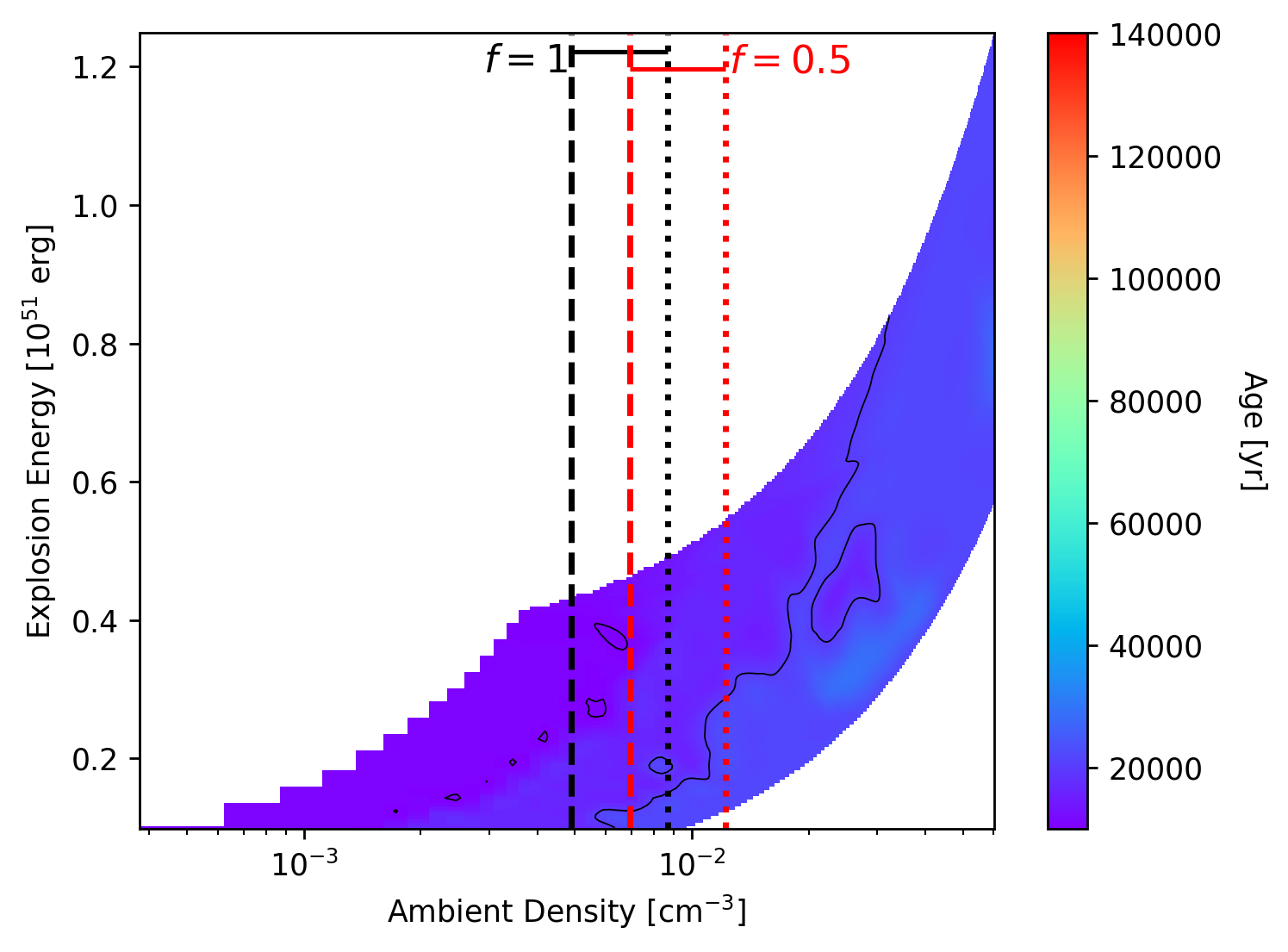}
		\caption{\label{fig:PKS_sedtay}SNR model parameters similar to \autoref{fig:mg_snr_estimates}, but for PKS 0646+06, using the Sedov-Taylor model. The assumed distance is $D=4000$\,pc, which yields a radius of 29 pc.}
\end{figure}
\section{Summary}
\label{sec:summary}
In this work, we performed a morphological and spectral analysis of the entire Gemini-Monoceros X-ray enhancement for the first time, using novel eROSITA data of the first four all-sky surveys. We performed a careful background analysis across the entire field to create a predictive background model for all source regions. The source regions were defined both statistically, as well as manually, and we extract and modeled spectra for each of the regions, obtaining detailed plasma model parameter maps with fine spatial resolution. Our most important results are as follows:
\begin{itemize}
\item Most of the diffuse emission attributed to the Monogem Ring can be described well with an unabsorbed thermal plasma in NEI with a low average temperature of $kT \sim 0.14\pm 0.03\,$keV. However, in the regions in the southeast (green arc, \autoref{fig:finding_chart}) of the Monogem Ring, a second, hotter plasma component is necessary to achieve good spectral fits. This component has a temperature of $kT > 0.28\,$keV, in addition to the low-temperature component of the Monogem Ring. The morphology indicates the presence of an arc-like temperature enhancement with harder (green) X-ray color, which is reproduced by the high temperature component when extending the two-plasma model to the entire field. Although statistics are limited, the abundances for the southeastern hot plasma are typical for a type Ia supernova, while the colder plasma component is consistent with solar abundances. A dust absorption features at $600-900\,$pc appears to be anticorrelated with the X-ray emission in the center. However, we found no significant $N_\mathrm{H}$ in the spectral fits, putting an upper limit of $D < 600\,$pc on the plasma.
\item We showed, that a second remnant at roughly the same distance as the Monogem Ring SNR ($D=300-500\,$pc) explains our results well, with the designated name G205.6+12.4. In this configuration, the center of the Monogem Ring would be shifted in a direction, which would lead to a more favorable agreement between the proper motion of PSR B0656+14 and the Monogem Ring SNR center. A second remnant would also explain the different plasma conditions near the ``green arc`` in the southeast. This candidate would have an age of $40,000-85,000\,$yr and explosion energy of $0.2-0.6\cdot 10^{51}\,$erg, depending on the distance. No compact object was found near the approximate center in the given distance range.
\item For the Monogem Ring, if we assume a single SNR we also obtain an age similar to the age of PSR B0656+14 and explosion energies typical for a CC SN, albeit on the lower end. However, with the presentet results, a scenario with two supernova remnants at $D\approx 300\,$pc, that is, the Monogem Ring and G205.6+12.4, is favored. With this new geomentry, we estimate an age of $\sim 1.1-1.2\cdot 10^5$\,yr for the Monogem Ring.
\item We found a new SNR candidate at $(l, b) \sim (190.4^{\circ}, 12.5^{\circ})$ with a radius of $\sim 3^{\circ}$, that shows up as an almost perfect circular diffuse structure for X-ray images with $E > 0.4\,$keV. We obtained significant foreground absorption to the plasma, and a temperature of $\sim 0.4\,$keV, with the plasma being in NEI. For a distance of $\sim 1.5\,$kpc we estimate an age of $\sim 40,000-60,000$\,yr. Due to relatively high uncertainties of the spectral fit results, we are able to constrain the explosion energy only to a lower limit of $> 0.2\cdot 10^{51}\,$erg.
\item We analyzed the Monoceros Loop and PKS 0646+06 SNRs and improved on previous studies with the new data. We were able to disentangle the X-ray emission in the line-of-sight into the low-temperature foreground emission by the Monogem Ring and higher-temperature emission by the SNRs. Previous studies did not take into account the contamination by the foreground emission, therefore, we obtained significantly higher temperatures for the SNRs' plasma. For the Monoceros Loop, we estimate a ``young'' age of $\sim 15,000$\,yr and an explosion energy of $>0.5\cdot 10^{51}\,$erg. The expansion appears to take place into a low-density environment, consistent with previous studies.
For the more distant PKS 0646+06 we find an age of only $15,000-20,000\,$yr, much lower compared to previous studies, and the remnant appears to be in the Sedov phase.
\end{itemize}
If the new SNR candidate G205.6+12.4 is confirmed, the association of the Monogem Ring SNR with PSR B0656+14 is strengthened. Our study provides the most detailed spectral analysis of the diffuse emission so far.
We studied the highly asymmetric morphology of the SNR and found a high temperature enhancement to the southeast. We found indications of a possible second SNR at a similar distance than the Monogem Ring SNR, both from the spectral analysis and multiwavelength data. This SNR candidate, dubbed G205.6+12.4, appears to have a similar explosion energy and an age of $40,000-55,000$\,yr, about half the age compared to the Monogem Ring if located at $D=300\,$pc. A distance beyond $D > 600$\,pc is rejected by our analysis. However, statistics are low to definitively claim two SNRs instead of only the "Monogem Ring". Additional deeper observations and multiwavelength studies are required to confirm or reject the newly discovered SNR candidate. 

Moreover, we analyzed the Monoceros Loop SNR and PKS 0646+06 for the first time with CCD resolution spectra and an accurate background model, which revealed different properties than derived in previous studies.
We also found a new SNR candidate (G190.4+12.5) at high Galactic latitudes and distances $> 1.5\,$kpc located at the edge of the Monogem Ring. While our estimation accuracy for this candidate is low, the energetics are consistent with a SN explosion. The remnant appears to expand into a very low density ambient medium with an age of $\sim 40,000-60,000$\,yr. Further dedicated pointed observations of this object might help to confirm this object as an SNR and determine its properties with higher precision.

The new faint old SNR candidates could explain part of the ``missing SNR'' problem in our Galaxy if more faint SNRs are discovered in the future. As concluded by \citet{snr_density_leahy}, even for low distances $D < 1\,$kpc around the Sun, about half of the expected SNRs are missing. Since those objects are too faint in the radio, full-sky X-ray surveys similar as performed by eROSITA might be able to explain the discrepancy for the nearby SNR density in the future.
%
\section*{Acknowledgments}
We thank the anonymous referee for their helpful and constructive comments.
We thank N. Locatelli and M. Sormani for providing the re-projected mass profiles of the Milky Way. We also thank Jeremy Sanders for helpful advice in applying the contour binning method. 
This work is based on data from eROSITA, the soft X-ray instrument aboard \textit{SRG}, a joint Russian-German science mission supported by the Russian Space Agency (Roskosmos), in the interests of the Russian Academy of Sciences represented by its Space Research Institute (IKI), and the Deutsches Zentrum für Luft- und Raumfahrt (DLR). The \textit{SRG} spacecraft was built by Lavochkin Association (NPOL) and its subcontractors, and is operated by NPOL with support from the Max Planck Institute for Extraterrestrial Physics (MPE). The development and construction of the eROSITA X-ray instrument was led by MPE, with contributions from the Dr. Karl Remeis Observatory Bamberg \& ECAP (FAU Erlangen-Nuernberg), the University of Hamburg Observatory, the Leibniz Institute for Astrophysics Potsdam (AIP), and the Institute for Astronomy and Astrophysics of the University of Tübingen, with the support of DLR and the Max Planck Society. The Argelander Institute for Astronomy of the University of Bonn and the Ludwig Maximilians Universität Munich also participated in the science preparation for eROSITA. The
eROSITA data shown here were processed using the eSASS/NRTA software
system developed by the German eROSITA consortium. 
Parts of the acknowledgements were compiled using the Astronomy Acknowledgement Generator. This research has made use of data and/or software provided by the High Energy Astrophysics Science Archive Research Center (HEASARC), which is a service of the Astrophysics Science Division at NASA/GSFC and the High Energy Astrophysics Division of the Smithsonian Astrophysical Observatory. This research has made use of the VizieR catalogue access tool, CDS, Strasbourg, France. This research made use of Astropy, a community-developed core Python package for Astronomy \citep{2018AJ....156..123A, 2013A&A...558A..33A} This research made use of matplotlib, a Python library for publication quality graphics \citep{Hunter:2007} This research made use of NumPy \citep{harris2020array} This research made use of XSPEC \citep{1996ASPC..101...17A} This research made use of ds9, a tool for data visualization supported by the Chandra X-ray Science Center (CXC) and the High Energy Astrophysics Science Archive Center (HEASARC) with support from the JWST Mission office at the Space Telescope Science Institute for 3D visualization. Some of the results in this paper have been derived using the healpy and HEALPix packages.

This work was supported by the Deutsche Forschungsgemeinschaft through the project
SA 2131/13-1.
M.S. acknowledges support from the Deutsche Forschungsgemeinschaft through the grants SA 2131/14-1 and SA 2131/15-1.  
GP acknowledges financial support from the European Research Council (ERC) under the European Union's Horizon 2020 research and innovation program ``HotMilk'' (grant agreement No. 865637) and support from Bando per il Finanziamento della Ricerca Fondamentale 2022 dell'Istituto Nazionale di Astrofisica (INAF): GO Large program and from the Framework per l’Attrazione e il
Rafforzamento delle Eccellenze (FARE) per la ricerca in Italia
(R20L5S39T9).

\bibliographystyle{aa}
\bibliography{literature}
\begin{appendix}
\section{Alternative geometry with a single SNR for the Monogem Ring}
\begin{figure}
	\centering
		\includegraphics[width=0.49\textwidth]{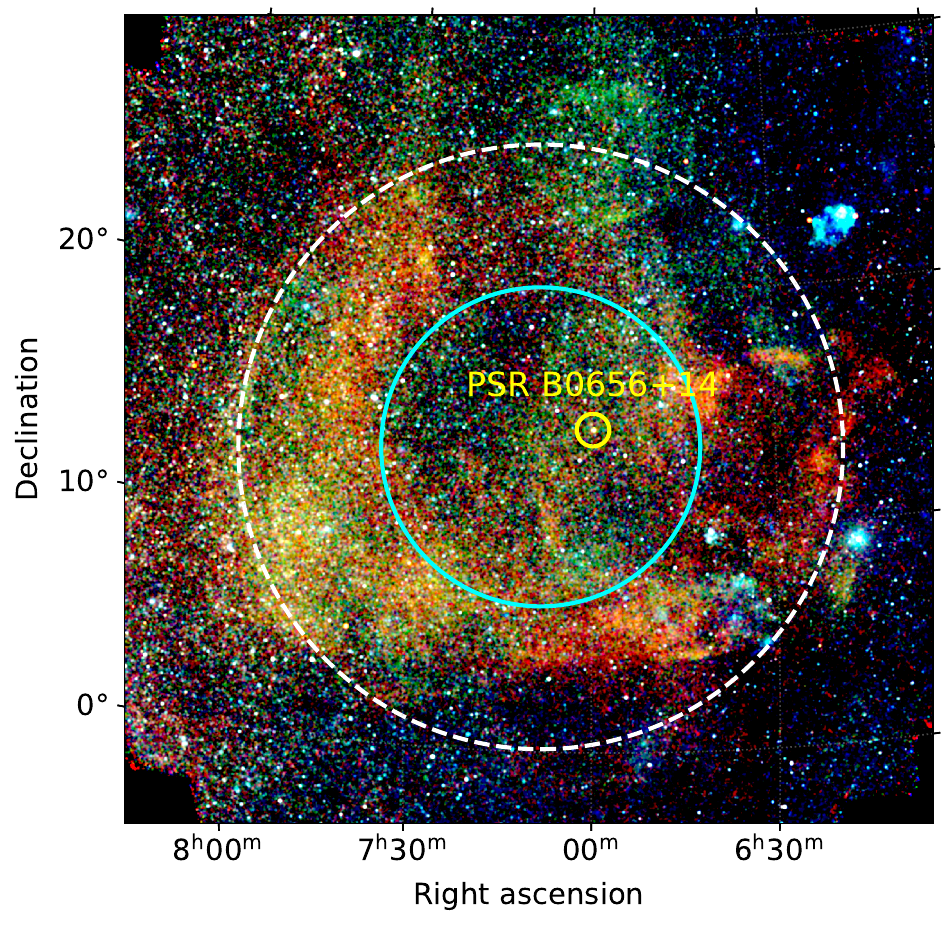}
		\caption{\label{fig:new_geometry_single_snr}eROSITA X-ray mosaic as in \autoref{fig:xray_rgb_mosaic}. The estimated shell extent of the Monogem Ring SNR is shown in dashed white, if assuming a single structure at $D=300\,$pc. The cyan circle highlights the darker interior of the remnant which was excluded for estimating the SNR properties, therefore only using the limb-brightened shell in an annulus. The current position of PSR B0656+14 is shown in yellow.}
\end{figure}
\section{Alternative CGM model}
\label{sec:alternative_bkg}
From the results of background analysis with ``Model B,'' we obtained normalizations of about one order of magnitude higher near the disk for the CGM component. While a certain increase in flux toward the disk is expected for a Galactic component, such a strong increase might hint toward inaccuracies in the underlying assumptions of the model. One possible explanation could be, that the high column density measured in the HI4PI \citep{hi4pi} toward the Galactic disk only partially affects the CGM component.

Since the exact nature of both the CGM and hot corona are not yet fully understood, we therefore introduced individual absorption models (\textsc{tbabs}) to both components. They were let free to fit, with the upper bound fixed to the HI4PI, since it should be impossible to obtain a stronger absorption. Using this modified model, the absorption for the hot corona was almost exclusively fit with the HI4PI value, while the CGM was fit with significantly lower absorption compared to the HI4PI values in the disk ($b \pm 10 ^{\circ}$). Therefore, we adopted the full absorption value reported from the HI4PI survey for both the hot corona and CXB, using a single TBabs like before, and a separate \textsc{tbabs} component for the CGM. In the following fits, we initially set the CGM absorption component to half of the HI4PI value, with an upper bound equal to the HI4PI value.

In addition, the modification to the CGM caused the ACX2 component normalizations to be consistent with zero for most regions. While the other model component at lower energies - the LHB component - appears to remain the same between the different models, there appears to be a degeneracy between the ACX2 model and the CGM component, if we assume it to be only weakly absorbed. Due to enhanced solar activities during eRASS 3 and 4, we expect at least a moderate amount of contamination by SWCX in the combined eRASS:4 data.
Therefore, we continue our analysis with ``Model B'' instead of this alternative model.
\begin{figure}
	\centering
		\includegraphics[width=0.49\textwidth]{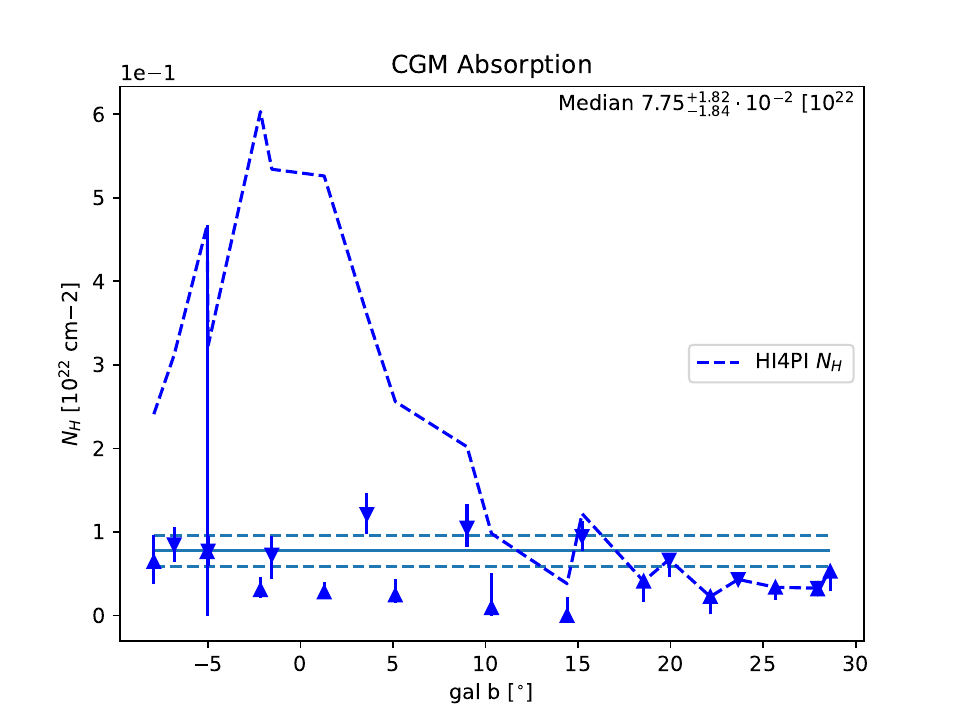}
		\caption{\label{fig:cgm_abs}CGM specific foreground absorption. We decoupled the absorption component from the Galactic $N_\mathrm{H}$ absorption in the spectral model. The solid lines show the median, while the dashed lines show the upper/lower limit median (90\% CI).}
\end{figure}
\section{Spectral fitting routine for source regions}
\label{sec:spectral_fitting_routine_source}
The detailed fitting routine of the source regions, both for the manual regions and contour bins, was performed as follows.
Similar to the background, we fit the FWC normalization constants from the hard-energy tail of the spectrum, before modeling the spectrum in the energy range $0.2-3.0$\,keV. The column density of the background \textsc{tbabs} component was fixed to the HI4PI survey value \citep{hi4pi}. Initial values for the \textsc{nei} component were chosen as $kT = 0.25$\,keV, fixed solar abundance, the ionization timescale $\tau = 1\cdot 10^{11}$ s cm$^{-3}$ and normalization of $\eta = 1\cdot 10^{-7}$ cm$^{-5}$\,arcmin$^{-2}$. We also used the cstat fit statistic for the source regions. If the threshold of $\text{cstat}/\text{d.o.f.} > 1.14$ was exceeded after the initial fit, we increased the background parameter uncertainties by $10\%$ and repeated the fit. In a last step, we checked the source component parameters for unrealistically high or low values and reset the respective parameters before performing another fit.

After the final fit, we calculated the uncertainties of the free-to-fit source component parameters. In order to reduce the computation time due to unnecessary free parameters, we froze the normalization constants for the FWC and TM calibration, since they are in any case constrained to a very narrow range.
We first attempted to calculate the uncertainties by using the \textsc{error} command for each parameter. If the diagnostic error string indicated problems, we instead calculated the uncertainties with the \textsc{steppar} command in an iterative automatic way. The initial \textsc{steppar} calculation was performed in an interval ($90\%-110\%$) of the best-fit value, using 20 steps. If no appropriate $90\%$ confidence interval limit was found, the calculation range was increased in the respective - or both - direction(s). This procedure was repeated up to a maximum of ten times. For the $\tau$ and \textsc{tbabs} $N_\mathrm{H}$ parameters, the relative change was chosen higher. Additionally, for $\tau$ and $N_\mathrm{H}$, the error was calculated with a logarithmic grid. 
We used this approach for both the contour bins and the manually defined large regions.

In addition, because we have higher statistics in the manually defined regions, we systematically tested SNR-characteristic elemental abundances. After the final fit with solar abundances, we tested the following elements: O, Ne, Mg, Si, S, and Fe. The elements were fit simultaneously. Next, we checked for unrealistically low or high values, and reset those elements back to solar abundances. Then, we performed another fit. Those two steps were repeated until all elements that were free to fit had realistic values. We confirmed if the free-to-fit abundances improved the model above the $3\sigma$ significance level, using the \textsc{ftest} command, and if not, reset the model to solar values. Again, we calculated the uncertainties as described above.
\section{Manually defined regions: Single component results}
\label{sec:manually_defined_analysis}
\begin{table*}
	\caption{\label{tab:abundance_fit_results}Spectral fit results of the manually defined regions.}
\renewcommand{\arraystretch}{1.25}
	\begin{center}
	\begin{adjustbox}{width=\textwidth,center}
			\begin{tabular}{l r r r r r r r}
					Region & $N_\mathrm{H}$& $kT$ & $\tau$ & O & Si & norm & cstat/dof \\
					& [$10^{22}$ cm$^{-2}$] & [keV] & [s/cm$^3$] & [solar] & [solar] & [cm$^{-5}$\,arcmin$^{-2}$] & \\ 
					\hline
					arc\_E & $-$ & $0.126_{-0.013}^{+0.011}$ & $(0.47_{-0.24}^{+0.37})\cdot 10^{11}$ & - & - & $(0.79_{-0.17}^{+0.17})\cdot 10^{-6}$ & 1.09 \\
					arc\_N1 & $-$ & $0.080_{}^{+0.006}$ & $(297.84_{-289.48}^{+202.16})\cdot 10^{11}$ & - & - & $(1.91_{-0.26}^{+0.11})\cdot 10^{-6}$ & 1.13 \\
					arc\_N2 & $-$ & $0.081_{-0.001}^{+0.013}$ & $(4.90_{-3.34}^{+18.70})\cdot 10^{11}$ & - & - & $(1.43_{-0.27}^{+0.34})\cdot 10^{-6}$ & 1.03 \\
					arc\_NE & $-$ & $0.090_{-0.090}^{+0.009}$ & $(3.98_{-2.19}^{+496.02})\cdot 10^{11}$ & - & - & $(1.25_{-0.32}^{+0.59})\cdot 10^{-6}$ & 1.09 \\
					arc\_S & $-$ & $0.149_{-0.009}^{+0.010}$ & $(0.43_{-0.15}^{+0.23})\cdot 10^{11}$ & - & - & $(0.73_{-0.10}^{+0.11})\cdot 10^{-6}$ & 1.16 \\
					arc\_SE & $-$ & $0.361_{-0.041}^{+0.040}$ & $(0.42_{-0.10}^{+0.18})\cdot 10^{11}$ & - & $5.650_{-0.998}^{+1.190}$ & $(0.15_{-0.02}^{+0.03})\cdot 10^{-6}$ & 1.38 \\
					arc\_outer\_E & $-$ & $0.080_{}^{+0.011}$ & $(7.02_{-5.03}^{+493.00})\cdot 10^{11}$ & $3.670_{-0.729}^{+0.330}$ & - & $(0.87_{-0.33}^{+0.09})\cdot 10^{-6}$ & 1.08 \\
					arc\_outer\_NE & $-$ & $0.127_{-0.007}^{+0.006}$ & $(285.84_{-272.90}^{+214.16})\cdot 10^{11}$ & - & - & $(1.27_{-0.10}^{+0.07})\cdot 10^{-6}$ & 1.10 \\
					central & $-$ & $0.080_{}^{+0.008}$ & $(3.72_{-1.93}^{+3.14})\cdot 10^{11}$ & - & - & $(0.75_{-0.19}^{+0.07})\cdot 10^{-6}$ & 1.14 \\
					central\_S1 & $0.003_{-0.003}^{+0.007}$ & $0.136_{-0.014}^{+0.015}$ & $(0.59_{-0.17}^{+0.28})\cdot 10^{11}$ & - & - & $(0.59_{-0.11}^{+0.07})\cdot 10^{-6}$ & 1.13 \\
					central\_S2 & $-$ & $0.144_{-0.013}^{+0.005}$ & $(2.79_{-0.69}^{+0.92})\cdot 10^{11}$ & - & - & $(1.04_{-0.08}^{+0.12})\cdot 10^{-6}$ & 1.07 \\
					central\_W1 & $-$ & $0.122_{-0.012}^{+0.008}$ & $(1.36_{-0.59}^{+1.42})\cdot 10^{11}$ & - & - & $(0.84_{-0.15}^{+0.16})\cdot 10^{-6}$ & 1.10 \\
					central\_W2 & $-$ & $0.153_{-0.010}^{+0.014}$ & $(0.31_{-0.13}^{+0.22})\cdot 10^{11}$ & - & - & $(0.50_{-0.09}^{+0.11})\cdot 10^{-6}$ & 1.06 \\
					central\_W\_halpha & $-$ & $0.140_{-0.014}^{+0.031}$ & $(0.63_{-0.36}^{+0.86})\cdot 10^{11}$ & - & - & $(0.53_{-0.05}^{+100})\cdot 10^{-6}$ & 0.99 \\
					circular\_halpha\_N & $-$ & $0.120_{-0.004}^{+0.007}$ & $(1.61_{-0.40}^{+1.24})\cdot 10^{11}$ & - & - & $(1.04_{-0.10}^{+0.13})\cdot 10^{-6}$ & 1.14 \\
					circular\_halpha\_S & $-$ & $0.136_{-0.017}^{+0.005}$ & $(1.57_{-0.55}^{+0.84})\cdot 10^{11}$ & - & - & $(0.83_{-0.11}^{+0.09})\cdot 10^{-6}$ & 1.17 \\
					circular\_hot\_spot & $-$ & $0.109_{-0.010}^{+0.003}$ & $(1.65_{-0.41}^{+1.27})\cdot 10^{11}$ & - & - & $(2.75_{-0.24}^{+0.65})\cdot 10^{-6}$ & 1.10 \\
					circular\_hot\_spot\_N & $-$ & $0.110_{-0.017}^{+0.011}$ & $(1.09_{-0.63}^{+1.88})\cdot 10^{11}$ & - & - & $(1.54_{-0.39}^{+0.83})\cdot 10^{-6}$ & 1.11 \\
					diffuse\_S & $-$ & $0.122_{-0.008}^{+0.006}$ & $(1.02_{-0.25}^{+0.34})\cdot 10^{11}$ & - & - & $(1.39_{-0.05}^{+0.40})\cdot 10^{-6}$ & 1.14 \\
			\end{tabular}
	\end{adjustbox}
	\tablefoot{
		The source model is \textsc{tbabs~$\times$~vnei}. The abundances were let free to fit and unconstrained or consistent with zero abundances were reset back to solar, as outlined in the text. The uncertainties are equal to the $90\%$ confidence interval (CI). 
	}
	\end{center}
\end{table*}
%
For the manually defined regions we used the model described in \autoref{eq:model_expr_source} ( \textsc{tbabs}~$\times$~\textsc{vnei}) for the source. The fit results are shown in \autoref{tab:abundance_fit_results}. After an initial fit with solar abundances, we used the procedure described in \autoref{sec:spectral_fitting_routine_source} to obtain possible nonsolar abundances. We show examples of the spectra and best-fit models in \autoref{fig:manual_example_spectra} for some of the more interesting regions.

With this model, we obtain very good fits for most regions, with cstat/d.o.f. ranging $\sim 0.99-1.17$, however, for ``arc\_SE'' we obtain only $1.38$. The spectra for this region is shown in \autoref{fig:example_arc_SE_one_apec}. 
For the foreground absorption by the \textsc{tbabs} component, we obtain values consistent with zero within uncertainties except for the region ``central S1,'' however, with small upper limits of only $\sim 10^{20}$\,cm$^{-2}$. The temperature appears to be consistent with the contour bin analysis results: low temperatures of $kT \sim 0.08-0.15$\,keV. In the region "arc\_SE`` we obtain a temperature higher by a factor of about two $kT \sim 0.36$\,keV. Interestingly, this feature is also obtained with the contour bin method for bins in the vicinity of this region. The normalization of the diffuse plasma component range $\eta \sim 0.5-2.8 \cdot 10^{-6}$\,cm$^{-5}$\,arcmin$^{-2}$. The two regions ``circular hot spot'' and ``circular hot spot N'' have one of the highest normalizations among all regions, together with a low temperature $kT \sim 0.11-0.12$\,keV, which explains the bright red morphology observed in \autoref{fig:xray_rgb_mosaic}.

For only two regions, we obtain nonsolar abundances with our initial approach. In the high-temperature region ``arc\_SE'' we obtain a significantly enhanced Si abundance of $5.65^{+1.19}_{-1.00}$ and in the ``arc\_outer\_E'' an enhanced O abundance of $3.67^{+0.33}_{-0.73}$, relative to solar. For many of the other regions we also obtain nonsolar abundances as best-fit values, however, with high uncertainties that do not rule out solar abundances. Either the plasma is indeed dominated by solar abundances for most parts by the surrounding ISM, or we lack the necessary statistics to measure the nonsolar abundances.

We also performed a detailed comparison between the contour bin method and the manually defined regions, which is discussed in \ref{sec:comparison_methods}. Both methods agree very well and are complementary to each other.
\section{Comparison of methods}
\label{sec:comparison_methods}
\begin{figure*}
	\centering
	\begin{subfigure}[t]{0.33\textwidth}
		\includegraphics[width=1.0\textwidth]{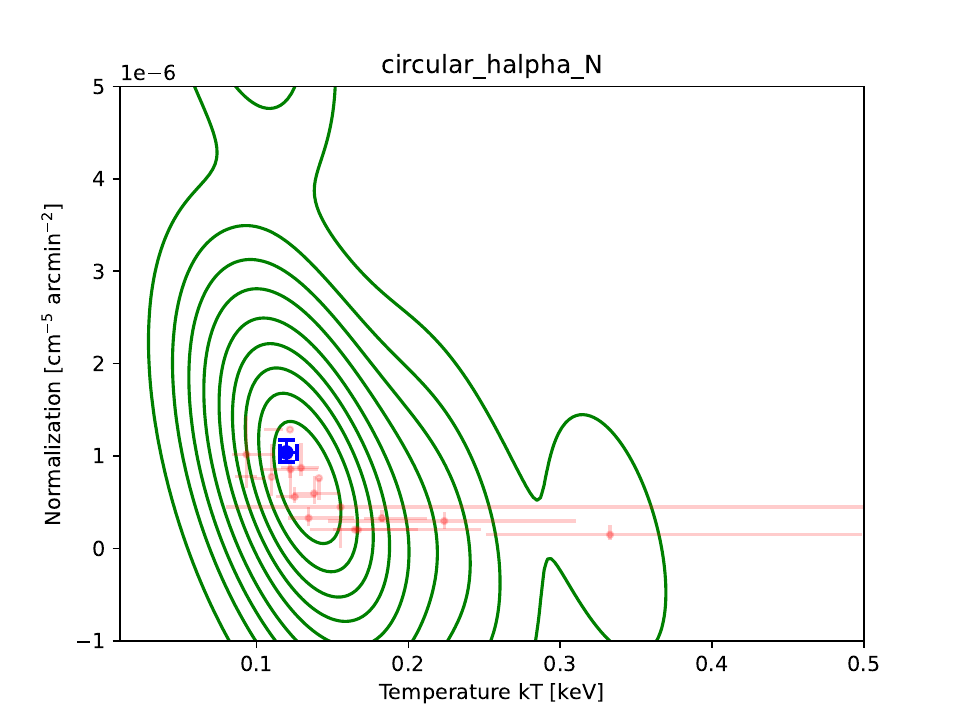}
		\caption{\label{fig:comparison_halpha_N}}
\end{subfigure}
\hfill
\begin{subfigure}[t]{0.33\textwidth}
	\centering
		\includegraphics[width=1.0\textwidth]{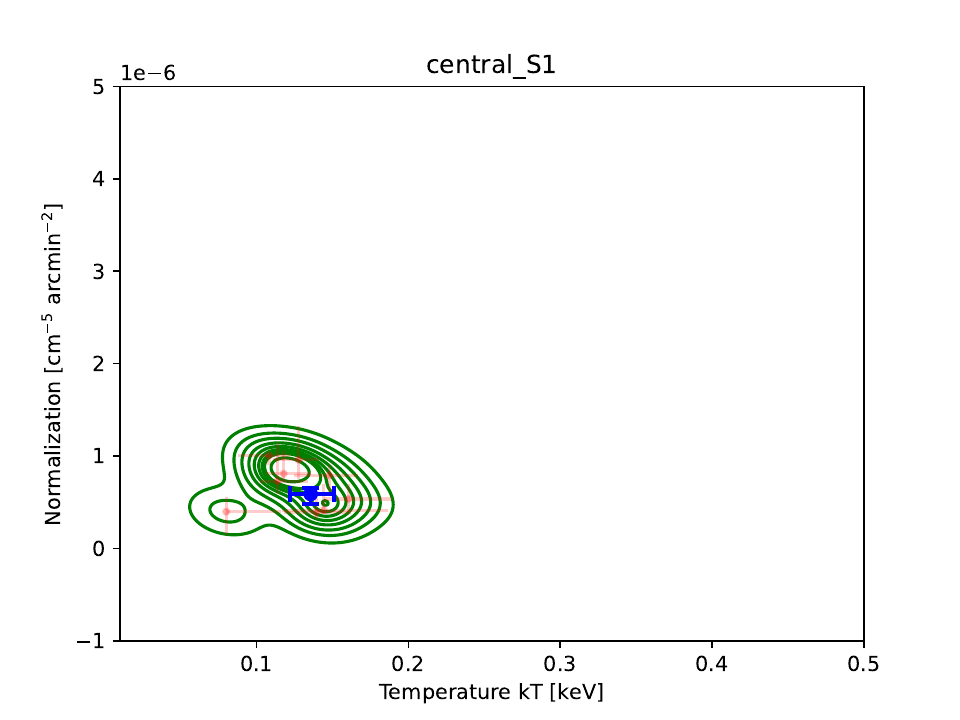}
		\caption{\label{fig:comparison_central_S1}}
		 \end{subfigure}
		 \hfill
		\begin{subfigure}[t]{0.33\textwidth}
		\includegraphics[width=1.0\textwidth]{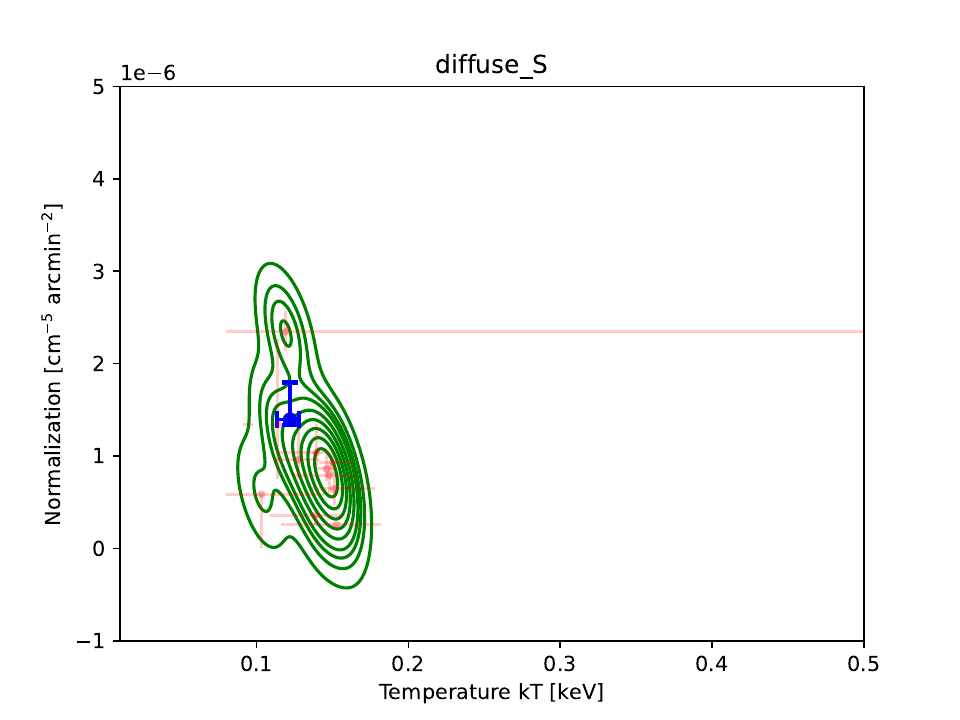}
		\caption{\label{fig:comparison_diffuse_S}}
\end{subfigure}
\\
	\begin{subfigure}[t]{0.33\textwidth}
		\includegraphics[width=1.0\textwidth]{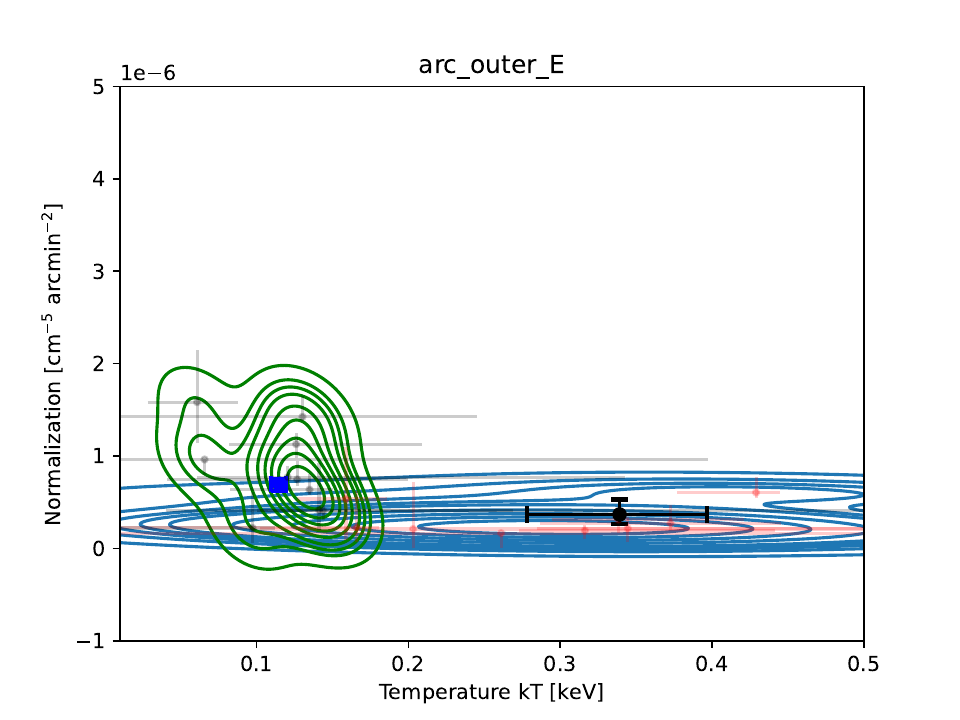}
		\caption{\label{fig:comparison_arc_outer_E}}
\end{subfigure}
\hfill
\begin{subfigure}[t]{0.33\textwidth}
	\centering
		\includegraphics[width=1.0\textwidth]{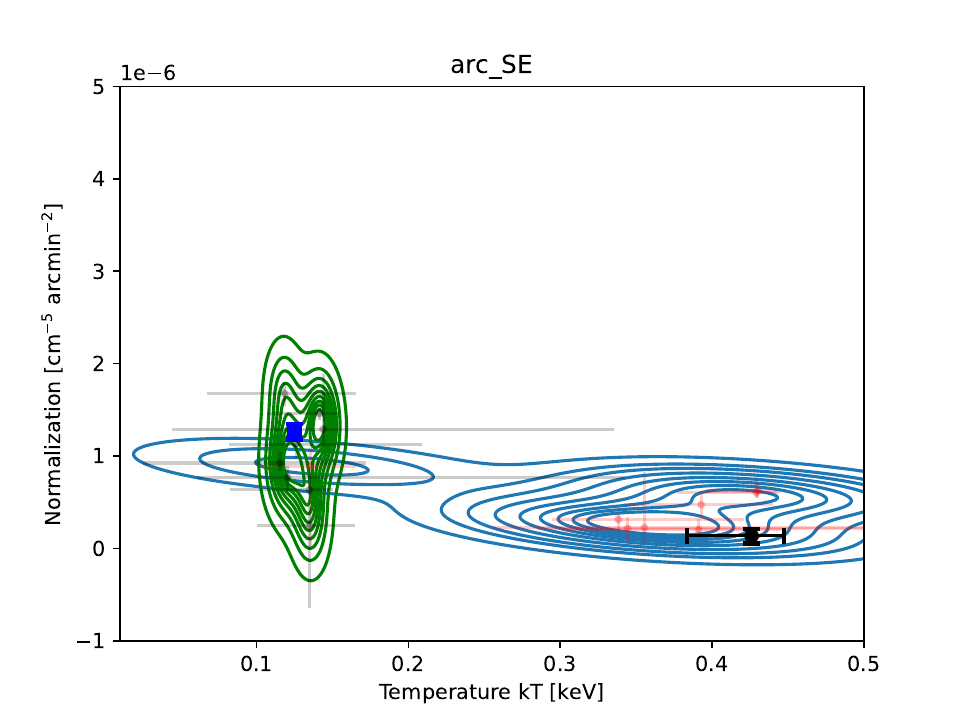}
		\caption{\label{fig:comparison_arc_SE}}
		 \end{subfigure}
		 \hfill
		\begin{subfigure}[t]{0.33\textwidth}
		\includegraphics[width=1.0\textwidth]{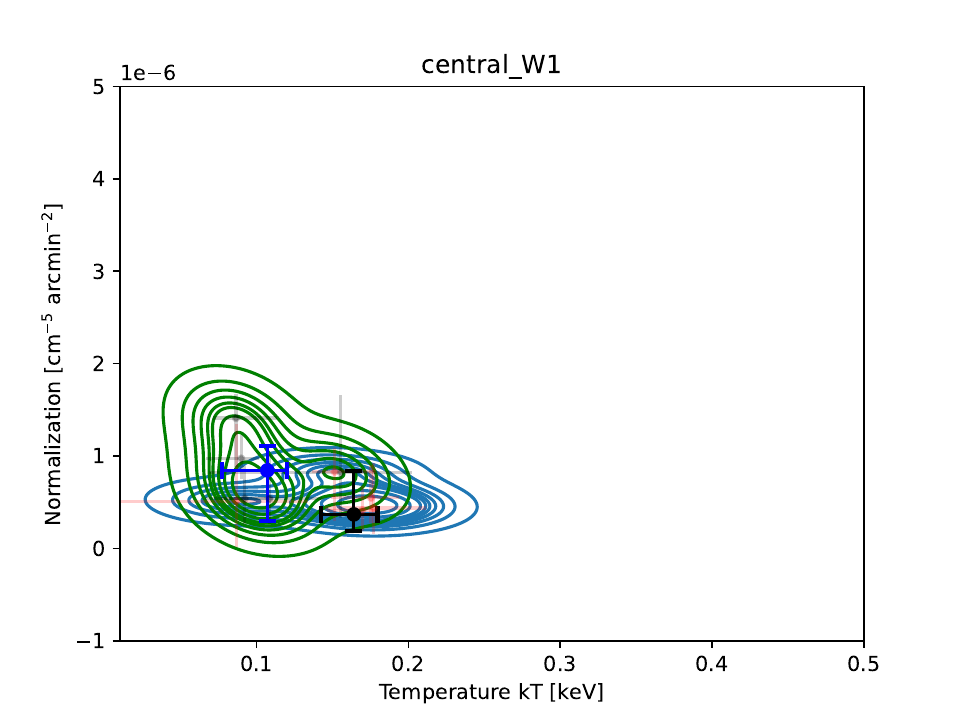}
		\caption{\label{fig:comparison_w1}}
\end{subfigure}
   \caption{\label{fig:comparison_methods}Comparison between the contour bin and manually defined region spectral fits. The contours show the density of the fit values from the contour bins, weighted by the area overlap with the respective manual region. In (a)-(c) we show examples using the simple \textsc{tbabs~$\times$~nei} model. The green contours and red markers show the best-fit values with uncertainties of the contour bins. The fit result from the manually defined region is shown with a blue marker. In (d)-(f) we show the \textsc{tbabs~$\times$~vapec$_1$+apec$_2$} model, where the green contours and black markers show the low temperature \textsc{apec}, and the blue contours and red markers the hotter \textsc{vapec} component best-fit values. The fit results from the manually defined region are shown with a blue and black marker.}
\end{figure*}

In order to asses our spectral analysis methods, we made a direct comparison of the fit results using the same model with both methods, which are contour binning and manually defining the regions. The comparison is shown in \autoref{fig:comparison_methods}. The fit results of the two methods appear to agree remarkably well, independent of the spectral model we used. This comparison also highlights the strengths and weaknesses of both methods. The contour binning allows to determine a distribution of plasma parameters while at the same time the individual uncertainties are higher. Contrary, the manually defined regions basically average over a range of plasma conditions which is demonstrated by the fit result often being located right in the middle of the distribution of plasma parameters we derived with the contour binning. At the same time, the uncertainties are smaller. However, this smaller error hides the underlying wrong physical assumption that we only have a singular plasma (or plasma state). Using both analysis methods, we are able to mitigate the weaknesses of both methods and we obtain a much more complete picture about the plasma in our sources.
\end{appendix}
\end{document}